\documentclass[11pt,letterpaper]{article}
\pdfoutput=1
\usepackage{jcappub}
\usepackage{blkarray, bigstrut}
\usepackage{epsfig,latexsym,cancel,amssymb,amsmath}
\usepackage{hyperref}
\usepackage{graphicx}
\usepackage{caption}
\usepackage{subcaption}
 \usepackage{comment}
\usepackage{color}
\usepackage{bm}
\usepackage[normalem]{ulem}
\usepackage{float}

\def\beq{\begin{equation}}
\def\eeq{\end{equation}}
\def\bea{\begin{eqnarray}}
\def\eea{\end{eqnarray}}

\def\Tcmb{ {T_{\textrm{CMB}}} }
\def\Teff{ {T_{\textrm{eff}}} }

\begin{document}
\title{A strong broadband 21 cm cosmological signal from dark matter spin-flip interactions}
\author[a,c]{Mansi Dhuria,}
\author[b,c]{Viraj Karambelkar,}
\author[c]{Vikram Rentala,}
\author[c]{and Priyanka Sarmah}

\affiliation[a]{Basic Sciences Department, Institute of Infrastructure, Technology, Research And Management, Near Khokhra Circle, Maninagar (East), Ahmedabad-380026, Gujarat, India}
\affiliation[b]{Cahill Center for Astronomy, California Institute of
Technology, 1216 E California Blvd. Pasadena, CA 91106, USA}
\affiliation[c]{Department of Physics, Indian Institute of Technology Bombay, Powai, Mumbai, Maharashtra, 400076, India}
\emailAdd{mansidhuria@iitram.ac.in}
\emailAdd{viraj@astro.caltech.edu}
\emailAdd{rentala@phy.iitb.ac.in}
\emailAdd{sarmahpriyanka07@gmail.com}

\abstract{
In the standard cosmology, it is believed that there are two relatively weak and distinct band-limited absorption features, with the first absorption minima near 20 MHz~($z\sim70$) and the other minima at higher frequencies between $50-110$~MHz ($z\sim 12-27$) in the global cosmological 21~cm signal, which are  signatures of collisional gas dynamics in the cosmic dark ages and \mbox{Lyman-$\alpha$} photons from the first stars at cosmic dawn, respectively. A similar prediction of two distinct band-limited, but stronger, absorption features is expected in models with excess gas cooling, which have been invoked to explain the anomalous EDGES signal. In this work, we explore a novel mechanism, where dark matter spin-flip interactions with electrons through a light axial-vector mediator could directly induce a 21~cm absorption signal which is characteristically different from either of these. We find generically, that our model predicts a strong, broadband absorption signal extending from frequencies as low as 1.4~MHz ($z\sim1000$), from early in the cosmic dark ages where no conventional signal is expected, all the way up to higher frequencies where star formation and $X$-ray heating effects are expected to terminate the absorption signal. We find a rich set of spectral features that could be probed in current and future experiments looking for the global 21~cm signal. In the standard cosmology and in excess gas cooling models it is expected that the gas spin temperature as inferred from the absorption signal is a tracer of the gas kinetic temperature. However, in our model we find in certain regions of parameter space that the spin temperature and kinetic temperature of the gas evolve differently, and the absorption signal only measures the spin temperature evolution. Large swathes of our model parameter space of interest are safe from existing particle physics constraints, however future searches for short range spin-dependent forces between electrons on the millimeter to nanometer scale have the potential to discover the light mediator responsible for our predicted signal.
 }
%\pacs{ }
\maketitle

\section{Introduction}

Cosmic Microwave Background (CMB) radiation serves as a backlight for the neutral hydrogen in the universe along any line of sight. At any redshift $z$, the Rayleigh-Jeans tail of the CMB spectrum has $21$ cm (1420 MHz) photons which can be absorbed by neutral hydrogen in the hyperfine singlet ground state, exciting it to the triplet state. Alternatively, these photons can also trigger stimulated emission via de-excitation of the triplet states to the ground state of hydrogen. One can measure the net absorption or emission from neutral hydrogen at a redshift $z$ by measuring the intensity of background radiation at frequencies $\nu = 1420/(1+z)$~MHz.

The observed intensity of 21 cm photons from any redshift $z$ can be expressed in terms of the differential brightness temperature $\delta T_b(\nu)$. The expected value of $\delta T_b$ can be related to the difference between the spin temperature $T_s(z)$ of neutral hydrogen and the CMB temperature $\Tcmb = 2.73 (1+z)$~K at the corresponding redshift $z$ as~\cite{2012RPPh...75h6901P},
\begin{equation}
\label{eq:brightness}
\delta T_b = 27 \, x_{HI}(z) \left(\frac{1+z}{10}\right)^{1/2} \left(\frac{\Omega_b h^2}{0.022}\right ) \left(\frac{0.14}{\Omega_m h^2}\right )^{1/2} \left( \frac{T_s(z) - \Tcmb(z)}{T_s(z)} \right) \textrm{\, mK}.
\end{equation}
Here, the spin-temperature $T_s$ is defined in terms of the relative populations of the triplet and singlet states, such that $n_1/n_0 = 3 e^{-\frac{\Delta}{T_s}}$, where $ \Delta = (h c)/( k_B \, 21 \, \textrm{cm}) =   69~\textrm{mK} = 5.9 \,\mu\textrm{eV} $ is a temperature/energy scale corresponding to the hyperfine-splitting and $n_1$ and $n_0$ are the number densities of the triplet and singlet states respectively.  In the formula above $\Omega_b$ and  $\Omega_m$ are the present-day cosmological baryon and matter density fractions, respectively, and $x_{HI}(z)$ is the redshift dependent fraction of neutral hydrogen, which is $\sim 1$ during the cosmic dark ages from $z\sim1000$ to $z\sim 10$.

In standard cosmology, the spin temperature is coupled to the CMB temperature and the gas temperature through the relation~\cite{1958PIRE...46..240F,Furlanetto:2006tf},
\begin{equation}
    T_s^{-1} = \frac{{\Tcmb}^{-1} + x_{C}T_{K}^{-1} + x_{\alpha} T_{c}^{-1} } {1 + x_C + x_{\alpha}},
\label{eq:spinT}
\end{equation}
where $T_K$ is the gas kinetic temperature and $T_c$ (which can be taken to be $\sim T_K$) is the ``color temperature'' of Lyman-$\alpha$ (Ly-$\alpha$) photons from the first stars~\cite{Barkana:2004vb}. The effective coupling to $T_K$ due to collisions between the gas molecules is denoted as $x_C$ and the coupling to $T_c$ due to Ly-$\alpha$ photons is denoted as $x_\alpha$.

At very high redshifts $z \gtrsim 200$, the gas is tightly kinetically coupled to the CMB temperature through the residual free electrons and the collisional coupling $x_C$ is dominant -- latching the spin temperature on to the gas temperature, resulting in no net absorption signal. However, near $z\sim 150$ ($\nu \simeq 10$~MHz), the gas decouples from the CMB and begins to cool adiabatically (as $\sim (1+z)^2$, which is faster than the CMB cooling rate which scales as $(1+z)$). For a while, the collisional coupling continues to keep the spin temperature coupled to the gas temperature, and this combined with the cold gas temperature results in an absorption dip in the brightness temperature near $z\sim150$. The collisional coupling becomes less dominant than the coupling to the CMB near $z\sim 50$ ($\nu\simeq30$~MHz), and the spin temperature rises back up to the CMB temperature once again, leading to a distinct absorption dip extending from approximately 10 to 30 MHz. Meanwhile, the gas continues to cool adiabatically. Somewhere between a redshift of $z \sim 10-30$, it is expected that Ly-$\alpha$ photons from the first stars once again couple the spin temperature to the gas temperature through the Wouthuysen-Field effect~\cite{1958PIRE...46..240F, 1952AJ.....57R..31W}, resulting in a second absorption dip. This absorption dip ends due to $X$-ray heating of the gas, which raises its temperature above the CMB temperature, which could also potentially give rise to an emission signal, depending upon the history of reionization due to the first sources.

Thus, there is a robust prediction of two distinct absorption dips in the global 21 cm brightness temperature. The first absorption feature extends from $\nu \sim 10-30$~MHz ($z\sim 46-140$) with a minima near $\nu \sim 20$~MHz ($z\sim 70$) and is due to gas collisional dynamics. This absorption feature is relatively shallow, since the gas has only just started cooling adiabatically at these redshifts. The second feature is expected at higher frequencies, with a minimum between $50-110$~MHz ($z\sim 12-27$) and is due to Ly-$\alpha$ photons from the first stars. This absorption feature is expected to be deeper, since the gas has cooled much more. However, the magnitude of the second absorption dip is also tightly constrained by the gas cooling history and is expected to lead to an absorption dip no stronger than $\delta T_b \sim - 300$~mK at 90 MHz. Both these absorption features are expected to be band-limited, since for the first feature, the collisional coupling turns off shortly after the gas starts cooling adiabatically, and for the second feature, $X$-ray heating is expected to accompany the turn on of the first stars.

While the exact location and depth of the high frequency minima is strongly dependent on the assumptions of the properties of the first sources of Ly-$\alpha$ photons and X-rays, the prediction of two distinct, band-limited, and relatively weak absorption signals is a fairly robust prediction of the standard cosmology\footnote{For a more detailed discussion of the physics of the spin temperature and predicted 21~cm signal in the standard cosmology, we direct the reader to refs.~\cite{Furlanetto:2006tf,Pritchard:2011xb,Barkana2016,Furlanetto2019}.}. The exact astrophysics of the first sources is currently a matter of speculation, however, it is expected that detailed observations of the 21~cm signal will allow us to characterize this physics.

A large number of radio telescope experiments such as EDGES~\cite{Bowman:2012hf}, PRIZM~\cite{doi:10.1142/S2251171719500041}, REACH~\cite{REACH}, SARAS~\cite{Patra_2013,Singh:2017syr}, SCI-HI~\cite{Voytek2014}, BIGHORNS~\cite{Sokolowski_2015} are looking for a global (isotropic, redshift-dependent) 21 cm signal up to a redshift $z \sim 15$, near the epoch of reionization. Other experiments are looking for a signal from higher redshifts going deeper into the cosmic dark ages such as LEDA~\cite{greenhill2012hi} (up to $z \sim 46$) and the proposed satellite constellation DAPPER~\cite{2019arXiv190710853C} (up to $z\sim 80$). Proposals for space-based missions operating on the far side of the moon such as FARSIDE~\cite{burns2019farside}, PRATUSH~\cite{Pratush}, DARE~\cite{Jones_2015}, and a possible future iteration of the pathfinder Netherlands-China Low-Frequency Explorer (NCLE)~\cite{Chen2020Netherlands} would ultimately have the best sensitivity and potential to probe all the way up to $z\sim 1000$ right up till the era of recombination~\cite{2019arXiv190804296K,Burns:2019zia}.

In addition to the global signal, several other experiments are looking for 21~cm anisotropies which are expected to unveil the detailed history of reionization such as LOFAR~\cite{Rottgering:2006ms}, PAPER~\cite{Parsons_2010}, MWA~\cite{Tingay_2013}, 21CMA~\cite{Zheng_2016}, OVRO-LWA~\cite{Eastwood2019The2C}, GMRT~\cite{10.1111/j.1365-2966.2009.14980.x} and HERA~\cite{DeBoer_2017}, with the future to be led by the Square Kilometer Array (SKA)~ \cite{Mellema2013,Koopmans:2015sua}.

The EDGES experiment has already claimed a detection of anomalous absorption in the global 21~cm signal with strength $\delta T_b \simeq - 500$~mK at 78 MHz ($z\simeq17$)~\cite{Bowman:2018yin}, which is twice as deep as the maximum expected absorption in the standard cosmology. This claim led to intense speculation of exotic physics beyond the Standard Model of particle physics that could explain the anomaly. These explanations were based on the following possibilities:  a)~the absorption dip is due to anomalous gas cooling due to interactions with dark matter ~\cite{Barkana:2018lgd,Munoz:2018pzp,Berlin:2018sjs,Barkana:2018cct,Kovetz:2018zan,Jia:2018mkc,Klop:2018ltd,
Liu:2019knx,PhysRevD.100.023528,Houston:2018vrf,Sikivie:2018tml,Johns:2020mmo}, b) there is a modification of the Rayleigh-Jeans portion of the CMB spectrum from the standard expectation~\cite{Pospelov:2018kdh,Moroi:2018vci,Fraser:2018acy,Bondarenko:2020moh,Brahma:2020tmk} or c) the free electron fraction is smaller than expected and thus the gas decouples from the CMB earlier, leading to a longer epoch of adiabatic cooling~\cite{Falkowski:2018qdj}.

All these models still predict the same \textit{band-limited} (albeit stronger) absorption features in the global signal, due to transition features which are still controlled by the same collisional gas dynamics and Ly-$\alpha$ couplings as in the standard cosmology.

However, it has been debated whether the observed EDGES signal is really a signature of the first stars or is a spurious detection. In ref.~\cite{Hills:2018vyr}, the authors have questioned whether the EDGES observation can be claimed to be an unambiguous detection of a cosmological 21 cm signal by arguing that attempts to fit the purported signal lead to unphysical fit parameters for the foreground emission model.

%\vspace{4 mm}

In light of these experiments (and whether or not the EDGES signal is really cosmological), it is worthwhile to ask if there are alternative predictions of the cosmological global 21~cm signal that could be tested in the near future. A recent attempt at such a prediction has been made in the literature in ref.~\cite{Caputo:2020avy}, which has predicted
spectral edges and endpoints in 21~cm measurements resulting from resonant dark-photon to visible-photon conversions which would alter the Rayleigh-Jeans tail of the CMB.

In this work, we explore a different possible modification of the global 21 cm signal in a model where dark matter interactions with electrons through an axial-vector mediator could lead to direct spin-flips of the hydrogen atoms through collisions with dark matter particles. This interaction would directly alter the \textit{spin-temperature of the gas}, rather than the distribution of CMB photons. We find generically, that such an interaction would lower the spin-temperature over a large redshift range, leading to a strong, broadband absorption signal from the cosmic dark ages. We will discuss the reasoning for this generic prediction below.

%\vspace{3mm}

The spin-flip interactions proposed in our work could lead to two effects. The first is a coupling of the spin temperature to a new effective temperature scale $\Teff$, which is in general colder than the gas temperature. We will show that the spin-flip interaction leads to  a modification of eq.~\ref{eq:spinT}  to
\begin{equation}
    T_{s}^{-1} = \frac{{\Tcmb}^{-1} + x_{C}T_{K}^{-1} + x_{\alpha} T_{c}^{-1} +  x_{D} T_{\textrm{eff}}^{-1}} {1 + x_C + x_{\alpha}+  x_{D}},
\label{eq:Ts}
\end{equation}
where $x_D$ is the effective coupling of the spin temperature to $T_{\textrm{eff}}\equiv \mu \left(  \frac{T_K}{m_H} +  \frac{T_\chi}{m_\chi} \right )$, where $\mu$ is the reduced mass of the dark matter (with mass $m_\chi$) and the hydrogen gas ($m_H$), and $T_\chi$ is the dark matter temperature. The second effect of the spin-flip interactions is a cooling of the gas, due to transfer of kinetic energy to the dark matter.

It is the interplay between the cosmological relevance of these two effects that leads to a distinct prediction for the absorption signal. We find a hierarchy between the spin-flip transition rate and the kinetic energy transfer rate which guarantees that the former is always significantly larger. The parameter space of the dark sector model which leads to distinct cosmological signatures can thus be split into three scenarios of a) strong coupling -- where both spin-flip interactions and kinetic energy transfer are cosmologically relevant b) weak coupling -- where kinetic energy transfer is irrelevant, but spin-flip interactions are important and c) intermediate coupling -- which represents a transition region between the other two scenarios.

The dominance of the spin-flip interaction rate is the key feature that distinguishes our model from other models with excess cooling of the gas. Models with only excess gas cooling, lead to deeper absorption dips, which are still band-limited since the transition features are due to the same dynamics (collisional couplings and Ly-$\alpha$ photons) as in the standard cosmology.

In our model, unlike in the excess cooling only models, the dominance of the spin-flip coupling $x_D$, and a low temperature scale $\Teff$, which is smaller than $\Tcmb$ through the bulk of the cosmic dark ages, lead to a prediction of a single, strong, broadband absorption trough. The absorption trough is predicted to begin at frequencies as low as 1.4~MHz ($z\sim 1000$)  -- a region of frequencies from which no absorption signal is expected in the standard cosmology -- and extends all the way up to high frequencies, where our signal merges with that of the standard absorption feature due to the first stars and $X$-ray sources. The signal is also expected to be very strong compared to standard predictions, since the new temperature scale $\Teff$ is much colder at a given redshift than the corresponding $T_K$ in the standard cosmology.

In the specific case of the weak coupling scenario, for sufficiently low DM masses, $\Teff \simeq T_\chi$ and the spin-temperature latches on to the low DM temperature. In the limit of extreme weak coupling, we find that kinetic energy transfer is negligible and the gas cools no more than usual as compared to the standard cosmology.
While in the standard cosmology and in excess gas cooling models, any absorption signal is expected to be a tracer of the gas kinetic temperature (as suggested by eq.~\ref{eq:spinT}), the absorption feature predicted in the weak coupling scenario of our model would only be a tracer of the gas spin temperature (and hence the DM temperature), and it would not measure the gas kinetic temperature.

%\vspace{3mm}
Extracting a cosmological 21 cm absorption signal requires careful removal of foregrounds which are typically many orders of magnitude larger than the expected signals. Although our predicted absorption signal is stronger than that of the standard cosmological models, the absorption feature is also smoother and spread over a larger frequency range. This smoothness could potentially complicate extraction of the cosmological signal. Other properties, such as uniformity of the cosmological signal over large scales and the unpolarized state of the redshifted 21 cm spectrum, which are distinct from the spatially varying and polarized emission from the bright foregrounds~\cite{Burns_2017}, could potentially be used to facilitate extraction of the signal. Experiments would also need to adopt new strategies for modelling or fitting the foregrounds, see for instance~\cite{Harker_2009,Liu_2009, Singh:2015pga,Tauscher:2017zsj}, to extract such a smooth broadband signal.

A few papers have previously attempted to explain the EDGES absorption signal using dark matter spin-flip interactions. A set of papers on axion Bose-Einstein condensate with spin-flip interactions was considered in~\cite{Lambiase:2018lhs,Auriol:2018ovo}. However, \cite{Auriol:2018ovo} claimed that such a condensate would be too weakly coupled to affect the spin-flip. A model similar to ours was studied in the context of the EDGES signal in ref.~\cite{Widmark:2019cut}, however in addition to our primary objective -- which is to explore the implications for the 21 cm signal, rather than explain the EDGES signal -- our calculational methods and parameter space of interest differ significantly from this work.

In our analysis, we focus on regions of parameter space that have strong absorption at $z\sim 17$, consistent with the magnitude of the signal $\delta T_b(z=17) \simeq -500$~mK claimed by the EDGES collaboration. However, this value is only taken as a benchmark point to which we ``pin'' our absorption signal, we make no demands on the shape of the absorption signal.

We list some of the key new methods of our paper below:
\begin{itemize}
\item We make an accurate Born level bound-state calculation of both the spin-flip scattering cross-section through a light-mediator, which determines the coupling $x_D$ and we also calculate the energy-transfer cross-section, which determines the temperature evolution of the gas.
\item Our calculation demonstrates that the thermally averaged bound state cross-section scales as $1/\Delta^2$, where $\Delta$ is the tiny hyperfine splitting, allowing for large spin-flip interactions even for relatively weak couplings between electrons and the new mediator.
\item We demonstrate the existence of a hierarchy between the spin-flip rate and the energy transfer rate, which leads to the three different predicted scenarios of strong, weak, and intermediate couplings discussed previously.
\item We find that several regions of the model parameter space which lead to a broadband absorption signal are safe from all laboratory constraints. We argue that various astrophysical and cosmological constraints may be evaded under simple extensions of the model. We find interesting regions of parameter space which present promising targets for searches for novel short-range spin-dependent interactions for electrons on the millimeter to nanometer scale.
\end{itemize}

This paper is organized as follows. In sec.~\ref{sec:model}, we present an effective Lagrangian for our dark matter model interacting with electrons via a light axial-vector mediator.  In sec.~\ref{sec:spinflip}, we show how eq.~\ref{eq:Ts} for the spin temperature evolution follows from our model, and present a relationship between the parameter $x_D$ and our Lagrangian parameters. The same spin-flip interaction leads to energy transfer between the dark matter and hydrogen and this in general leads to a coupled temperature evolution of both fluids in the post-recombination era. We discuss the temperature evolution equations in sec.~\ref{sec:tempevolution}. In sec.~\ref{sec:results}, we present our main results. We classify our parameter space into three distinct regions and discuss the signatures of the global 21~cm signal in each regime. In sec.~\ref{sec:discussion}, we clarify some of our assumptions and discuss how changes in these assumptions would change our results. We also discuss the validity of our approximations. In sec.~\ref{sec:constraints}, we discuss a variety of constraints on our model and show that there is a large region of our benchmark parameter space that is as yet unconstrained by experimental searches. Finally, we summarize our findings and discuss some future directions in sec.~\ref{sec:conclusions}. We have shown the details of our calculation of the cross-section for the spin-flip interaction, energy transfer cross-section and some other cross-sections of interest in the appendices.

\section{Dark matter model and spin-flip interactions}
\label{sec:model}
We present here our model which consists of a dark matter particle $\chi$, which we take to be a Dirac fermion, and a light pseudo-vector mediator particle, which we denote as $V$, which couples to both the dark matter and electrons with coupling strengths $g_\chi$ and $g_e$, respectively. The relevant interaction terms in our effective Lagrangian are given by,
\begin{equation}
\mathcal{L} = i g_\chi \overline{\chi} \gamma^\mu \gamma^5 \chi V_\mu +  i g_e \overline{e} \gamma^\mu \gamma^5 e V_\mu.
\label{eq:lagrangian}
\end{equation}
Thus the model has four new parameters, the two coupling constants and the masses of the dark matter and mediator particles, which we denote as $m_\chi$ and $m_V$, respectively.

We assume that the dark matter particle $\chi$ is produced asymmetrically in the early universe, with no $\overline{\chi}$ particles surviving as relics.
We allow for the possibility that the particle $\chi$ that we have considered here makes up only a fraction $f$ of the total DM relic density, which introduces a fifth parameter into our theory.

The interaction between dark matter and neutral hydrogen atoms proceeds through \mbox{$t$-channel} exchange of the mediator and gives rise to the spin-flip reactions,
 \begin{align}
\chi+H_0\rightleftarrows \chi +H_1,
\end{align}
where $H_0$ denotes neutral hydrogen in the singlet ground state and $H_1$ denotes the triplet state. The forward reaction is for excitations and the reverse for the de-excitation process.
In addition to flipping the spin of the hydrogen atoms, these reactions also transfer energy between the gas and the dark matter particles and could modify the evolution of the gas and DM kinetic temperatures (and hence $\Teff$).

\section{Spin-flip interaction rate and the coupling $x_D$}
\label{sec:spinflip}
At any redshift the balance between excitation and de-excitation processes for neutral hydrogen leads to the relationship between the reaction rates~\cite{Lambiase:2018lhs,Auriol:2018ovo, Widmark:2019cut},
\begin{equation}
    \label{eq:detbal}
 n_0( B_{01} + C_{01} + P_{01} +  D_{01}) =  n_1( A_{10} + B_{10} +  C_{10}  +  P_{10} +  D_{10})
\end{equation}
where on the left-hand side $B_{01}$, $C_{01}$,  $P_{01}$ and $D_{01}$  are redshift dependent rate coefficients (with units of inverse time) for excitation through stimulated interaction with CMB photons, collisional excitations,
\mbox{Ly-$\alpha$} excitations and dark matter scattering induced excitations, respectively. Similarly, the coefficients with subscripts reversed on the right hand side denote the corresponding de-excitation rates.  $A_{10} = 2.85 \times 10^{-15} \textrm{\, s}^{-1} = 1.88 \times 10^{-39} \textrm{\, GeV} $ is the decay width for the spontaneous electromagnetic decay process of the triplet state into the singlet state.
 All coefficients other than $A_{10}$ are a result of scattering and can be written as the product of a number density of scattering targets and the thermally averaged cross-section times velocity for the corresponding excitation or de-excitation process. For example, $D_{10} = n_\chi \langle \sigma_{10} v \rangle$, where $n_\chi$ is the number density of dark matter particles, and $\langle \sigma_{10} v \rangle$ is the thermal velocity averaged cross-section for the process $\chi + H_1 \rightarrow  \chi + H_0 $.

The principle of detailed balance can be invoked to determine the ratio between the coefficients for excitation and de-excitation for individual processes. Applying this to the CMB induced processes, the collisional processes (due to collisions of neutral hydrogen with either other $H$ atoms, electrons or protons) and to Ly-$\alpha$ processes, one can obtain the relations,
\begin{align}
               \label{eq:photonrate}
B_{10} &= 3 B_{10}  \simeq A_{10} \frac{\Tcmb}{\Delta}
                         = 1.26 \times 10^{-12} \left(\frac{1+z}{1+10}\right)~\textrm{s}^{-1},\\
\frac{C_{01}}{C_{10}} &= 3 e^{-\Delta/T_K}\simeq 3\left( 1 - \frac{\Delta}{T_K} \right),  \\
\frac{P_{01}}{P_{10}} &= 3 e^{-\Delta/T_c} \simeq 3\left( 1 - \frac{\Delta}{T_c} \right),
\end{align}
where $T_K$ is the kinetic temperature of the gas, and $T_c$ is the color temperature of Ly-$\alpha$ radiation, both of which are typically much larger than the hyperfine splitting $\Delta$.

We cannot simply use the principle of detailed balance to get the ratio of rate coefficients for dark matter excitations $D_{01}$  and de-excitations $D_{10}$, because the kinetic temperature of the DM and gas are different in general. We explicitly work out these rates in the appendix and we show that the ratio is given by,
\begin{equation}
\label{eq:DM_detbal}
\frac{D_{01}}{D_{10}} = 3 e^{-\Delta/T_{\textrm{eff}}} \simeq 3\left( 1 - \frac{\Delta}{T_{\textrm{eff}}} \right),
\end{equation}
where $T_{\textrm{eff}}= \mu \left(  \frac{T_K}{m_H} +  \frac{T_\chi}{m_\chi} \right )$, and $\mu = \frac{m_H m_\chi}{m_H + m_\chi}$ is the reduced mass of the dark matter and the hydrogen gas. Intuitively, the threshold energy needed for excitation reactions leads to the exponential Boltzmann suppression factor of the excitation rate relative to the de-excitation rate.

Using all of the above ratios of rate coefficients, we find an expression for the spin-temperature,
\begin{equation}
    T_s^{-1} = \frac{{\Tcmb}^{-1} + x_{C}T_{K}^{-1} + x_{\alpha} T_{c}^{-1} +  x_{D} T_{\textrm{eff}}^{-1}} {1 + x_C + x_{\alpha}+  x_{D}},
\label{eq:tsmod}
\end{equation}
where $x_C = \frac{P_{10}}{B_{10}}$, $x_{\alpha} = \frac{C_{10}}{B_{10}}$, and $x_D =\frac{D_{10}}{B_{10}}$ are effective couplings of the spin-temperature to the gas, Ly-$\alpha$ photons and dark matter respectively. The coupling $x_D$ to the temperature scale $\Teff$ is a direct consequence of the spin-dependent interaction between electrons and dark matter. Thus, we see that there is another temperature scale that the spin temperature can couple to when DM spin-flip interactions dominate the hyperfine transitions.

If we know how the gas and DM temperatures evolve with time (or redshift) and if we also know how the effective couplings $x_C$,  $ x_{\alpha}$  and $ x_{D}$ change with redshift, then we can determine the spin temperature at any epoch.

We have evaluated the cross-section for the excitation and de-excitation processes $\chi+H_0\rightleftarrows \chi +H_1$ in appendix~\ref{sec:appendixb}.
This scattering process has a large cross-section which arises because of the usual forward scattering divergence of a light mediator exchanged in the \mbox{$t$-channel}. For sufficiently light mediator masses, the divergence is cut-off, not by the mediator mass, but rather by the tiny inelastic mass-splitting $\Delta$ between the singlet and triplet states. Our detailed calculation of the cross-section in the regime of light mediator mass shows that the thermally averaged bound state cross-section is of the form $\sigma_{01} \sim \frac{3}{4\pi} \frac{g^2_\chi g^2_e}{\Delta^2}$, which leads to a large interaction rate even for relatively weak couplings. In the appendix, we show that the de-excitation rate
\begin{align}
   \label{eq:deex_sigv}
D_{10} &= 3.01\times 10^{-12}\left( \frac{f}{0.1} \right)   \left( \frac{0.1~\textrm{GeV}}{m_\chi} \right)  \left( \frac{\alpha_\chi}{10^{-2}} \right) \left(\frac{\alpha_e}{10^{-14}}\right)   \left(\frac{0.1~\textrm{GeV}}{\mu} \right)^{\frac{1}{2}} \left(\frac{\Teff}{10~\textrm{K}} \right)^{\frac{1}{2}} \left( \frac{1+z}{1+10} \right)^3~\textrm{s}^{-1},
\end{align}
where $\alpha_\chi = \frac{g^2_\chi}{4\pi}$ and $\alpha_e = \frac{g^2_e}{4\pi}$. Taking the ratio of the expression for $D_{10}$ above, with the rate $B_{10}$ in eq.~\ref{eq:photonrate}, we find an expression for the coupling of the spin temperature to the effective temperature as,
\begin{align}
 x_D =2.4 \left( \frac{f}{0.1} \right)   \left( \frac{0.1~\textrm{GeV}}{m_\chi} \right) \left( \frac{\alpha_\chi}{10^{-2}} \right) \left(\frac{\alpha_e}{10^{-14}}\right) \left(\frac{0.1~\textrm{GeV}}{\mu} \right)^{\frac{1}{2}} \left(\frac{\Teff}{10~\textrm{K}} \right)^{\frac{1}{2}} \left( \frac{1+z}{1+10} \right)^2.
\label{eq:xd}
\end{align}

\section{Temperature evolution equations}
\label{sec:tempevolution}

The evolution of the dark matter and neutral hydrogen gas temperatures are given by  the following coupled Boltzmann equations (appendix ~\ref{sec:appendixc}):
\begin{align}
\label{eq:tempevolutionDM}
\frac{d T_\chi}{d \log (1+z)} & = +2  T_\chi - \frac{2}{3}  \frac{\Gamma_\chi}{H }  \left(T_K - T_\chi \right), \\
\label{eq:tempevolutiongas}
\frac{d T_K}{d \log (1+z)} &= +2 T_K -\frac{\Gamma_c}{H} (\Tcmb-T_K) - \frac{2}{3} \frac{\Gamma_H}{H} (T_\chi-T_K).
\end{align}
These equations are valid in the epoch prior to the cosmic dawn, before $X$-ray heating from the first sources turns on. The first terms on the right hand side of these equations correspond to adiabatic non-relativistic cooling of the dark matter and the gas, respectively,  due to the expansion of the universe. The rate $\Gamma_c$ is the compton rate which couples the gas to the CMB temperature through the residual free electron fraction $x_e(z)$ and is given by~\cite{2010PhRvD..82f3521A,Ma:1995ey},
\begin{align}
\Gamma_c
&= \frac{8 a_r\sigma_T   }{3 m_e c}T^4_{\textrm{CMB}}(z)\frac{x_e(z)}{ (1+0.08 + x_e(z))} \\
 &=7.4 \times 10^{-20} \left(\frac{1+z}{1+10}\right)^4 \times \frac{x_e(z)}{2\times10^{-4}} \frac{1.08}{ (1+0.08 + x_e(z)) } s^{-1}.
\end{align}
In this expression $m_e = 0.511$~MeV/$c^2$ is the electron mass, $a_r=  4.72 \times 10^{-3}$~MeV~m$^{-3}~$K$^{-4}$ is the radiation constant and $\sigma_T= 6.65 \times 10^{-29}$~m$^2$ is the Thomson scattering cross-section for free-electrons with the CMB. The compton rate depends on the free electron fraction $x_e(z)$ which must be solved for in a given cosmology. Since we have assumed that our DM is asymmetric, it does not self-annihilate. Therefore, the physics of recombination or the ionization fraction history $x_e(z)$ are both unaltered from the standard cosmology. In our numerical simulations in sec.~\ref{sec:results} we use the code HyRec~\cite{2010PhRvD..82f3521A} to compute the free electron fraction.

The rates $\Gamma_\chi$ and $\Gamma_H$ are the heating rates for DM and the gas respectively, which can be thought of as the inverse time-scale for transfer of an $\mathcal{O}(1)$ fraction of their kinetic energy to one another. Schematically these rates are of the form $\Gamma \sim n \langle \overline{\sigma} v \rangle$, where $n$ is the number density of targets and $\langle \overline{\sigma} v \rangle$ is thermally averaged ``energy-transfer cross-section''. In appendix~\ref{sec:appendixc} we evaluate the energy transfer cross-sections and derive the temperature evolution eqs.~\ref{eq:tempevolutionDM} and \ref{eq:tempevolutiongas}. We argue that the energy transfer cross-section is dominated by the inelastic scattering process $\chi + H_0 \leftrightarrows  \chi + H_1 $ rather than by elastic scattering processes. This is because the forward divergence of the inelastic scattering, which is cut-off by the hyperfine mass-splitting, diverges as $\frac{1}{\Delta}$, whereas the divergence of elastic scattering, which is cut-off by the mediator mass, is only enhanced as $\textrm{Log } m_V$. The detailed expressions for the rates $\Gamma_\chi$ and $\Gamma_H$ are of the form (appendix~\ref{sec:appendixc}),
\begin{align}
\Gamma_H
=& \, 3.11\times 10^{-15} \Biggl[\left( \frac{f}{0.1} \right) \left( \frac{0.1~\textrm{GeV}}{m_\chi}\right ) \left( \frac{\alpha_\chi}{10^{-2}} \right) \left(\frac{\alpha_e}{10^{-14}}\right)  \left( \frac{1~\textrm{GeV}}{M} \right) \left(\frac{\mu}{0.1~\textrm{GeV}} \right)^{\frac{1}{2}}  \left(\frac{10~\textrm{K}}{\Teff} \right)^{\frac{1}{2}}\nonumber \\
&{\hskip 0.75in} \times   \left( \frac{1+z}{1+10} \right)^3\Biggr]~\textrm{s}^{-1},    \label{eq:gammaHmain} \\
\Gamma_\chi =& \,  6.02 \times 10^{-16} \Biggl[ \left( \frac{\alpha_\chi}{10^{-2}} \right) \left(\frac{\alpha_e}{10^{-14}}\right) \left( \frac{1~\textrm{GeV}}{M} \right) \left(\frac{\mu}{0.1~\textrm{GeV}}\right)^{\frac{1}{2}}  \left(\frac{10~\textrm{K}}{\Teff}\right)^{\frac{1}{2}}
  \left( \frac{1+z}{1+10} \right)^3 \Biggr]~\textrm{s}^{-1}, \label{eq:gammachimain}
\end{align}
where $M = m_\chi + m_H$. Since the inelastic scattering process is dominated by forward scattering in which the energy transfer per collision is small, this leads to a suppression of the energy-transfer rate $\Gamma_H$ relative to the standard excitation/de-excitation rate $D_{01}$~$(\simeq 3 D_{10})$  by a factor of $S\equiv \left(\frac{\Delta \mu}{2 M \Teff}\right) = 3.42\times 10^{-4} \left(\frac{10 \, \textrm{K}}{\Teff} \right ) \left(\frac{\mu}{0.1\textrm{~GeV}}\right) \left( \frac{1\textrm{~GeV}}{M} \right )$. Also, the energy transfer rate to the gas $\Gamma_H$, is enhanced relative to the energy transfer rate to the DM $\Gamma_\chi$, by a factor of $R$, where
\begin{equation}
R\equiv\frac{n_\chi}{n_H} =\frac{ f\Omega_{\textrm{DM}} /m_\chi}{ \Omega_b /m_H}  =\, 5.16  \left( \frac{f}{0.1} \right) \left( \frac{0.1~\textrm{GeV}}{m_\chi}\right ).
\end{equation}
Here, we have used the present-day total DM density fraction $\Omega_{\textrm{DM}} =  0.26$ and the baryon density fraction $\Omega_b= 0.05$. For simplicity, we have also assumed that all the baryons are in the form of hydrogen.

The Hubble rate $H$ in our temperature evolution equations is as usual given by,
\begin{equation}
H(z) = H_0\sqrt{ \Omega_\Lambda+ \Omega_m (1+ z)^3 + \Omega_r (1+z)^4 },
\end{equation}
where $H_0 = 67$~km/s/Mpc is the Hubble constant and $\Omega_m = 0.31$, $\Omega_\Lambda = 0.69$ are the present-day total matter and dark energy density fractions, respectively. All values of the cosmological parameters are taken from Planck data~\cite{Aghanim:2018eyx}. We take $\Omega_r= 9.6\times 10^{-5}$ to be the present day radiation density fraction. This value includes the contributions of photons and neutrinos, but does not include possible contributions from the light pseudo-vector mediator to the radiation density in the early universe. The exact value of $\Omega_r$ is not so important in the temperature evolution, since for the redshifts of interest $10 \lesssim z \lesssim 1000$ the universe is matter dominated to a good approximation. However, there are observational constraints on extra radiation species, usually parameterized in terms of extra number of effective neutrinos (or $\Delta N_\textrm{eff}$). We will discuss this constraint and the implications in more detail in sec.~\ref{sec:constraints}.

In the matter dominated era, the Hubble rate has the approximate form,
\begin{align}
H &= H_0 \Omega_m^{1/2} (1+z)^{3/2}, \\
    &=4.41 \times 10^{-17} \left (\frac{ 1+ z}{1+10}\right)^{\frac{3}{2}}~\textrm{s}^{-1}.
\end{align}

In the standard cosmology it is not possible to obtain very strong differential brightness temperatures.  To see this, we can start by examining the rates $\Gamma_c(z)$ and $H(z)$ which are plotted in fig.~\ref{fig:rates_weak_coupling}. In the figure, we see that $\Gamma_c$ drops below $H$ at a decoupling redshift of $z_d\sim130$. Without exotic interactions with the DM, the gas would kinetically decouple from the CMB at a redshift $z_d$ and begin to adiabatically cool. This would lead to a gas temperature of,
\begin{align}
T_K(z=17) &= \Tcmb(z_d) \left(\frac{1+17}{1+z_d} \right)^2
                   = 6.8 \left(\frac{1+130}{1+z_d}\right)\textrm{K}.
\end{align}
If one assumes a strong Ly-$\alpha$ induced coupling at $z=17$ to the color temperature (which can be taken to be the same as the gas temperature $T_K$), then this leads to the lowest possible  prediction of the spin temperature $T_s (z= 17)= 6.8$~K, or equivalently a differential brightness temperature $\delta T_b \simeq -227$~mK. This is clearly not enough of an absorption dip to explain the maximum strength of the EDGES absorption signal $\delta T_b(z=17) \simeq -500$~mK.

In order to obtain $\delta T_b(z=17) \simeq -500$~mK, we would need a much colder spin temperature of $T_s(z=17) = 3.32$~K. Such a low spin temperature can not be obtained in the standard cosmology, but can be obtained in models with excess gas cooling, and also in our model, as we shall see in the next section.

\section{Method and results}
\label{sec:results}
\subsection{Predicting the differential brightness temperature evolution with spin-flip interactions}
Our goal is to predict the global differential brightness temperature evolution $\delta T_b(z)$ for different parameters of our model. We will then contrast the predictions of our model with those of the standard cosmology, as well as models with excess gas cooling that attempt to explain the EDGES absorption dip.

The inputs to our prediction rely on the following parameters: the coupling product $\alpha_e \alpha_\chi$, the DM mass $m_\chi$ and the fraction of DM  $f$ that couples to the pseudo-vector mediator. We do not specify the value of the mediator mass $m_V$, but we assume that it is light enough that the forward scattering divergence of the cross-section for inelastic scattering between the hyperfine states has a cut-off which is dominated by the inelastic mass-splitting parameter $\Delta$ rather than by the mediator mass. Under this assumption, our calculations are independent of the mediator mass.

For a given point in the parameter space of our model, our procedure to predict the global 21~cm brightness temperature is as follows:

\begin{enumerate}
\item First, we solve the coupled temperature evolution equations~\ref{eq:tempevolutionDM} and \ref{eq:tempevolutiongas} for $T_K(z)$ and $T_\chi(z)$  given an assumption of the initial conditions on these temperatures.
\item
Our fiducial choice of initial conditions will be to assume $T_K=T_\chi = \Tcmb= 2.73 \times (1+1000)$~K at $z=1000$, near recombination.
 We will discuss the effect of different choices for the initial conditions and the assumptions they correspond to about pre-recombination physics in sec.~\ref{sec:discussion}.
\item Once we know the temperature evolution history of $T_K$ and $T_\chi$, we can predict $\Teff$ at all redshifts.
\item In order to predict the spin-temperature evolution using eq.~\ref{eq:tsmod}, we need the evolution of the couplings $x_C$ - the collisional coupling of the spin temperature to the gas kinetic temperature (which is well known, see for e.g. eq.~10 in ref~\cite{2012RPPh...75h6901P}), $x_\alpha$ - the Ly-$\alpha$ coupling which depends on the astrophysical assumptions of the formation of first stars and galaxies (see for e.g.~\cite{Mittal:2020kjs} and references therein), and $x_D$ - the coupling to the effective temperature $\Teff$ (which we have derived, see eq.~\ref{eq:xd}).

\item We will assume for simplicity that the astrophysical processes from cosmic dawn, namely Ly-$\alpha$ photons and $X$-ray heating of the gas, are only relevant at redshifts $z\lesssim 15$, and we will set $x_\alpha = 0$ at higher redshifts. Our predictions of the absorption signal will be limited to the redshift range $15 \lesssim z \lesssim 1000$, or 21 cm frequencies between $1.4$ and $89$~MHz, to avoid complications of modelling the uncertain astrophysics of the cosmic dawn.
\item After solving the temperature evolution equations, and with knowledge of the coupling coefficients of the spin temperature, we solve for the spin temperature and brightness temperature over the redshift range of interest.
\end{enumerate}

Since strong absorption signals are a generic feature of our model, as a benchmark, we focus our analysis on the region of parameter space of $f$, $\alpha_e \alpha_\chi$ and $m_\chi$ which lead to a differential brightness temperature
$\delta T_b (z=17) \simeq - 500$ mK ($T_s(z=17) = 3.32$~K), consistent with the magnitude of the signal claimed by the EDGES collaboration at this redshift. In our numerical scan, we allow for some flexibility in this constraint, by relaxing the requirement to
$-600$~mK~$\lesssim\delta T_b (z=17)\lesssim$~-400~mK $(2.7~\textrm{K} \lesssim T_s(z=17) \lesssim 3.9~\textrm{K})$.
We reiterate that our goal is not to explain the EDGES signal, but rather, this point is a useful place to ``pin'' our predicted absorption spectrum, without making additional demands on the shape of the absorption signal.

\begin{figure}[H]
\begin{subfigure}{.5\textwidth}
 \centering
  \includegraphics[width=\linewidth,, height=0.9 \linewidth]{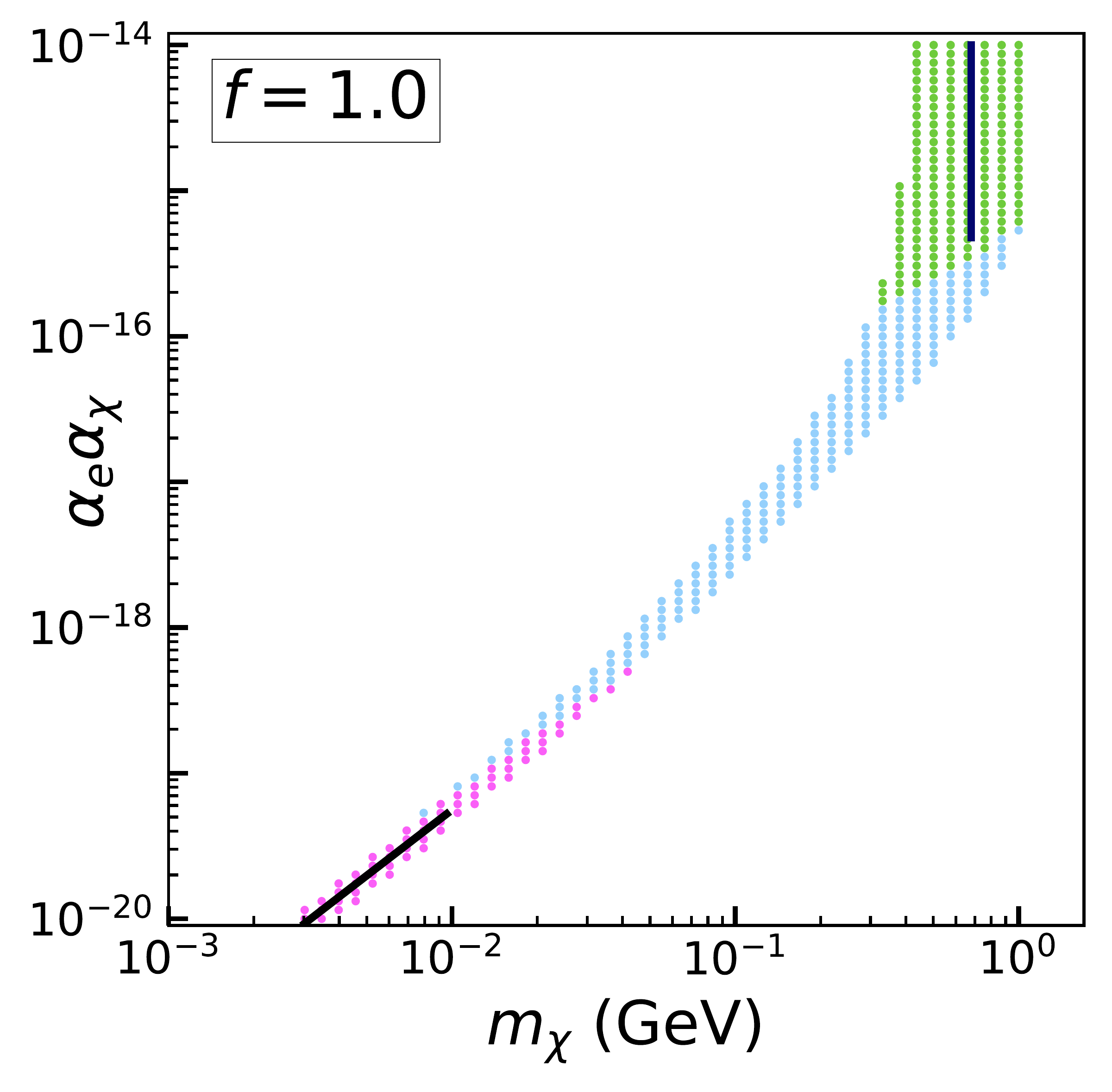}
  %\caption{f=1}
  \label{fig:paramspace_f_1}
\end{subfigure}%
\begin{subfigure}{.5\textwidth}
  \centering
  \includegraphics[width=\linewidth, height=0.9 \linewidth]{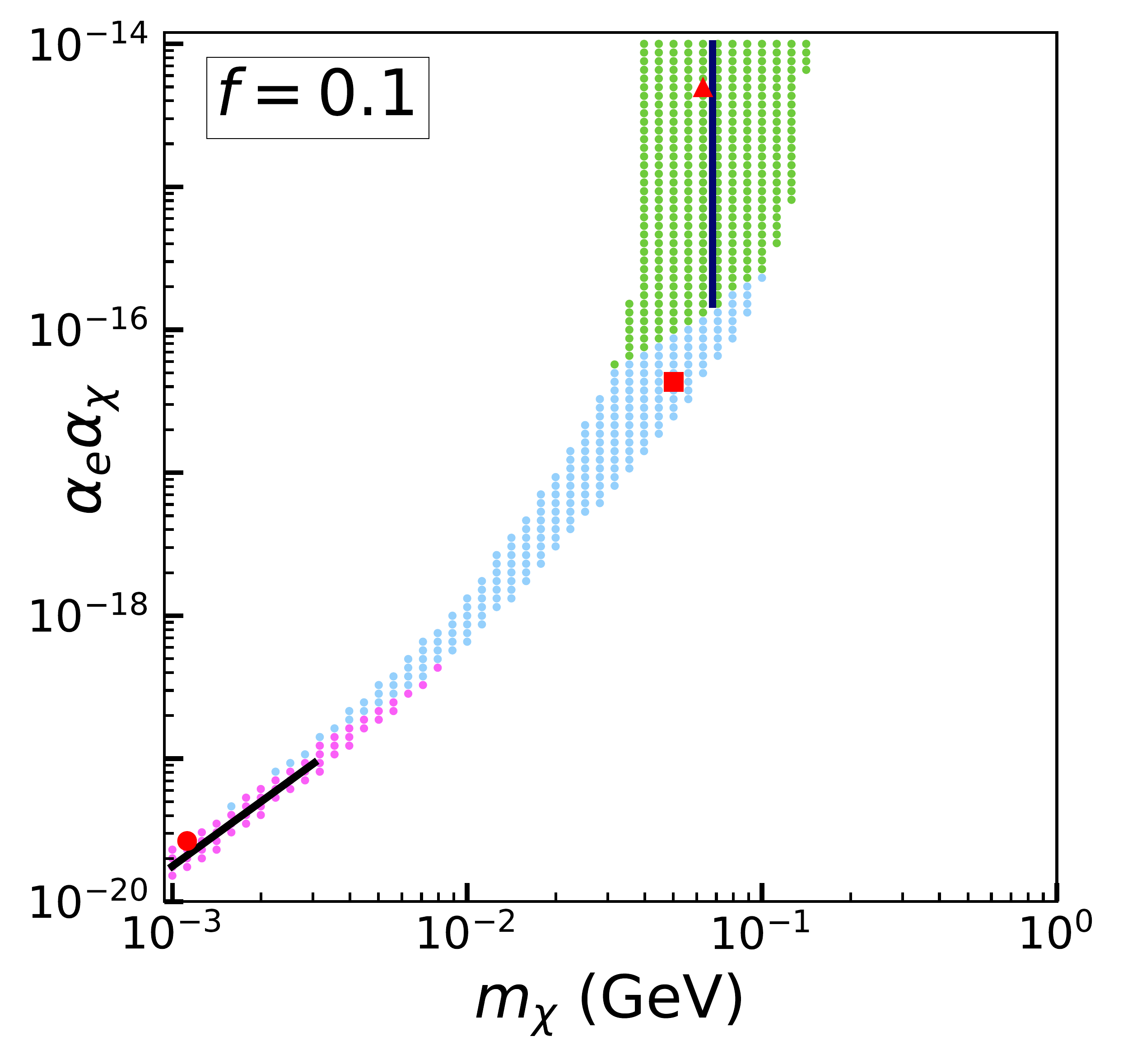}
 %\caption{f=0.1}
  \label{fig:paramspace_f_01}
\end{subfigure}
\begin{subfigure}{.5\textwidth}
 \centering
  \includegraphics[width=\linewidth,, height=0.9 \linewidth]{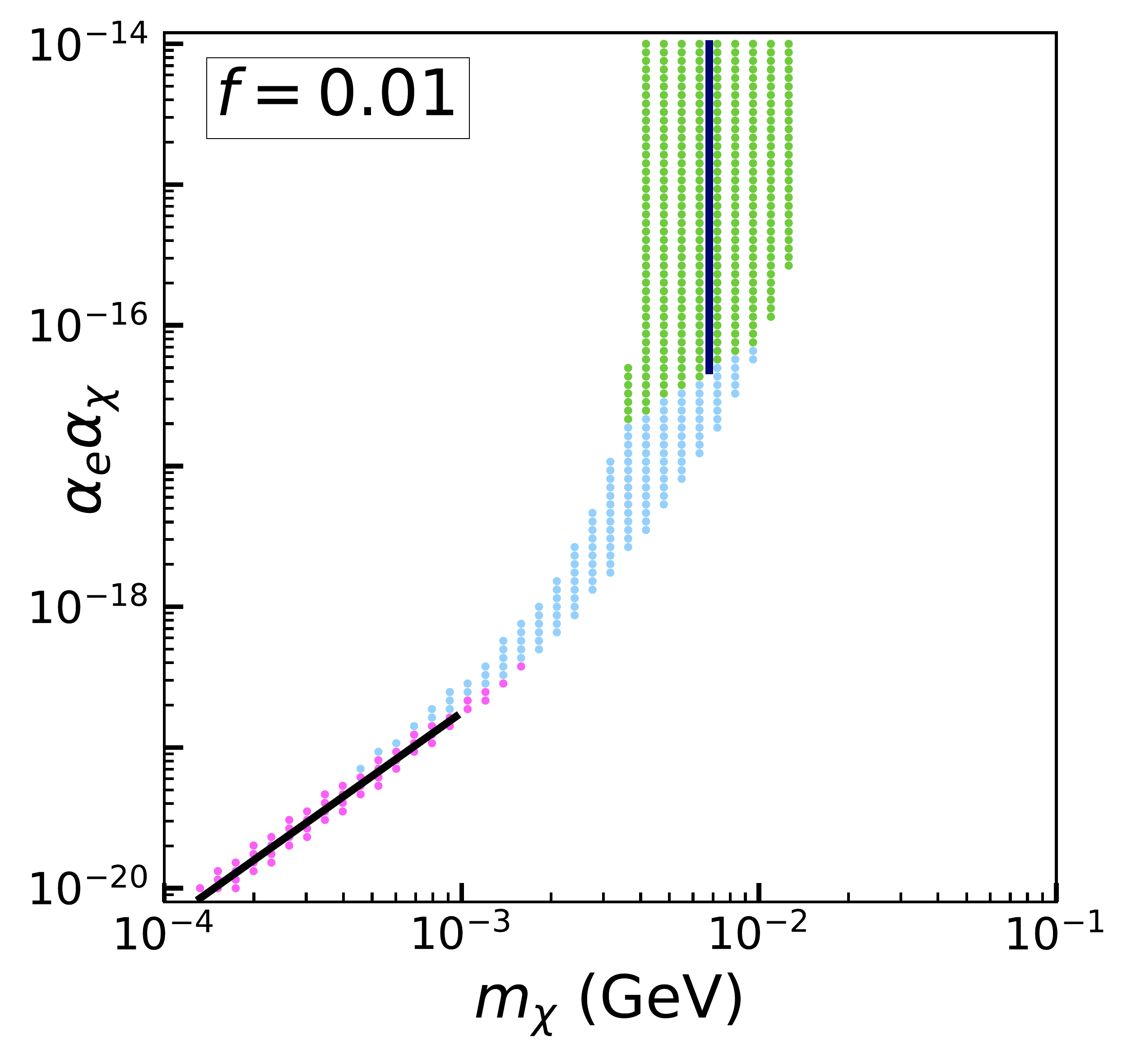}
  %\caption{f=0.01}
  \label{fig:paramspace_f_001}
\end{subfigure}%
\begin{subfigure}{.5\textwidth}
  \centering
  \includegraphics[width=\linewidth,, height=0.9 \linewidth]{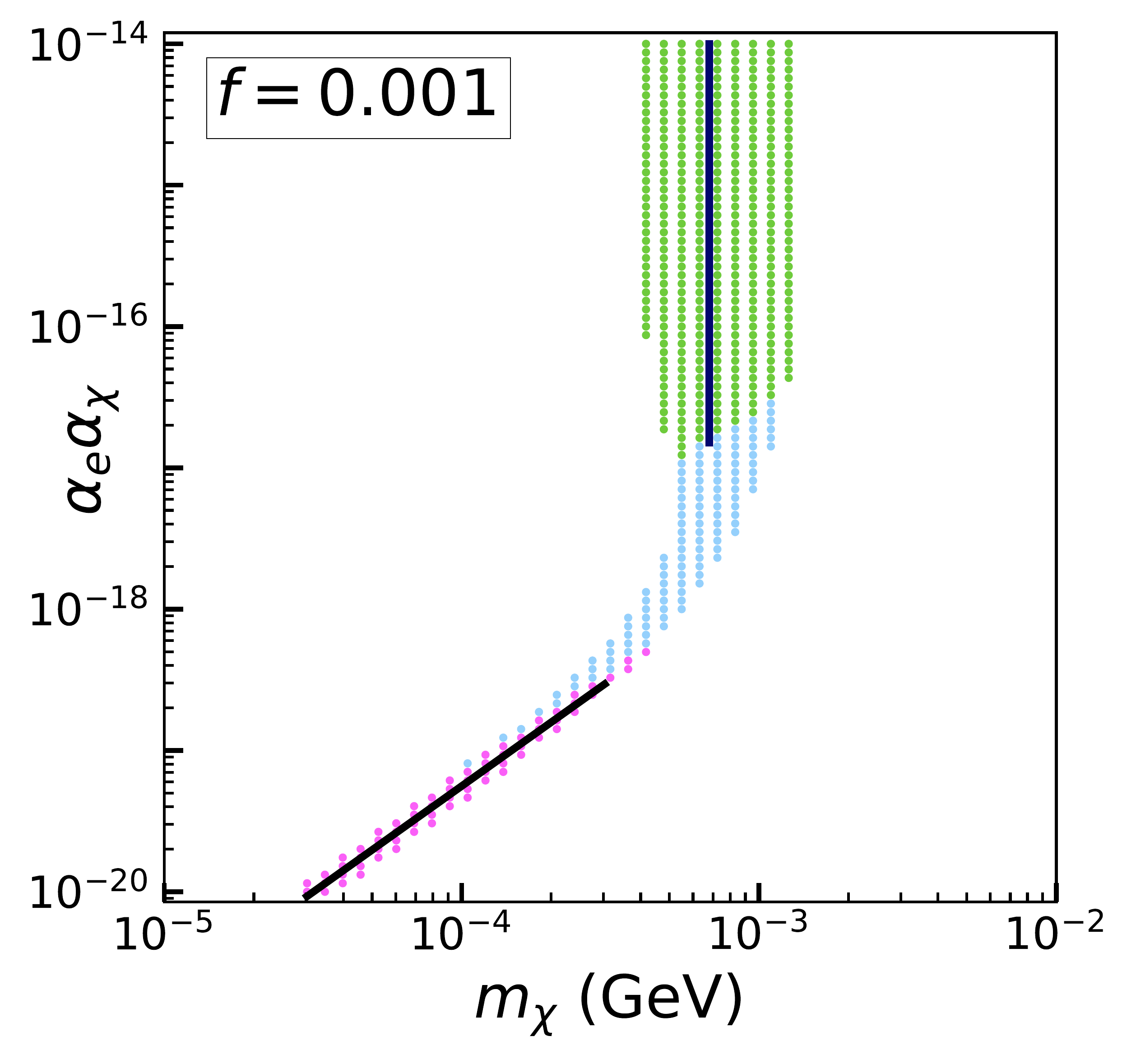}
 % \caption{f=0.001}
  \label{fig:paramspace_f_0001}
\end{subfigure}
\caption{Allowed parameter space of the coupling product $\alpha_e \alpha_\chi$ and the dark matter mass $m_\chi$ that leads to a predicted value of $\delta T_b(z=17)$ between -400 mK and -600 mK. Each panel shows the allowed parameter space for different values of the dark matter density fraction $f= 1, 0.1, 0.01, 0.001$ made up of the dark matter candidate in our model. The allowed parameter space was found by a numerical scan whose procedure is described in the text. The colors green, blue, and magenta of the different regions correspond to strong coupling, intermediate coupling, and weak coupling scenarios, respectively. For comparison, we also show an analytic prediction of the allowed parameter space that leads to $\delta T_b (z=17) = - 500$~mK at strong and weak couplings with black lines. The analytic prediction is in good agreement with the results of our numerical scan.  Also for $f=0.1$ we have shown three reference points (in red), one for each scenario,  which will be used when presenting detailed numerical results. }
\label{fig:parameterspaceplot}
\end{figure}

The result of a numerical scan, using the procedure outlined above, yields the parameter space of interest, which  is shown in fig.~\ref{fig:parameterspaceplot}. In each panel of the figure we show, for different values of $f = 1, 0.1, 0.01, 0.001$ (the dark matter density fraction made up by our dark matter candidate), the allowed values of the coupling product $\alpha_e \alpha_\chi$ and the dark matter mass $m_\chi$ that lead to a predicted value of $\delta T_b(z=17)$ between -400 mK and -600 mK. For each value of $f$, we find three distinct categories of solutions which are shown in different colors in the figure -- 1) strong coupling (green), 2) weak coupling (magenta), and 3) intermediate coupling (blue).

For the strong and weak coupling scenarios, we are also able to obtain approximate analytic solutions to the temperature evolution equations and therefore we can analytically determine the expected parameter space regions which give rise to the benchmark $\delta T_b (z=17) = - 500$~mK. The analytically determined parameter space points are plotted with black lines in fig.~\ref{fig:parameterspaceplot}, and are in good agreement with our numerical results.

In the rest of this section, we will discuss each of the scenarios in turn. As we discuss each scenario, we will first discuss our analytic understanding of the solutions, and then show the comparison with our numerical results. For concreteness, we will take three reference points with $f=0.1$ (shown in red in the top-right panel of fig.~\ref{fig:parameterspaceplot}) for strong/weak/intermediate couplings when discussing our numerical results.

For each reference point we will show the predicted global 21~cm signal, and compare this to the predictions of the standard cosmology and also to the predictions of a model with excess gas cooling without spin-flip interactions~\cite{Mirocha:2018cih}, with parameters of the latter model chosen to explain the EDGES anomalous absorption dip. At the end of this section, we will summarize and characterize the difference between each of the three regimes of the parameter space of our model.
\vskip -0.11in

\subsection{Scenario 1: Strong Coupling}

\subsubsection{Analytic understanding of strong coupling solutions}
When the coupling product $\alpha_\chi \alpha_e$ is sufficiently large, both the spin-flip coupling and kinetic energy transfer rate are large. In such a limit we expect the gas and the DM to behave like a single tightly coupled fluid with a common temperature  $\Teff=T_\chi = T_K$.
We can then multiply the temperature evolution equations for the dark matter and the gas in eq.~\ref{eq:tempevolutionDM} and eq.~\ref{eq:tempevolutiongas} by $n_\chi$ and $n_H$, respectively, and add them together and then take the limit of a common temperature. In this limit we find the following equation for the evolution of $\Teff$:
\begin{align}
\frac{d \Teff}{dz}  &= +\frac{2  \Teff}{1+z} -\frac{1}{1+ R}\frac{\Gamma_c}{H(1+z)} (\Tcmb-\Teff),
\end{align}
where $R = \frac{n_\chi}{n_H} = \frac{f\Omega_{\textrm{DM}}}{\Omega_b}\frac{ m_H}{m_\chi} = 5.16 \frac{f}{0.1}\frac{0.1 \, \textrm{GeV}}{m_\chi}$. Thus, the combined fluid has an effective coupling rate to the CMB which is given by $\Gamma^\prime_c \equiv \frac{1}{1+R}\Gamma_c$, which is reduced by a factor $\frac{1}{1+R}$ relative to the coupling  rate $\Gamma_c$ of baryons alone. This is because the kinetic energy transferred from the photons to the baryons is redistributed over both the DM and baryons.

In addition, at strong coupling, we have a large spin-flip rate $D_{10}$ which dominates the CMB spin-flip rate $B_{10}$ (i.e. $x_D \gg 1$). The dominance of the DM spin-flip rate over the CMB spin-flip rate ensures that the spin temperature $T_s = \Teff$ at $z=17$. In order to obtain an absorption dip with $\delta T_b(z=17) \simeq - 500$~mK, we would need $T_s(z=17) = \Teff(z=17) \simeq 3.32$~K in this scenario (using eq.~\ref{eq:brightness}).

Now the combined DM-gas fluid is tightly coupled to the CMB at early times when $\Gamma^\prime_c  > H$ and decouples from the CMB when $\Gamma^\prime_c $ drops below $H$ at a redshift denoted as $z_d$. Thereafter, the combined fluid continues to cool adiabatically. Thus, the temperature evolution of $\Teff$ in this scenario is as follows,
\begin{align}
\Teff(z) &= \Tcmb(z=1000) \left( \frac{1+z}{1+1000} \right ) \, \, \, \textrm{for $z> z_d$} \\
&= \Tcmb(z=1000)\left(\frac{1+z_d}{1+1000}\right) \left( \frac{1+z}{1+z_d} \right )^2 \, \, \, \textrm{for $z< z_d$}.
\end{align}
In order to obtain $T_s = \Teff = 3.32$~K at $z =17$, we need $z_d= 265$.
We can solve for the DM model parameters needed to obtain this redshift of decoupling by equating $\Gamma_c^\prime = H$ at $z=265$. Since the rate evolution of $\Gamma_c$ and $H$ are well known and do not depend on the dark matter physics, this condition determines the value of $R$, which depends only on $m_\chi$ and $f$, as $R=7.6$.
Using the expression for $R$, we have the following criteria on the parameter space for solutions with strong coupling:

 \textbf{1. Criteria needed for decoupling of the combined gas-DM system from the CMB at $z_d = 265$}
\begin{equation}
\label{eq:scenario1criteria1}
m_\chi = 0.68 f~\textrm{GeV}
\end{equation}
Thus, for a given value of the dark matter fraction $f$, which has spin-flip interactions, the mass of the dark matter for this class of solutions is \textit{uniquely determined}.

In order to ensure that the spin temperature of the gas is coupled to $\Teff$, we also need the coupling $x_D$ to be large, i.e. $x_D \gtrsim 10$ at $z=17$. This criteria can be expressed as:

\textbf{2. Criteria for large spin-flip coupling rate at $z = 17$}
\begin{equation}
\alpha_\chi\alpha_e > 8.6 \times 10^{-16}\frac{1}{f} \frac{m_\chi}{\textrm{GeV}} \left(\frac{\mu}{\textrm{GeV}}\right)^{1/2},
\end{equation}
For small values of $f$, the first criteria implies $m_\chi \ll 1$~GeV. Using this limit, we can set $\mu \rightarrow m_\chi$ and substitute the value of $m_\chi$ from criteria 1 in the second criteria, this gives us:

\textbf{2$^\prime$. Modified Criteria for large spin-flip coupling rate at $z = 17$ (valid for small $f$)}
\begin{equation}
\alpha_\chi\alpha_e > 4.8 \times 10^{-16}\sqrt{f}.
\end{equation}
These two criteria 1 and 2 (or 2$^\prime$) give us an analytic range of parameters $m_\chi$ and $\alpha_\chi\alpha_e$ for a given value of $f$. We plot this analytic solution in each of the panels of fig.~\ref{fig:parameterspaceplot}.
Note that in addition to the criteria on large spin-flip rates, we also need a self-consistency criteria for this solution to ensure tight coupling between the gas and the dark matter fluid. The criteria for self-consistency are given by $\Gamma_H ,\Gamma_\chi > H $ at $z =265$, i.e. the gas and the DM remain tightly coupled to each other till the combined fluid kinetically decouples from the CMB\footnote{Thereafter, for small enough $f$ or $m_\chi$, $\Teff$ is dominated by $T_\chi$, so it is sufficient that the DM fluid decouples from the gas any time after $z=265$.}. Since, $\Gamma_H= R \Gamma_\chi$, and since $R > 1$, it is sufficient to look at the self-consistency criteria $\Gamma_\chi > H$ at $z=265$. Thus, the condition on couplings that we obtain from self-consistency is:

\textbf{3. Self-consistency criteria for tight kinetic coupling till $z=265$}
\begin{equation}
\alpha_\chi\alpha_e > 1.8\times10^{-19} \left(  \frac{M}{\textrm{GeV}} \right ) \left(\frac{\textrm{GeV}}{\mu}\right)^{1/2},
\end{equation}
where $M = m_\chi + m_H$. Once again, for small $f$ and substituting criteria 1 for the mass $m_\chi$ in terms of $f$, we obtain:

\textbf{3$^\prime$. Modified self-consistency criteria for tight kinetic coupling till $z=265$ (valid for small $f$)}
\begin{equation}
\alpha_\chi\alpha_e > 2.13 \times10^{-19}  \frac{1}{\sqrt{f}}
\end{equation}
This criteria is a weaker constraint than criteria 2 (or 2$^\prime$) for $f > 0.001$, so self-consistency is automatically satisfied in such cases by requiring tight spin-flip coupling.

\subsubsection{Numerical results for strong coupling reference point}
\label{sec:strongcouplingnumerical}
\begin{figure}[H]
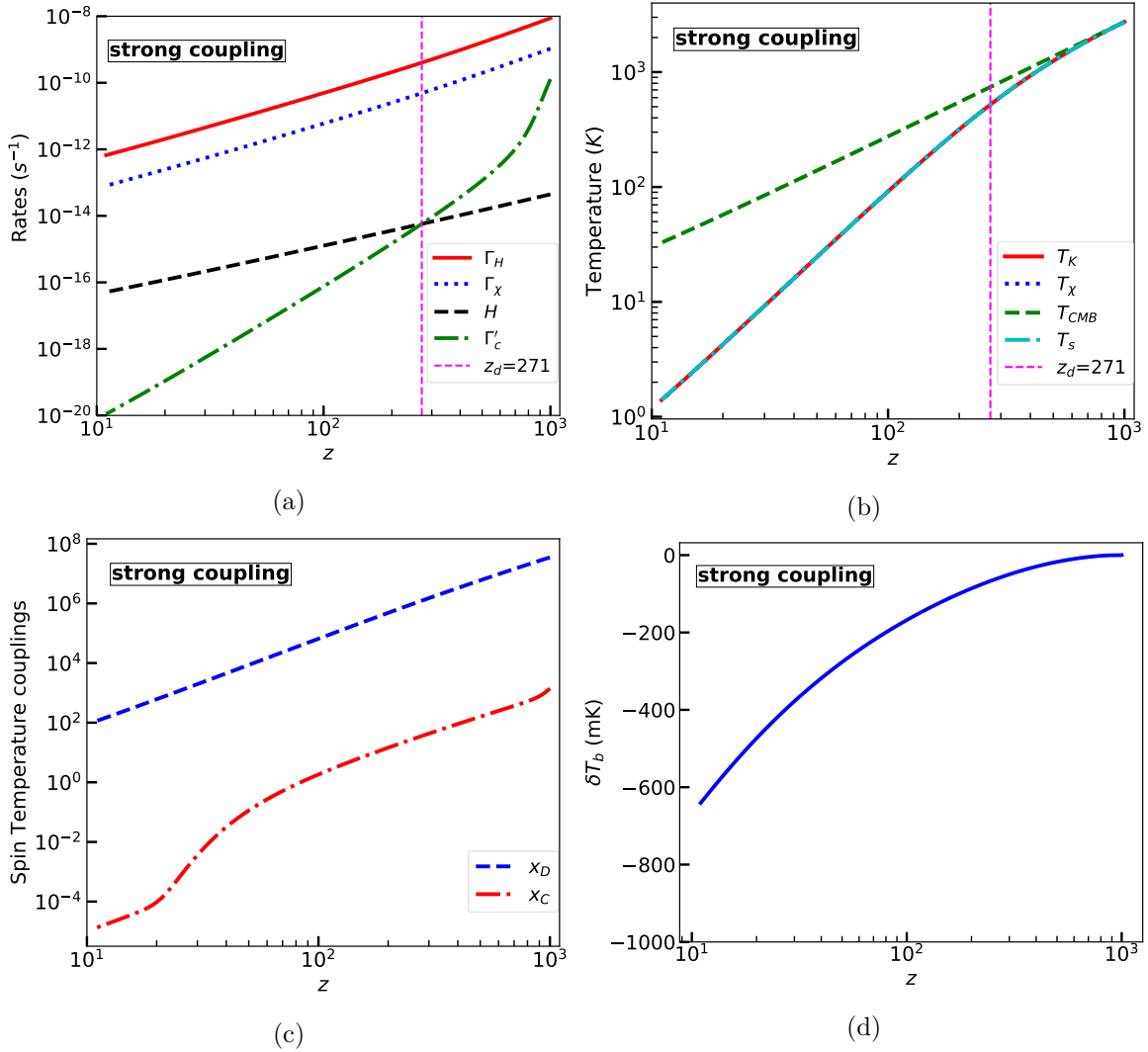

\begin{subfigure}{.5\textwidth}
 \centering
   \includegraphics[width=\linewidth]{{strong_Rates_vs_z}.pdf}
 \caption{}
  \label{fig:rates_strong_coupling}
\end{subfigure}%
\begin{subfigure}{.5\textwidth}
  \centering
 \includegraphics[width=\linewidth]{{strong_temperature_vs_z}.pdf}
  \caption{}
  \label{fig:tempevolution_strong_coupling}
\end{subfigure}
\begin{subfigure}{.5\textwidth}
 \centering
 \includegraphics[width=\linewidth]{{strong_coupling_vs_z}.pdf}
  \caption{}
  \label{fig:xdxc_strong_coupling}
\end{subfigure}%
\begin{subfigure}{.5\textwidth}
  \centering
  \includegraphics[width=\linewidth]{{strong_T21_vs_z_new}.pdf}
  \caption{}
  \label{fig:Tb_strong_coupling}
\end{subfigure}
\caption{ Plots for the strong coupling scenario with parameters $f=0.1$, $m_\chi =0.06$~GeV and $\alpha_e\alpha_\chi = 5\times 10^{-15}$. (a) Rate evolution of the DM ($\Gamma_\chi$) and gas ($\Gamma_H$) kinetic coupling to each other. Also shown is the effective rate of coupling of the combined DM-gas fluid to the CMB ($\Gamma^\prime_c$) indicating a decoupling at $z=271$. (b) Temperature evolution of the DM and gas kinetic temperatures, compared to the CMB temperature. The gas and DM are tightly kinetically coupled to each other and have a common temperature evolution. Adiabatic cooling of the tightly coupled DM-gas fluid begins at $z=271$ and leads to a low $\Teff$ at $z=17$. The spin temperature is tightly coupled to the DM-gas temperature through a large spin-flip rate~$x_D$. (c) The collisional coupling $x_C$ to the gas kinetic temperature $T_K$, and the spin-flip coupling $x_D$ to the effective temperature $\Teff$. The DM induced spin-flip coupling rate dominates both the collisional and CMB induced spin-flip rates at all redshifts in this scenario. (d) Predicted differential brightness temperature $\delta T_b(z)$ as a function of redshift. The rise at low redshifts due to standard expected astrophysics is not shown.}
\label{fig:scenario1strongstrongcouplingbenchmark}
\end{figure}

 In our numerical scan, in addition to the basic criteria that we had imposed to identify regions of parameter space of interest in the beginning of this section, we impose the requirement that $x_D(z=17)> 10$ to identify solutions in the strong coupling scenario. We will discuss a particular reference point with $f=0.1$, $m_\chi =0.06$~GeV and $\alpha_e\alpha_\chi = 5\times 10^{-15}$.
In fig.~\ref{fig:rates_strong_coupling}, we show the energy transfer rates $\Gamma_H$, $\Gamma_\chi$ and the Hubble rate $H$ as functions of $z$ for this reference point. We can see from the figure that the rates $\Gamma_H$ and $\Gamma_\chi$ are both larger than the Hubble rate throughout the redshift range $10 \lesssim z \lesssim 1000$, indicating that the DM and gas are tightly kinetically coupled to each other. This behaviour is clearly seen in fig.~\ref{fig:tempevolution_strong_coupling}, in which we show the temperature evolution of $T_K$ and $T_\chi$, where both temperatures track each other very closely. In fig.~\ref{fig:rates_strong_coupling}, we also plot the effective ``compton coupling rate'' of the DM-gas fluid to the CMB ($\Gamma_c^\prime$), and we see that this rate decouples (drops below the Hubble rate) at $z=271$ (near $z=265$ as predicted by our analytic estimate). In fig.~\ref{fig:tempevolution_strong_coupling}, we can see the common temperature evolution of the DM-gas switches from tracking the CMB temperature (and scaling as $(1+z)$) from $z= 1000 $ till $z=271$, to adiabatic cooling (scaling as $(1+z)^2$) from $z=271$ onwards to lower redshifts.

In fig.~\ref{fig:xdxc_strong_coupling}, we show the evolution of the spin-temperature coupling $x_D$ to the temperature $\Teff$ (which is just the same as the common temperature of the DM-gas fluid for strong coupling), and the collisional coupling $x_C$ to the gas temperature $T_K$. At all redshifts, we see that the coupling $x_D\gg x_C$ and also $x_D\gg 1$. This indicates that the DM spin-flip coupling reaction rate $D_{10}$ is the most dominant spin-flip rate over both the collisional coupling rate, as well as the CMB induced spin-flip rate $B_{10}$. Thus, at all redshifts, and in particular, at $z=17$, $T_s$ tracks the common temperature of the gas-DM fluid (also shown in fig.~\ref{fig:tempevolution_strong_coupling}). The early decoupling of the gas-DM fluid at $z=271$ leads to a value of $T_s = \Teff = 3.32$~K at $z = 17$, which yields a differential brightness temperature dip $\delta T_b(z=17) \simeq -516$~mK.

In fig.~\ref{fig:Tb_strong_coupling}, we show the differential brightness temperature as a function of redshift $z$, inferred from the spin temperature evolution history. Since, $\delta T_b \propto T_s-\Tcmb$ and $T_s$ drops faster than $\Tcmb$ at low redshift, we see that $\delta T_b$ becomes more negative at lower redshifts, indicating greater absorption at these redshifts. We expect that below some redshift, $z\lesssim 15$, Ly-$\alpha$ coupling to $T_K$ would turn on due to star formation and be the dominant determining coupling of the spin-temperature to the color temperature (or $T_K$). Moreover $X$-ray heating would raise $T_K$, thus leading to a rise in the spin temperature, and also the brightness temperature below some redshift. However, since the details of this would depend on the astrophysical model, we do not show this rise in our figure, but rather our figure is to be taken seriously only above redshifts $z \gtrsim 15$.

\begin{figure}
\begin{center}
  \includegraphics[width=0.7\linewidth]{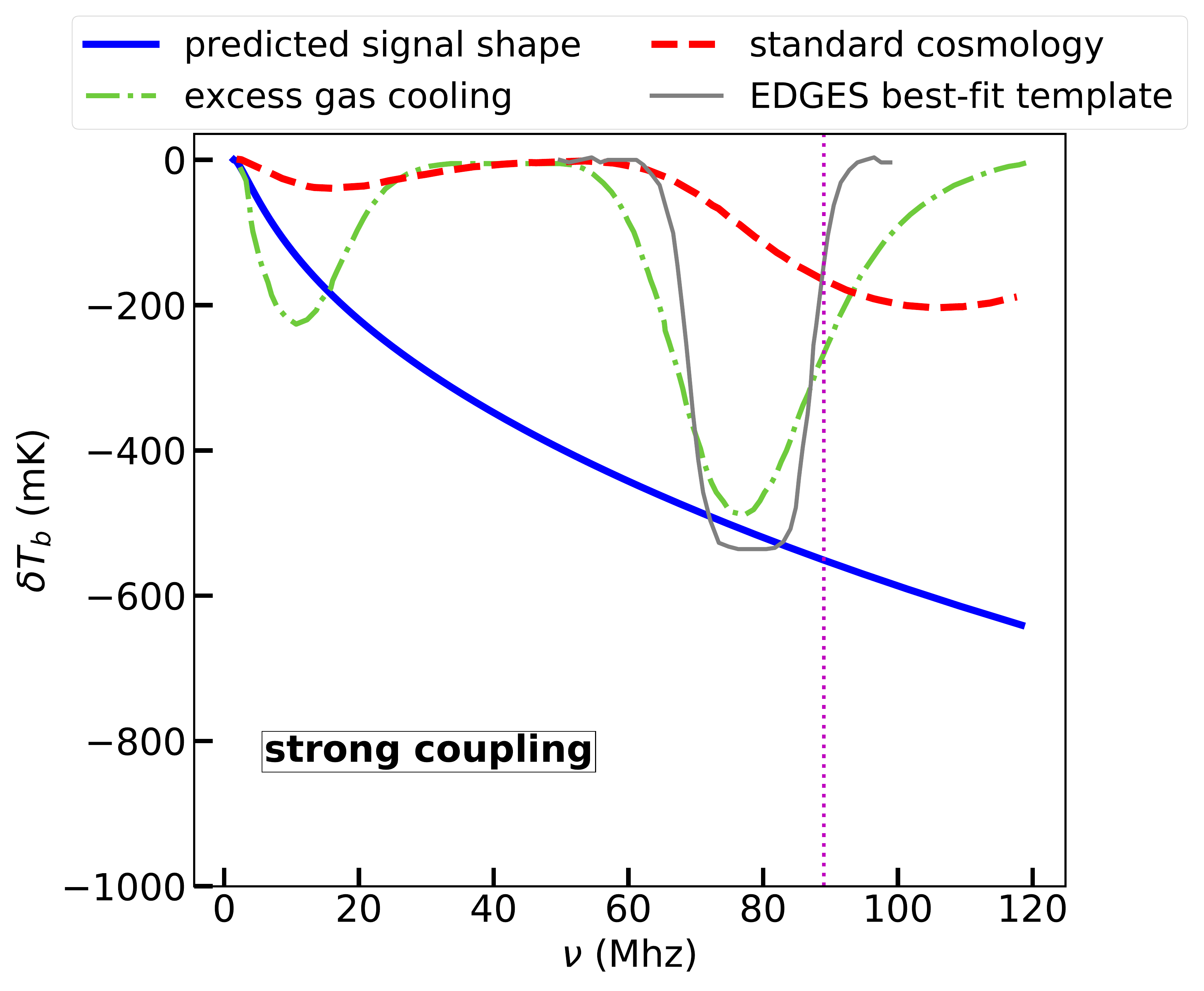}
\caption{Predicted differential brightness temperature $\delta T_b(\nu)$ as a function of frequency (blue solid curve) with parameters $f=0.1$, $m_\chi =0.06$~GeV and $\alpha_e\alpha_\chi = 5\times 10^{-15}$. Our model predicts a single, strong broadband absorption feature that begins at $5.3$~MHz~$(z=271)$ and extends all the way to high frequencies. Above $\sim 89$~MHz $(z=15)$, indicated by the vertical dotted magenta line, we expect that Ly-$\alpha$ couplings are expected to become important, leading to a rise in $\delta T_b$ due to standard expected astrophysics. This rise is not shown in the figure. We have also shown for comparison the prediction of the standard cosmology~\cite{10.1093/mnras/stw2412} (red dashed) and for an excess gas cooling model~\cite{Mirocha:2018cih} (green dot-dashed) with parameters chosen to explain the EDGES best-fit absorption dip (gray)~\cite{Bowman:2018yin}. We see that, unlike our model, both these models have band-limited absorption features with transitions induced due to collisional couplings and Ly-$\alpha$ photons.
}
\label{fig:Tbvsnu_strong_coupling}
\end{center}
\end{figure}

In fig.~\ref{fig:Tbvsnu_strong_coupling} we show, for our strong coupling reference point, the predicted differential brightness temperature evolution as a function of redshifted 21~cm frequency. We also show for comparison, in the same figure, the expected signal in a model of standard cosmology~\cite{Burns_2017, 10.1093/mnras/stw2412} and also in a phenomenological model with excess gas cooling without spin-flip interaction as obtained in ref.~\cite{Mirocha:2018cih}, where the specific curves in our figure have been taken from ref.~\cite{Burns:2019zia}.

In the standard cosmology model, the low frequency absorption feature near $\sim$ 20 MHz in the cosmic dark ages has a well predicted frequency and magnitude at the minimum, but the higher frequency absorption dip (at $\sim$ 110 MHz in the figure) has a much larger uncertainty on both of these features because of dependence on the assumptions of the unknown properties of the first sources. A key assumption of the model is that it assumes that the UV and X-ray emission properties of the stars in the first sources are consistent with those of Pop II stars. The model parameters are calibrated to match measurements of the high-$z$ galaxy luminosity function (LF) and further tuned to match the Thomson scattering optical depth of the cosmic microwave background. Current uncertainties in the faint-end of the LF, binary populations in star-forming galaxies, and UV and X-ray escape fractions introduce a $\sim$ 20 MHz uncertainty in the location of the high frequency minima and $\sim$ 50 mK uncertainty in the maximum absorption depth~\cite{10.1093/mnras/stw2412}. Furthermore, violation of the basic assumptions, such as by assuming that the Pop III stars of the first sources have significantly different properties from Pop II stars can change the location of the minima of the high frequency absorption feature to frequencies as low at 50 MHz~\cite{Burns_2017}.

The excess cooling model (shown in  fig.~\ref{fig:Tbvsnu_strong_coupling}) is consistent with high-$z$ luminosity functions inferred from the Hubble Space Telescope, but introduces a phenomenological faster-than-adiabatic (or super-adiabatic) cooling~\cite{Mirocha:2018cih} at a redshift earlier than that of the standard cosmological adiabatic transition at $z\sim 130$. The parameters of this model are tuned to explain the best-fit EDGES anomalous absorption signal~\cite{Bowman:2018yin}.
We see that in both the standard cosmology and in the excess cooling model, there are two distinct, band-limited absorption features, one at high redshifts ($\sim$~20~MHz), which corresponds to the dominance of collisional coupling of the spin temperature to the gas temperature, and the second at lower redshifts/higher frequencies (with a minima at $\sim$~$110$ MHz for the standard cosmology model and at $\sim$~$80$ MHz for the excess cooling model), which corresponds to dominance of Ly-$\alpha$ couplings. The model with excess gas cooling, has deeper absorption dips than that of the standard cosmology, but the absorption features are still distinct.

In contradistinction to these signals, our strong coupling reference point predicts a single, strong, broadband absorption feature that extends from 5.3~MHz ($z=271$) all the way up to 89~MHz ($z=15$). At all frequencies in this range, the dip in brightness temperature is caused by spin-flip interactions between the dark matter and neutral hydrogen.

Moreover, in general in the strong coupling scenario, since the spin temperature is controlled by the adiabatically cooling DM-gas fluid temperature, the shape of the absorption signal is very precisely predicted in this frequency range, independently of the exact dark matter mass and couplings and also independently of our assumptions on the initial conditions. In particular, given the strength of the absorption dip at any frequency, for e.g. given $\delta T_b ($78$~\textrm{MHz}, z=17) = -516$~mK, we can precisely predict the location of the start of the absorption feature at $\nu = 5.3$~MHz $(z=271)$.

At larger frequencies than 89 MHz the signal is expected to rise back up due to a combination of Ly-$\alpha$ photons and $X$-ray heating. We have not shown this astrophysical model dependent rise in our figure.

\subsection{Scenario 2: Weak Coupling}

\subsubsection{Analytic understanding of weak coupling solutions}

When the coupling product $\alpha_\chi \alpha_e$ is sufficiently weak, the DM and gas are kinetically decoupled from each other. In this scenario, the DM begins to cool adiabatically from the assumed initial condition, which is set at recombination. However, as we will show, for sufficiently low DM masses, \textit{it is possible for the DM and gas to be kinetically decoupled (small $\Gamma_\chi$), while still having a strong spin-flip interaction rate (large $D_{10}$).} The spin-flip rate of the gas due to DM interactions can be large enough so that the spin temperature couples to the cold dark matter temperature and we can get a large absorption dip in the differential brightness temperature in this scenario.

Since we have decided to anchor our absorption spectrum to a fixed value at $z=17$, we are interested in looking for regions of parameter space that give rise to a large spin-flip rate at this particular redshift.

Assuming that $\Gamma_\chi \ll H$ at $z=1000$, would guarantee decoupling of the DM from the gas. Since we have assumed the initial condition to be $T_\chi = T_K= \Tcmb$ at $z=1000$, we would thus have the following temperature evolution of the DM temperature
\begin{align}
\label{eq:tempevolutionweak}
T_\chi(z) =  \Tcmb(z=1000)\left (\frac{1+z}{1+1000} \right)^2.
\end{align}
For small DM mass, $\Teff(z) \simeq  T_\chi(z)$.
At our anchor point $z=17$, this would lead  to $\Teff(z=17) = 0.9$~K, which would be the coldest possible $\Teff$ at this redshift, given our initial conditions.

The hierarchy between the DM induced spin-flip rate of neutral hydrogen $(D_{10})$ and the kinetic coupling rate of DM to the gas $(\Gamma_\chi)$  is possible because $\Gamma_\chi = {\frac{3S}{ R}}D_{10}$, where $S= \left(\frac{\Delta \mu}{2 M \Teff}\right)\ll1$, and $R = \frac{n_\chi}{n_H} \gg 1$, for sufficiently low DM masses (see discussion following eq.~\ref{eq:gammachimain}). In particular, given the scaling of $\Teff$ as $(1+z)^2$, we can see that
$D_{10} \propto \frac{1}{m^{3/2}}(1+z)^4$~(see eq.~\ref{eq:deex_sigv})  and $\Gamma_\chi \propto m^{1/2}(1+z)^2$~(see eq.~\ref{eq:gammachimain}).
Thus, although $D_{10}$ drops rapidly with redshift, for sufficiently low DM masses it is still possible to obtain a large spin-flip rate $D_{10}\sim B_{10}$ at $z=17$, while insisting that the DM decouples at $z=1000$, $H \gg \Gamma_\chi$ at $z=1000$.

Since we have assumed that Ly-$\alpha$ couplings have not yet turned on at $z=17$, and also since the collisional couplings can be neglected at this redshift, we have,
\begin{equation}
    T_s^{-1} \simeq \frac{{\Tcmb}^{-1} +  x_{D} T_{\textrm{eff}}^{-1}} {1 +x_{D}}.
\label{eq:tsmod2}
\end{equation}
In order to obtain $\delta T_b(z=17) = - 500$~mK, we would need $T_s(z=17) \simeq 3.32$~K. Plugging this desired value into the left-hand side of the equation above, and for the lowest possible value of $\Teff=0.9$~K at $z=17$, this would imply that we need $x_D(z=17) = 0.34$, i.e. $D_{10} = 0.34 B_{10}$; moderate, but not tight spin-flip coupling at $z=17$ is needed since $\Teff$ is much colder than the desired spin temperature\footnote{For the range of $-600~\textrm{mK}< \delta T_b(z=17) <-400~\textrm{mK}$ or $2.7~\textrm{K}< T_s(z=17) <3.9~\textrm{K}$, we need $0.27<x_D(z=17) <0.47$. In our numerical scan, we take the criteria $x_D(z=17)< 0.47$ to define the weak coupling regime of the parameter space.}.

This condition on $x_D$ gives us,

\textbf{1. Criteria for moderate spin-flip coupling $\mathbf{x_D(z=17) = 0.34}$:}
\begin{equation}
\label{eq:moderatespincouplingcrit1}
\alpha_e \alpha_\chi  = \frac{5.6 \times 10^{-17}}{f} \left(\frac{m_\chi}{\textrm{GeV}}\right)^{3/2}.
\end{equation}
For self-consistency of our solution, we need the coupling to be sufficiently weak so that we are in the regime where the DM decouples at $z=1000$, this criteria is given by the requirement that $\Gamma_\chi \lesssim H/10$ at $z=1000$, or in terms of the couplings,

\textbf{2. Self-consistency criteria for weak kinetic coupling of DM:}
\begin{equation}
\label{eq:selfconsistencyweak}
\alpha_e \alpha_\chi <5.1\times10^{-21}\sqrt{\frac{\textrm{GeV}}{m_\chi}}.
\end{equation}
Comparing this with criteria 1 above, we see that our weak coupling solution is valid only up to a maximum mass (or maximum coupling). We can thus write down criteria 2 in the form below,

\textbf{2$^\prime$. Modified self-consistency criteria for weak kinetic coupling of DM:}
\begin{equation}
m_\chi < 9.5 \times 10^{-3} \sqrt{f}~\textrm{GeV},
\end{equation}
or equivalently,
\begin{equation}
\alpha_e \alpha_\chi < \frac{5.2 \times 10^{-20}}{f^{1/4}}.
\end{equation}

\vspace{2mm}

These two criteria 1 and 2 (or 2$^\prime$) give us an analytic range of parameters $m_\chi$ and $\alpha_\chi\alpha_e$ for a given value of $f$, which define the weak coupling region of our parameter space. We have plotted this analytic solution in each of the panels of fig.~\ref{fig:parameterspaceplot}.

\vspace{4mm}

\textbf{What happens to the baryons in this scenario?}

There are two possible histories for the baryonic temperature evolution which depend upon the mass of the DM particle in our parameter space of interest. We will discuss each of these possibilities in turn.

\textbf{Possibility 1: No excess cooling as compared to the standard cosmology for very low DM masses}

 If $\Gamma_H < H$ at $z = 1000$, the gas will just decouple from the DM and cool at $z = 130$ adiabatically as in the standard scenario with no exotic DM couplings. This criteria can be reexpressed as,
\begin{equation}
\alpha_e \alpha_\chi  < \frac{9.9 \times 10^{-21}}{f} \left(\frac{m_\chi}{\textrm{GeV}}\right)^{1/2}.
\end{equation}
This condition is satisfied for parameters that satisfy criteria 1, as long as $m_\chi < 1.8 \times 10^{-4}$~GeV. For such low values of DM mass, the baryon temperature evolution is completely unaffected and the gas kinetic temperature would be large compared to the spin temperature, i.e. the gas would have a temperature $T_K(z=17) = 6.8$~K, as in the standard cosmology without exotic DM interactions, whereas the spin temperature $T_s(z=17) = 3.32$~K. Despite the large gas \textit{kinetic} temperature, the gas \textit{spin} temperature is low enough to lead to a strong absorption signal at $z=17$ because of a combination of the large DM spin-flip rate $D_{10}$ at $z=17$ and the low DM temperature.

%\newpage

\textbf{Possibility 2: Super-adiabatic cooling}

For larger values of the dark matter mass,  $ 1.8 \times 10^{-4}$~GeV~$< m_\chi <9.5 \times 10^{-3} \sqrt{f}~\textrm{GeV}$\footnote{This range exists for all the values of $f$ that we are considering. For smaller values of $f$ this interval shrinks to zero size, and this second type of weak coupling solution does not exist; the baryons would just cool as in the standard cosmology.}, the baryons have two competing rates which determine their temperature evolution, on the one hand the coupling to DM which tries to lower the baryon temperature, and the coupling to CMB which tries to keep the baryons at the CMB temperature. $\Gamma_H \propto (1+z)^2$ and $\Gamma_c \propto (1+z)^4 x_e(z)$, so at some point it is conceivable that if the coupling is not too small, $\Gamma_H$ can become the dominant rate that determines the evolution of the gas temperature. If $\Gamma_H > H$, the gas will cool \textit{super-adiabatically} (faster than $(1+z)^2$) in a bid to latch on to the low DM temperature. If at some redshift $\Gamma_H < H$, then the gas will switch to cooling adiabatically from this point onwards with $T_K \propto (1+z)^2$. In general, the evolution of the gas temperature in this range of DM masses is complicated and must be solved for numerically.

\vspace{3mm}
Note that for both possibilities in the weak coupling scenario, the evolution of the gas temperature is practically irrelevant for (and distinct from) the evolution of its spin temperature (since $\Teff \sim T_\chi$ for low DM masses), as long as the coupling $x_D$ is the dominant spin-flip coupling. \emph{The absorption signal in this scenario would trace only the spin-temperature and not the kinetic temperature of the gas for most of the cosmological history, as opposed to the standard cosmology or even excess gas cooling models where the absorption signal is expected to be a tracer of the gas kinetic temperature.} At low redshifts, if Ly-$\alpha$ coupling becomes strong due to star formation, then the gas temperature becomes relevant for determining the further evolution of the spin temperature and hence could be tracked using the differential brightness temperature.

\subsubsection{Numerical results for weak coupling reference point}
In our numerical scan, we identify the weak coupling regions of our parameter space by the criteria $x_D(z=17)< 0.47$. We will discuss a particular reference point with $f=0.1$, $m_\chi =1.1 \times 10^{-3}$~GeV and $\alpha_e\alpha_\chi = 2.7\times 10^{-20}$. This reference point satisfies the criteria of moderate spin-flip coupling in eq.~\ref{eq:moderatespincouplingcrit1} and also the mass lies in the relatively larger range for this scenario, $ 1.8 \times 10^{-4}$~GeV~$< m_\chi <9.5 \times 10^{-3} \sqrt{f}~\textrm{GeV}$.

\begin{figure}[H]
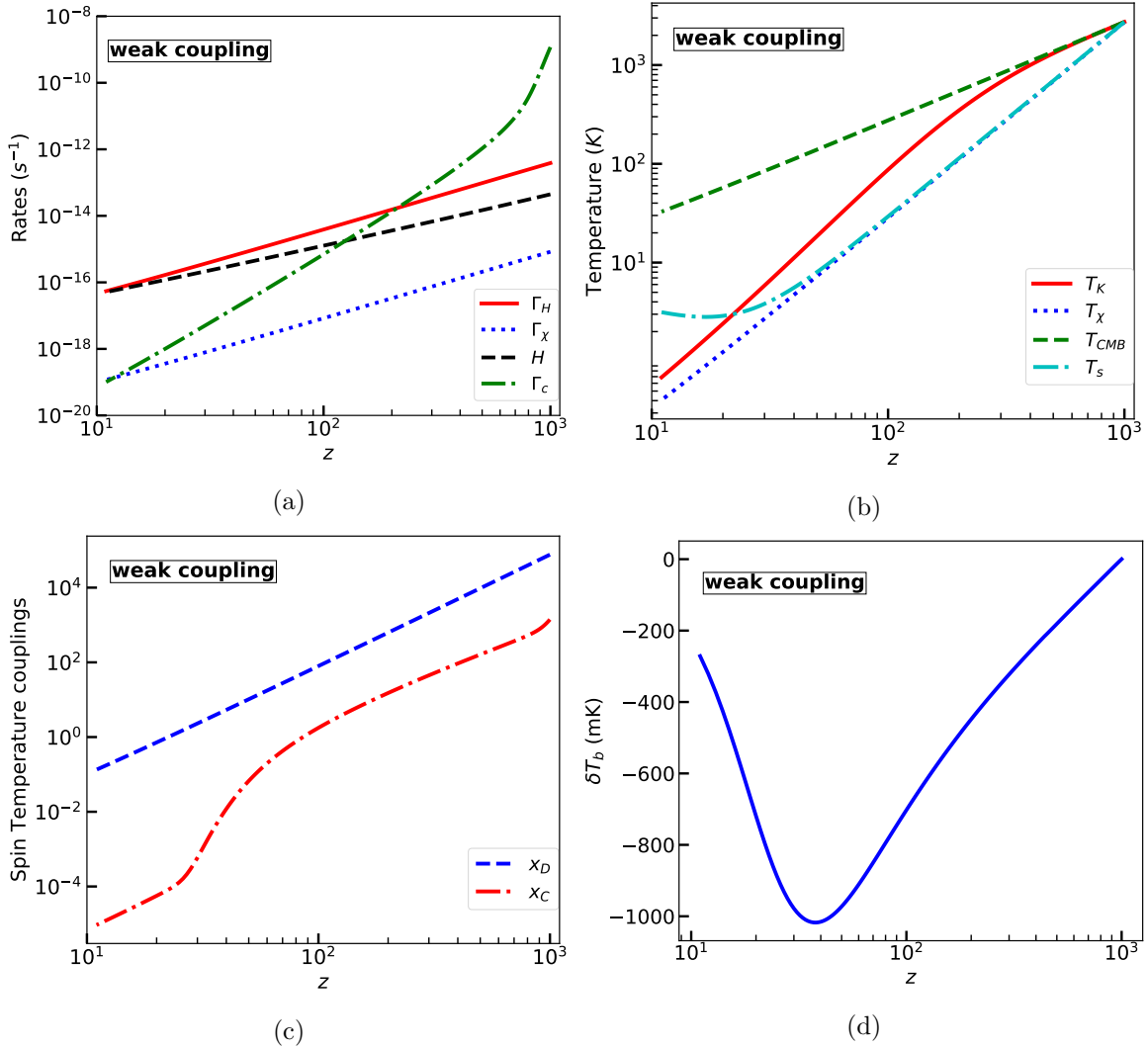

\begin{subfigure}{.5\textwidth}
 \centering
   \includegraphics[width=\linewidth]{{weak_Rates_vs_z}.pdf}
 \caption{}
  \label{fig:rates_weak_coupling}
\end{subfigure}%
\begin{subfigure}{.5\textwidth}
  \centering
 \includegraphics[width=\linewidth]{{weak_temperature_vs_z}.pdf}
  \caption{}
  \label{fig:tempevolution_weak_coupling}
\end{subfigure}
\begin{subfigure}{.5\textwidth}
 \centering
 \includegraphics[width=\linewidth]{{weak_coupling_vs_z}.pdf}
 \caption{}
  \label{fig:xdxc_weak_coupling}
\end{subfigure}%
\begin{subfigure}{.5\textwidth}
  \centering
  \includegraphics[width=\linewidth]{{weak_T21_vs_z_new}.pdf}
 \caption{}
  \label{fig:Tb_weak_coupling}
\end{subfigure}
\caption{Plots for the weak coupling scenario with parameters $f=0.1$, $m_\chi =1.1 \times 10^{-3}$~GeV and $\alpha_e\alpha_\chi = 2.7\times 10^{-20}$. (a) Rate evolution of the DM ($\Gamma_\chi$) and gas ($\Gamma_H$) kinetic coupling to each other and the rate coupling $\Gamma_c$ of the gas to the CMB. For this scenario, the dark matter is decoupled from both the gas and the CMB, and adiabatically cools from our initial conditions at $z=1000$. (b) Temperature evolution of the DM and gas kinetic temperatures compared to the CMB temperature. The effective temperature is given by $\Teff \simeq T_\chi$ for low DM masses. The spin temperature $T_s$ is initially tightly coupled to $\Teff$ through a large spin-flip coupling rate $x_D$, but then starts to heat up to the CMB temperature at $z\sim 40$, when $x_D \sim 5$. The gas kinetic temperature and the spin temperature evolve almost completely independently in this scenario. (c)~The collisional coupling $x_C$ to the gas kinetic temperature $T_K$ and the spin-flip coupling $x_D$ to the effective temperature $\Teff$. The rate $D_{10}$ dominates over both the collisional and the CMB spin-flip rates till $z\sim40$, but below this redshift the coupling of $T_s$ to the CMB temperature begins to dominate and the spin temperature attempts to rise towards $\Tcmb$. (d) Predicted differential brightness temperature $\delta T_b(z)$ as a function of redshift. The low redshift behavior ($z\lesssim 15$) due to the astrophysics of Ly-$\alpha$ coupling and $X$-ray heating is not included in the figure.}
\label{fig:scenario2weakcouplingbenchmark}
\end{figure}

In fig.~\ref{fig:rates_weak_coupling}, we show the reaction rates $\Gamma_H$, $\Gamma_\chi$, $\Gamma_c$ and the Hubble rate $H$ as functions of $z$ for this reference point. We can see from the figure that the rate  $\Gamma_\chi$ is less than the Hubble rate throughout the redshift range $10 \lesssim z \lesssim 1000$, indicating that the DM decouples from the gas (and the CMB) and begins cooling adiabatically from the initial condition assumed at $z= 1000$. Meanwhile, the rate $\Gamma_H$ is larger than the Hubble rate throughout the redshift range $10 \lesssim z \lesssim 1000$, indicating that the gas is kinetically coupled to the adiabatically cooling DM. However, the gas is also coupled to the CMB and at high redshifts, and it is this interaction which dominantly determines the evolution of the gas kinetic temperature at early times.

The evolution of the gas and DM kinetic temperatures are shown in fig.~\ref{fig:tempevolution_weak_coupling}. At low DM masses, as at this reference point, the effective temperature $\Teff\simeq T_\chi$, and since the DM is cooling adiabatically, $\Teff$ cools as $(1+z)^2$. At $z=17$, $\Teff\simeq0.9$~K, which is the coldest possible value for the effective temperature. The gas is initially tightly coupled to the CMB and tracks the CMB temperature from $z\sim1000$ down to $z\sim 300$. Near $z\sim 300$, the kinetic coupling of the gas to the CMB and the coupling of the gas to the DM become comparable, indicating a competition between these two rates in determining the evolution of $T_K$. The coupling to the DM makes the gas cool below the CMB temperature. Finally, at low redshifts $(z\lesssim200)$, the only relevant coupling is of the gas to the DM, and the gas attempts to cool super-adiabatically down to the DM temperature.

In fig.~\ref{fig:xdxc_weak_coupling}, we show the evolution of the spin-temperature coupling ($x_D$) to the temperature $\Teff$, and the collisional coupling $(x_C)$ to the gas temperature $T_K$. At all redshifts, we see that the coupling $x_D \gg x_C$, which indicates that the DM spin-flip coupling reaction rate $D_{10}$ dominates over the collisional coupling rate. Moreover, from $40\lesssim z\lesssim 1000$, $x_D  \gtrsim1$, which implies that the DM spin-flip rate $D_{10}$  is larger than the CMB induced spin-flip rate $B_{10}$ over this redshift range. Thus, the spin temperature $T_s$ is tightly coupled to $\Teff$ from $z\sim 1000$ till a redshift $z\sim40$. At $z\sim 40$, $x_D\sim 1$ and the CMB induced spin-flip rate $B_{10}$ begins to compete with the DM induced spin-flip rate $D_{10}$. We can see this behaviour in the evolution of $T_s$ shown in fig.~\ref{fig:tempevolution_weak_coupling}. $T_s$ initially tracks the dark matter temperature till $z\sim40$, but then as the coupling $x_D$ drops to $\mathcal{O}(1)$ values, the spin temperature begins to cool slower than the adiabatically cooling DM and eventually attempts to rise back up towards the CMB temperature. For the reference set of parameters we have chosen, $T_s  = 2.8$~K at $z=17$ -- a value which is intermediate between the DM temperature of $0.9$~K and the CMB temperature of $49$~K at this redshift.

In fig.~\ref{fig:Tb_weak_coupling}, we show the differential brightness temperature as a function of redshift $z$, inferred from the spin temperature evolution history. Since $\delta T_b \propto T_s-\Tcmb$ and $T_s$ drops and stays below $\Tcmb$ at high redshifts (from $40\lesssim z \lesssim 1000$), we see a dip in $\delta T_b$ which begins at our initial conditions of $z\sim1000$. We get a minimum of $\delta T_b$ at $z=40$, where $x_D\sim 5$. As the coupling $x_D$ drops further from this point onwards, the spin temperature attempts to rise back up to the CMB temperature, we get a rise in $\delta T_b$ till lower redshifts. Since $T_s = 2.8~K$ at $z=17$, this leads to a differential brightness temperature dip $\delta T_b(z=17) \simeq -580$~mK.

In this scenario, we see that it is possible to get \textit{both a dip and a rise} in the differential brightness temperature, purely from a combination of the DM temperature evolution and the spin-flip coupling $x_D$, \textit{even without considering Ly-$\alpha$ couplings.} However, we expect that below some redshift, $z\lesssim 15$, Ly-$\alpha$ coupling to $T_K$ would turn on due to star formation, and $X$-ray heating would raise $T_K$. Since the details of this would depend on the astrophysical model we do not show the effects of this in our figure.

\begin{figure}
\begin{center}
  \includegraphics[width=0.7\linewidth]{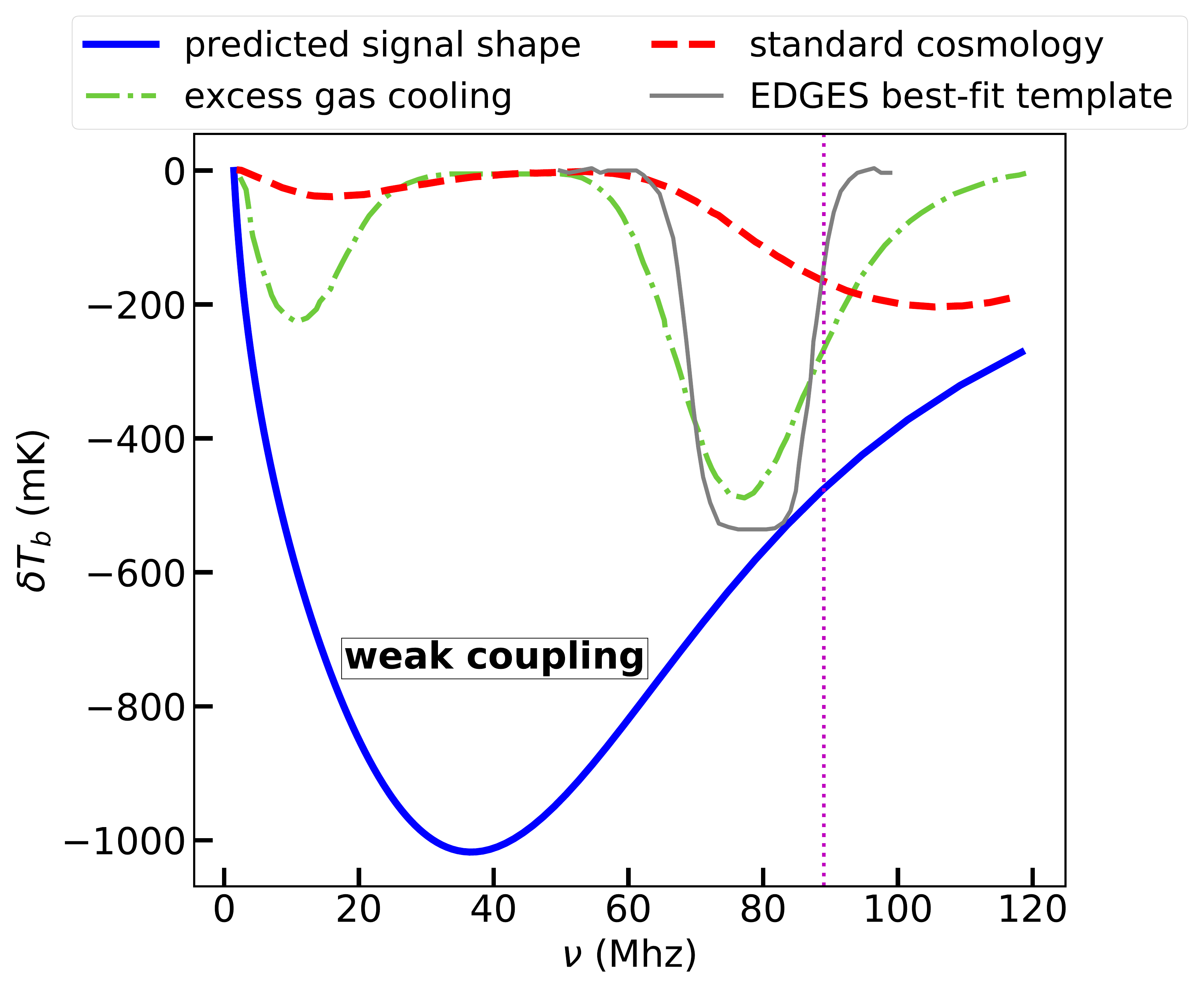}
 \caption{Predicted differential brightness temperature $\delta T_b(\nu)$ as a function of frequency (blue solid curve) with parameters $f=0.1$, $m_\chi =1.1 \times 10^{-3}$~GeV and $\alpha_e\alpha_\chi = 2.7\times 10^{-20}$. Our model predicts a single, strong broadband absorption feature that begins at $1.4$~MHz~($z\sim1000$) and extends all the way to high frequencies where Ly-$\alpha$ couplings are expected to become important. Unlike the strong coupling scenario, in this scenario we get a deep minima from the cosmic dark ages at $35$~MHz, ($z\simeq 40$), as the spin temperature attempts to rise up towards the CMB temperature. At frequencies above $89$~MHz ($z=15$), indicated by the vertical dotted magenta line, Ly-$\alpha$ couplings and $X$-ray heating could lead either to a sharper rise or a possible second dip, but these features are not shown in the figure. Unlike our model where the absorption feature is due to DM spin-flip interactions, both the standard cosmology (red dashed) and excess gas cooling models (green dot-dashed) have distinct band-limited absorption features with transitions due to collisional couplings and Ly-$\alpha$ photons.
}
\label{fig:Tbvsnu_weak_coupling}
\end{center}
\end{figure}

In fig.~\ref{fig:Tbvsnu_weak_coupling}, we show the differential brightness temperature evolution as a function of redshifted 21~cm frequency. We also show for comparison, in the same figure, the expected band-limited signals from the standard cosmology and from the excess gas cooling model, both of which were discussed in sec.~\ref{sec:strongcouplingnumerical}, where the exact curves have been taken from ref.~\cite{Burns:2019zia}. Once, again we see in our model a prediction of a single, strong, broadband absorption feature which extends from $\nu = 1.4$~MHz ($z=1000$)\footnote{This value depends on our assumed initial conditions, since we assumed $T_\chi = \Tcmb$ at $z=1000$, no absorption is expected at higher redshifts/lower frequencies.} all the way up to the cosmic dawn, unlike the two distinct band-limited absorption features of the standard cosmology and excess cooling models. The key observational feature of the weak coupling absorption spectrum that distinguishes it from the strong coupling scenario, is the existence of a deep minimum of $\delta T_b$ at low frequencies originating from the cosmic dark ages -- near 35~MHz $(z=40)$ for our reference point. From this point onwards, $\delta T_b$ rises up and at $78$~MHz $(z=17)$, the spectrum reaches its ``pinned'' value of $\delta T_b \simeq -580$~mK.

For frequencies larger than those corresponding to the cosmic dawn, above 89~MHz ($z=15$) by our assumption, Ly-$\alpha$ photons and $X$-ray heating could lead to two distinct possibilities: a) if X-ray heating is rapid and precedes strong Ly-$\alpha$ coupling, then the gas temperature would quickly rise up towards the CMB temperature, and the spin temperature would later attempt to latch on to this large $T_K$. In this case, the rise in $\delta T_b$ at higher frequencies would continue at a rate faster than what is shown in the figure, b) if strong Ly-$\alpha$ coupling turns on before $X$-ray heating, then the spin-temperature would attempt to cool again to the low gas temperature, leading to a second, relatively weaker dip in the absorption spectrum before rising again. Unlike the excess cooling models, this second dip would be less well separated from the broadband signal. Since, the details of these high frequency features depend on the astrophysical model, we have not shown them in our figure.

\subsection{Scenario 3: Intermediate Coupling}
\subsubsection{Qualitative understanding of intermediate coupling solutions}
This scenario, as the name suggests, lies between the scenarios of strong and weak coupling. In this scenario, the gas and dark matter are initially tightly kinetically coupled to each other (similar to the strong coupling scenario), but as they evolve, they might either cool together adiabatically, or the dark matter might decouple from the baryons and begin to cool adiabatically on its own. In either case, the effective temperature $\Teff$ at $z=17$ is less than the desired spin temperature of 3.32~K, as required by our pinning of the absorption spectrum. With moderate values of the spin-flip coupling (similar to the weak coupling scenario) it is possible to obtain $T_s(z=17) \simeq 3.32$~K, and thus, $\delta T_b(z=17) = - 500$~mK.

The intermediate coupling parameter space can be further subdivided into two distinct behaviors at larger and smaller couplings which we discuss below.

For larger couplings in the intermediate region, the DM and the gas remain tightly coupled throughout their evolution, which is similar to the tight kinetic coupling of the strong coupling scenario. However, the DM masses are slightly lower than $m_\chi \lesssim 0.68f$ (see discussion of scenario~1 on criteria needed for decoupling of the combined gas-DM system from the CMB and eq.~\ref{eq:scenario1criteria1}). This implies that $R = \frac{n_\chi}{n_H} > 7.6$, which in turn implies that the effective coupling rate of the gas-DM fluid to the CMB, $\Gamma_c^\prime = \frac{1}{1+R}\Gamma_c$, is weaker than in scenario 1 and thus the combined fluid decouples \textit{earlier} than $z=265$, i.e. between $z=1000$ and $z=265$. This leads to an effective temperature $\Teff(z=17)$ that is \textit{lower} than the value of 3.32~K, which is our desired value of the spin-temperature $T_s$ at $z=17$. We can still obtain the desired $T_s$ value at $z=17$ for low $\Teff$, if the spin-flip coupling $x_D$ is low enough so that the coupling rates of the spin temperature to $\Tcmb$ $(B_{10})$ and to $\Teff$ $(D_{10})$ are comparable. In this respect, the situation is similar to the weak coupling scenario with respect to moderate spin-flip couplings.

For smaller couplings in the intermediate region, the DM is initially tightly coupled to the gas (and in this way the scenario is distinct from the weak kinetic coupling of scenario 2), but it may decouple earlier than $z=265$. However, for low DM masses, $\Teff$ is once again dominated by $T_\chi$ and the spin temperature couples to the cold dark matter kinetic temperature. In this case, as in the weak coupling scenario, for moderate values of the spin-flip coupling $x_D$, it is possible to obtain $T_s = 3.32$~K. The baryons continue to be tightly kinetically coupled to the DM, since $\Gamma_H = R \Gamma_\chi$ and $R>1$, and eventually once the kinetic coupling to the CMB becomes irrelevant, they cool super adiabatically towards the DM temperature. This regime is very similar to the behavior of the weak coupling scenario except for the initial tight coupling of the DM to the gas, which leads to a slightly hotter $\Teff$ (or $T_\chi$) at $z=17$, than in the weak coupling scenario.

In either situation, of relatively larger or smaller coupling, $\Teff$ is lower than 3.32~K at $z=17$ for all parameter space points in the intermediate scenario.
At the redshift where $x_D\sim1$, we get a switch in the behaviour of $T_s$ from tracking $\Teff$ towards attempting to track the CMB temperature. Near the redshift where $x_D \sim 1$, the differential brightness temperature $\delta T_b$ is at its minimum. In the intermediate coupling regime it is possible to find points in parameter space where not only is the spin temperature $T_s(z=17) \simeq 3.32$~K, thus giving us the desired magnitude of the absorption dip in the brightness temperature at 78~MHz, but it is also possible to find a point among these, where $z=17$ $(\nu =78$~MHz) is also a \textit{minimum} of the differential brightness temperature. We will discuss such a reference point in the next subsection.

We note that like the weak coupling scenario, even in the intermediate coupling regime, the absorption signal is only a tracer of the spin temperature, which can differ from the gas kinetic temperature. The deviation between the $T_s$ and $T_K$ is most significant as we go towards the region of parameter space with smaller couplings. However, for couplings near the upper end of the intermediate coupling regime (including at our reference point), the deviation between the spin temperature and the kinetic temperature of the gas is only significant over a small redshift range prior to the cosmic dawn.

\subsubsection{Numerical results for intermediate coupling reference point}

In our numerical scan, we identify the region in parameter space for this scenario by demanding that $0.47 \lesssim x_D \lesssim 10$ at $z=17$. Now we discuss a particular reference point with $f=0.1$, $m_\chi =0.05$~GeV and $\alpha_e\alpha_\chi = 4.3\times 10^{-17}$. This point is in the relatively more strongly coupled regime of the intermediate coupling scenario.

In fig.~\ref{fig:rates_int1_coupling}, we show the reaction rates $\Gamma_H$, $\Gamma_\chi$ and the Hubble rate $H$ as functions of $z$ for this reference point. We can see from the figure that the rates  $\Gamma_\chi$ and $\Gamma_H$ are larger than the Hubble rate throughout the redshift range $10 \lesssim z \lesssim 1000$, indicating that the DM and the gas are tightly coupled to each other, as in the strong coupling scenario. This behaviour is clearly seen in fig.~\ref{fig:tempevolution_int1_coupling}, in which we show the temperature evolution of $T_K$ and $T_\chi$, where both temperatures track each other very closely. In fig.~\ref{fig:rates_int1_coupling}, we also plot the effective ``compton coupling rate'' of the DM-gas fluid to the CMB $\Gamma_c^\prime$, and we see that this rate decouples (drops below the Hubble rate) at $z=290$. In fig.~\ref{fig:tempevolution_int1_coupling}, we can see the common temperature evolution of the DM-gas switches from tracking the CMB temperature (and scaling as $(1+z)$) from $z= 1000 $ till $z=290$, to adiabatic cooling (scaling as $(1+z)^2$) from $z=290$ onwards to lower redshifts.

In fig.~\ref{fig:xdxc_int1_coupling}, we show the evolution of the spin-temperature coupling $x_D$ to the temperature $\Teff$ (which is just the same as the common temperature of the gas-DM fluid for strong kinetic coupling), and the collisional coupling $x_C$ to the gas temperature $T_K$. At all redshifts from $25\lesssim z \lesssim 1000$, we see that the coupling $x_D \gg x_C$, which indicates that the DM spin-flip coupling reaction rate $D_{10}$ is dominant over the collisional coupling spin-flip rate.  However, at $z=25$, $x_D=13\sim \mathcal{O}(10)$ and the CMB induced spin-flip rate $B_{10}$ becomes comparable to the rate $D_{10}$ of DM induced spin-flip interactions. Thus, over the redshift range $25\lesssim z \lesssim 1000$, $T_s$ tracks the common temperature of the gas-DM fluid. However, at $z\sim25$, the spin-temperature begins heating up towards the CMB temperature (or at least cools more slowly than the adiabatically cooling DM-gas fluid). The evolution of the spin-temperature of the gas is shown in fig.~\ref{fig:tempevolution_int1_coupling}. We can see that this scenario is intermediate between scenario 1 (strong coupling) and scenario 2 (weak coupling), in that while the DM and gas are tightly kinetically coupled to each other (as in scenario 1), the DM induced spin-flip rate drops below (or becomes comparable to) the CMB induced spin-flip rate at low redshifts (as in scenario 2).

The early decoupling of the gas-DM fluid at $z=290$ leads to a value of $\Teff =2.9$~K at $z = 17$. This combined with the moderate spin-flip coupling $x_D(z=17)=4.5$, leads to a spin temperature $T_s(z=17)=3.5$~K, intermediate between $\Teff$ and the CMB temperature at this redshift. This can be seen in fig.~\ref{fig:tempevolution_int1_coupling}, where we have also plotted the evolution of the spin temperature with redshift.

In fig.~\ref{fig:Tb_int1_coupling}, we show the differential brightness temperature as a function of redshift $z$, inferred from the spin temperature evolution history. Since, $\delta T_b \propto T_s-\Tcmb$ and $T_s$ drops faster than $\Tcmb$ at high redshifts, we see that $\delta T_b$ becomes more negative as we go from a  redshift of $z \sim 1000$ to $z\sim 25$. From this point on, the spin temperature  begins to heat up towards the CMB temperature. Thus, we expect $\delta T_b$ to attain a minimum for some lower redshift. For our particular reference point $\delta T_b$ is at a minimum at $z=17$.
Also, as for all points in our parameter space of interest, the low spin temperature leads to a dip in  $\delta T_b(z=17) \simeq -457$~mK, near our ``pinned'' value.

Once again, we expect that below some redshift $z\lesssim 15$, Ly-$\alpha$ coupling to $T_K$ would turn on due to star formation, and $X$-ray heating would raise $T_K$, altering the behavior of the differential brightness temperature for lower redshifts. However, since the details of this would depend on the astrophysical model, we do not show these effects in our figure.

\begin{figure}
\begin{center}
  \includegraphics[width=0.7\linewidth]{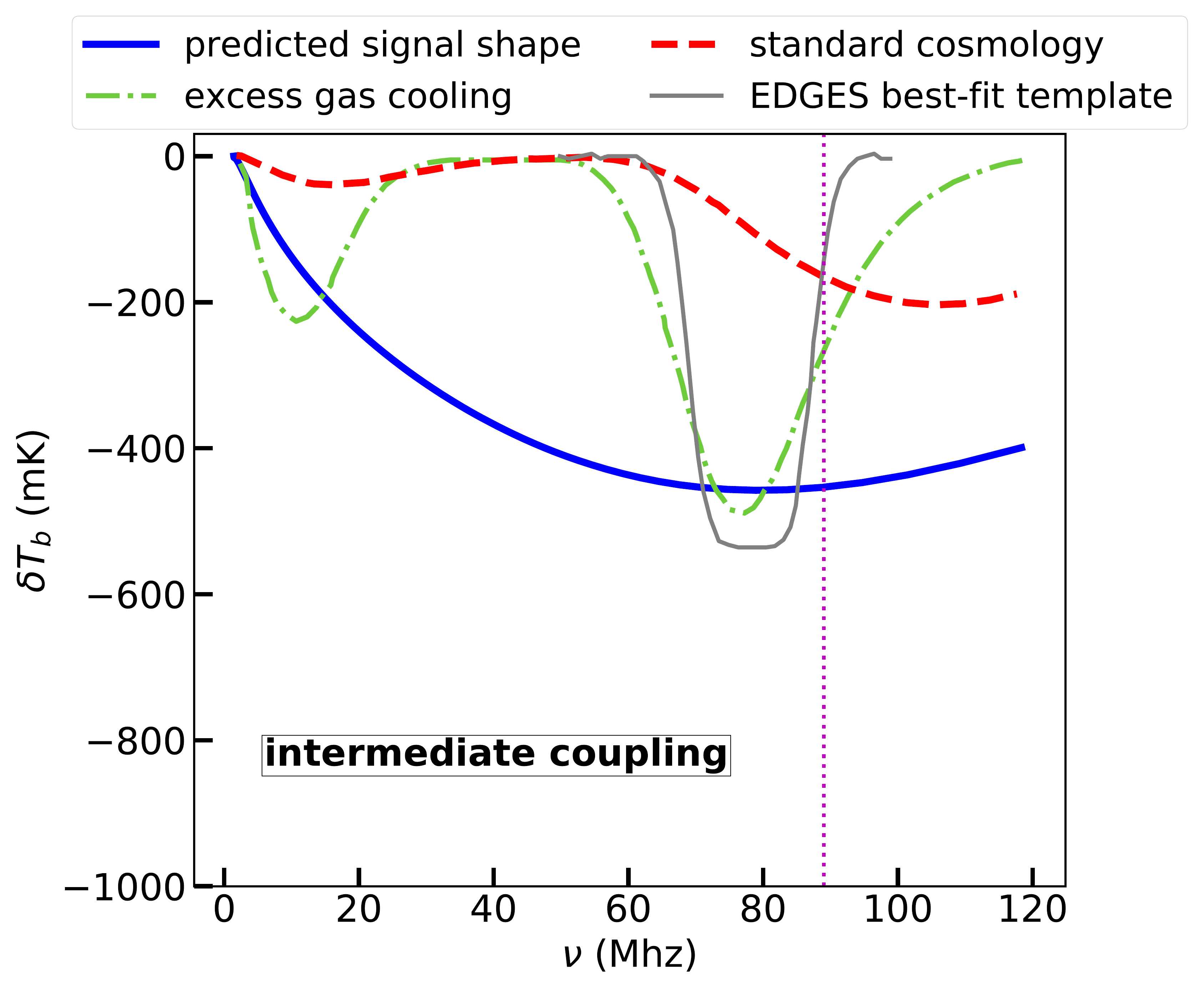}
\caption{Predicted differential brightness temperature $\delta T_b(\nu)$ as a function of frequency (blue solid curve) with parameters $f=0.1$, $m_\chi =2.5 \times 10^{-2}$~GeV and $\alpha_e\alpha_\chi = 1.1\times 10^{-17}$. Our model predicts a single, strong broadband absorption feature that begins at $4.9$~MHz ($z=290$) and extends all the way to high frequencies where Ly-$\alpha$ couplings are expected to become important . For this particular reference point we get a minima at $78$~MHz~($z=17$) with a value $\delta T_b(\nu=78~\textrm{MHz})\simeq -457$~mK. The minima arises as the spin temperature attempts to rise up towards the CMB temperature. Above $89$~MHz ($z=15$), indicated by the vertical dotted magenta line,  Ly-$\alpha$ couplings and $X$-ray heating could lead to either a sharper rise or a possible second dip, but these features are not shown. Unlike our model, both the standard cosmology (red dashed) and excess gas cooling models (green dot-dashed) have band-limited absorption features with transitions due to collisional couplings and Ly-$\alpha$ photons.}
\label{fig:Tbvsnu_int1_coupling}
\end{center}
\end{figure}

In fig.~\ref{fig:Tbvsnu_int1_coupling}, we show the differential brightness temperature evolution as a function of redshifted 21~cm frequency. We also show for comparison, in the same figure, the expected band-limited signals from the standard cosmology and from the excess gas cooling model, both of which were discussed in sec.~\ref{sec:strongcouplingnumerical}, where the exact curves have been taken from ref.~\cite{Burns:2019zia}. Again, we see the contrast between these signals, and our model which predicts a single, strong, broadband absorption feature that extends from 4.9~MHz ($z=290$) all the way up to high frequencies. At $78$~MHz $(z=17)$, the spectrum in our model attains the ``pinned'' value of $\delta T_b \simeq - 457 $~mK. The minima of the absorption spectrum for this reference point is also at 78~MHz, a frequency value which is intermediate between the weak coupling minima at 35~MHz and the strong coupling minima at $\nu > 89$~MHz, (compare fig.~\ref{fig:Tbvsnu_int1_coupling} with figs.~\ref{fig:Tbvsnu_weak_coupling}~and~\ref{fig:Tbvsnu_strong_coupling}).

For frequencies larger than those corresponding to the cosmic dawn, 89~MHz ($z=15$) by our assumption, we expect that Ly-$\alpha$ photons from the first stars and $X$-ray heating would determine the behavior of the absorption spectrum. Depending upon the time order in which these effects come into play, we could either get a sharp rise in the predicted spectrum above 89~MHz, or a possible second dip before the rise. In the latter case, unlike the excess cooling models, the second absorption dip would be less well separated from the first absorption dip in the broadband signal. Since the details depend on the astrophysical model, we have not shown these possible effects in the figure.

\begin{figure}[H]
\begin{subfigure}{.5\textwidth}
 \centering
   \includegraphics[width=\linewidth]{{int_Rates_vs_z}.pdf}
  \caption{}
  \label{fig:rates_int1_coupling}
\end{subfigure}%
\begin{subfigure}{.5\textwidth}
  \centering
 \includegraphics[width=\linewidth]{{int_temperature_vs_z}.pdf}
  \caption{}
  \label{fig:tempevolution_int1_coupling}
\end{subfigure}
\begin{subfigure}{.5\textwidth}
 \centering
 \includegraphics[width=\linewidth]{{int_coupling_vs_z}.pdf}
 \caption{}
  \label{fig:xdxc_int1_coupling}
\end{subfigure}%
\begin{subfigure}{.5\textwidth}
  \centering
  \includegraphics[width=\linewidth]{{int_T21_vs_z_new}.pdf}
  \caption{}
 \label{fig:Tb_int1_coupling}
\end{subfigure}
\caption{Plots for the intermediate coupling scenario with parameters $f=0.1$, $m_\chi =2.5 \times 10^{-2}$~GeV and $\alpha_e\alpha_\chi = 1.1\times 10^{-17}$. (a) Rate evolution of the DM ($\Gamma_\chi$) and gas ($\Gamma_H$) kinetic coupling to each other and the effective coupling of the combined DM-gas fluid to the CMB ($\Gamma_c^\prime$) indicating a decoupling at $z=290$. (b) Temperature evolution of the DM and gas kinetic temperatures compared to the CMB temperature. Adiabatic cooling begins at $z=290$ and leads to $\Teff = 2.9$~K at $z=17$.  The spin temperature is tightly coupled to the DM-gas temperature through a large spin-flip rate $x_D$ at high redshifts, but at redshifts below $z\sim25$, the spin temperature is also heated up by the CMB temperature (away from $\Teff$) as $x_D$ drops below $\sim 10$ (c) The collisional coupling $x_C$ to the gas kinetic temperature $T_K$ and the spin-flip coupling $x_D$ to the effective temperature $\Teff$. The rate $D_{10}$ dominates over both the collisional rate and the CMB rate till $z\sim 25$, but for lower redshifts the coupling of $T_s$ to the CMB temperature begins to dominate and the spin temperature rises (or falls slower than the temperature of the adiabatically cooling DM-gas fluid). (d) Predicted differential brightness temperature $\delta T_b(z)$ as a function of redshift. Additional contributions from astrophysics at low redshifts $z \lesssim 15$ are not shown.}
\label{fig:scenario3int1couplingbenchmark}
\end{figure}

\subsection{Summary of the absorption signal expected in different regions of the parameter space}
We have seen that in all scenarios, our model predicts a single, strong broadband absorption signal starting from early in the cosmic dark ages. This signal is unlike the two distinct band-limited absorption features expected in the standard cosmology and excess gas cooling models. In any of the panels in fig.~\ref{fig:parameterspaceplot}, we can imagine following a curve through the parameter space of interest starting at strong coupling and proceeding through intermediate coupling, on to weak coupling. The predicted absorption spectra as we follow this curve can be tracked by comparing figs.~\ref{fig:Tbvsnu_strong_coupling}, \ref{fig:Tbvsnu_int1_coupling}, \ref{fig:Tbvsnu_weak_coupling} in order. In all cases, by our benchmark choice, our predicted absorption spectrum is pinned at 78 MHz, to have a value $\delta T_b \simeq  - 500 $~mK. However, as we go from strong coupling to intermediate coupling to weak coupling, the minima of the absorption signal shifts from frequencies greater than 89 MHz at strong coupling, to $\sim 78$~MHz at intermediate coupling, to even lower frequencies at weak coupling. In the latter two scenarios, it is possible that there might be a second absorption dip at cosmic dawn due to Ly-$\alpha$ photons and $X$-ray heating, which is not well separated from the broadband signal which is due to the DM spin-flip coupling. The start of the absorption dip also drifts slightly to lower frequencies, starting at 5.3~MHz for the strong coupling scenario, moving to 1.3~MHz for the weak coupling scenario. This low frequency region is a regime of frequencies from which no signal is expected at all in the standard cosmology or in excess cooling models.

For the weak (and to some extent in the intermediate) coupling scenario, another prediction which is in stark contrast to the expectation of the standard cosmology or excess cooling models, is that the gas spin temperature history in our model is almost independent of the gas kinetic temperature history over our redshift range of interest. In this scenario, we have seen that the strong spin-flip coupling would relate $T_s$  to $\Teff\simeq T_\chi$, but the weak kinetic coupling would leave the gas kinetic temperature $T_K$ history unchanged from that of the standard cosmology.

What would happen to our predicted signal if we move slightly away from the benchmark parameter space shown in fig.~\ref{fig:parameterspaceplot}? We can imagine following a contour parallel to our parameter space towards the upper left (smaller $m_\chi$ near strong coupling, and larger couplings for a given $m_\chi$ near weak coupling). From our discussion of the analytic predictions of the spin temperature, it is easy to see that such a change would lead to a stronger absorption signal at $z=17$, thus altering our pinning point. Similarly, if we shift our preferred parameter space to the lower right, this would lead to a weaker absorption signal at $z=17$. For such small deviations in the parameter space, the shape of the signal would remain mostly unaltered. However, for larger variations across the parameter space, especially towards weaker coupling, we would expect to see significantly different behavior in the predicted absorption signal, as the coupling $x_D$ becomes small enough that the collisional coupling of the spin temperature to the gas kinetic temperature becomes cosmologically relevant once again.

\section{Some clarifications about assumptions}
\label{sec:discussion}
\subsection{Initial conditions}
Now we discuss the choice of initial conditions that we had assumed for solving the temperature evolution equations. Our default choice was to choose our initial conditions near recombination, and we assumed $T_\chi= T_K = \Tcmb$ at  $z=1000$.

The pre-recombination behavior that determines the initial conditions depends on the thermal history of the DM particle and the dark sector in general. The initial conditions are therefore model dependent, because interactions of the DM other than the spin-flip interactions considered here are also important for determining the kinetic coupling of the DM either to the visible sector plasma or to a possible dark radiation bath.

However in all models, the spin-flip interaction with electrons that we have considered would lead to a minimum level of elastic scattering between DM and electrons, which can kinetically couple the DM to the visible sector plasma in the pre-recombination era. The energy transfer cross-section for DM-electron scattering is of the form (see appendix~\ref{sec:appendixd}),
\begin{align}
\overline{\sigma}_{e\chi} &= \frac{3}{8 \pi} \frac{g_\chi^2 g_e^2}{\mu^2_{\chi e}v^4} \textrm{Log} \left( \frac{4 \mu^2_{\chi e}v^2 }{m_V^2}\right),
\end{align}
where $\mu_{e\chi}$ is the reduced mass of the dark matter and electron and $v$ is the relative velocity between the incoming DM and the electron. From this we can estimate the energy transfer rate coefficient from the electrons to DM as,
\begin{align}
\Gamma^e_{\chi} \sim  n_e \frac{3}{8 \pi }  \frac{g_\chi^2 g_e^2}{ T^2_{\textrm{eff}} } \left(\frac{ \mu_{e\chi}}{ M_{e\chi}} \right) \textrm{Log} \left( \frac{4\mu_{e\chi} \Teff}{2 m_V^2} \right )
\end{align}
where $M_{e\chi}$ is the sum of the DM and electron masses, and we have estimated the thermal average cross-section by substituting $\mu_{e\chi} v^2 \rightarrow \Teff$, where $\Teff = \mu_{e\chi}\left (  \frac{T_\chi}{m_\chi} + \frac{T_e}{m_e} \right )$, and $T_e$ is the electron/visible sector plasma temperature. We have also included an additional suppression factor of $\frac{ \mu_{e\chi}}{ M_{e\chi}}$ expected for energy transfer between particles of unequal masses.

This energy transfer rate is log enhanced because the forward scattering divergence of the elastic scattering is cut-off by the small mediator mass $m_V$ (rather than the mass splitting parameter $\Delta$ of inelastic scattering of DM with hydrogen). The energy transfer rate of inelastic scattering of DM with neutral hydrogen ($\Gamma_\chi$) in the post-recombination era scales as $\frac{1}{\Teff\Delta}$ (see eq.~\ref{eq:gammchianalytic} in appendix~\ref{sec:appendixc}), since the forward scattering divergence is cut-off by the inelasticity parameter. Thus, we see that the pre-recombination energy transfer rate, $\Gamma^e_{\chi}$ has a net suppression by a factor of $\sim \frac{\Delta}{\Teff} \textrm{Log} \left( \frac{4\mu_{e\chi} \Teff}{2 m_V^2} \right )$ relative the post-recombination rate, $\Gamma_\chi$.

We can estimate the typical size of this factor at recombination by taking $\Teff = 0.25~$eV which is the temperature of the plasma at recombination. In that case $\Delta/\Teff$ is $\sim \mathcal{O}(10^{-5})$. The log enhancement factor is $\mathcal{O} (10-20)$ depending on the mediator mass, thus the overall suppression of the DM-electron energy transfer rate prior to recombination is $\mathcal{O} (10^{-3})$ relative to the DM-hydrogen rate post-recombination. This implies that the energy transfer rate is slightly lower in the pre-recombination era.

In the region of parameter space which we dubbed as our strong coupling regime, this would not significantly alter the initial conditions since the DM would be tightly coupled to the plasma both before and after recombination. Thus, it is expected, that in the strong coupling scenario, the DM has a temperature equal to the plasma temperature at recombination as we have assumed, even if other interactions are present in the dark sector. We will discuss constraints on such tight coupling of DM to the plasma at recombination in section~\ref{sec:constraints}.

In the weak coupling scenario, the DM is kinetically decoupled post-recombination. Since the coupling is weak, we would also expect the DM to be decoupled prior to recombination. Thus, we would expect that the DM in the pre-recombination era was adiabatically cooling relative to the plasma, and we might expect it to be colder than the plasma temperature at recombination. However, if other interactions of the DM other than those considered in our model are important, it is possible that they could keep the DM at (or near) the same temperature as the plasma even in this scenario, thus providing a justification for our initial conditions.

Alternatively, one could ask what would happen in the weak coupling scenario if the dark matter is colder than the plasma temperature at recombination. If the dark matter is colder than the gas temperature by a factor of $\epsilon < 1$, then in this scenario, the DM would begin adiabatically cooling from this initial value and thus $T_\chi$ (or equivalently $\Teff$) would be a factor of $\epsilon$ colder at $z=17$ as well, compared to our expectation of eq.~\ref{eq:tempevolutionweak}.
This would lower the value of $x_D$ needed at $z=17$ to attain the pinned value of $\delta T_b$ by a factor of $\epsilon$. Since, $x_D \propto  \alpha_\chi \alpha_e\Teff^{1/2}$, this would lower the value of $\alpha_\chi\alpha_e $ by a factor of $\sqrt{\epsilon}$ compared to the expectation of eq.~\ref{eq:moderatespincouplingcrit1}. Thus, the allowed couplings of the weak coupling scenario in our parameter space of interest in fig.~\ref{fig:parameterspaceplot} would be lowered by a factor of $\sqrt{\epsilon}$ for every value of the DM mass in the weak coupling regime\footnote{This argument works for moderate $\epsilon$  say $1/5, 1/10$ or $1/100$. However, for much smaller values of $\epsilon$, the rates $\Gamma_H$ and $\Gamma_\chi$ are enhanced at the initial conditions by the small effective temperature (see Eqs.~\ref{eq:gammaHmain},~\ref{eq:gammachimain}) and we expect that this would lead to rapid energy exchange between the DM and gas till the DM temperature is heated up, such that $\epsilon$ is back in the moderate regime.}.

\subsection{Allowed range of mediator masses}
\label{sec:mediatormass}
The Born approximation that we have used in our cross-section calculations when determining electron DM scattering amplitudes is valid when $\alpha_\chi \alpha_e \mu^2 \ll m_V^2$ (see for e.g. ref.~\cite{Tulin:2013teo}), where $\mu$ is the reduced mass of DM and hydrogen (which is approximately the same as the DM mass for most of our benchmark parameter space).

On the other hand, we have also assumed an upper bound on the mediator mass in eq.~\ref{eq:medmass_upper} in appendix~\ref{sec:appendixb}, in order to ensure that the forward scattering divergence of the spin-flip cross-section is dominantly cut-off by the inelastic splitting $\Delta$ between the hyperfine states.

These two conditions together imply a mass range for the mediator which is,
\begin{align}
\label{eq:medmass}
 0.1~\textrm{eV} \sqrt{\frac{\alpha_\chi \alpha_e}{10^{-18}}} \left(\frac{\mu}{\textrm{0.1~GeV}}\right) \lesssim m_V \lesssim 2.3~\textrm{eV} \sqrt{\left(\frac{1000~\textrm{K}}{\Teff}\right)\left( \frac{\mu}{\textrm{0.1~GeV}}\right)}.
\end{align}
The allowed range of mediator masses is narrow in the strong coupling scenario that we have considered, but the lower bound relaxes for weaker couplings. In general mediator masses of a few eV are allowed in all scenarios, and mediator masses as low as $10^{-5}$~eV are allowed by the parameters of our weak coupling scenario, for small enough $m_\chi$.

\subsection{Relative velocity between baryons and dark matter}
The dark matter and baryons in standard cosmology are expected to have a relative velocity due to the baryons being dragged with the photons before decoupling and then falling back into DM potential wells~\cite{2010PhRvD..82h3520T,Ali-Haimoud:2013hpa}. This velocity is usually assumed to be Gaussian distributed, with a standard deviation of 29~km/s at $z=1010$. The dissipation of this relative velocity could lead to a heating of both the baryons and dark matter in the post-recombination era~\cite{Munoz:2015bca}.

In our model, the pre-recombination physics that sets the initial conditions would determine the relative velocity (if any) between the DM and baryons. If the DM is tightly kinetically coupled to the baryons till recombination (as is expected to happen in the strong and intermediate coupling scenarios), then the DM would also be dragged with the baryons and we would not expect a relative velocity between the DM and the gas\footnote{Such a scenario is ruled out for $f \gtrsim 0.01$ by CMB observations as we will discuss in the next section.}. If the DM is not kinetically coupled to the baryons pre-recombination (as could happen in the weak coupling scenario), then it is possible to have a relative velocity between the DM and the gas. However, as long as the DM is sufficiently cold, this would still lead to a low $\Teff$ and hence also a strong broadband absorption signal.

We have not taken into account the relative velocity between DM and baryons in our calculations of the temperature evolution equations.

\section{Constraints on our model}
In this section we will discuss  several different categories of constraints on our model.  A number of constraints that we discuss are on the mediator mass $m_V$, which does not enter directly into our computation of the predicted global 21 cm signal. However, these constraints indirectly rule out regions of the allowed parameter space of our model based on the assumptions of the allowed mediator mass discussed in sec.~\ref{sec:mediatormass}. At the end of this section we will attempt to synthesize these constraints and compare with the parameter space of interest in fig.~\ref{fig:parameterspaceplot}.

\label{sec:constraints}

\begin{enumerate}
\item{\bf Laboratory experiments:}
The axial-vector boson $V$ generates an effective long-range spin-dependent interaction potential between electrons (and/or positrons). This potential has the form $V(r) = \frac{\alpha_e}{r} \left(\vec{S}_1 \cdot \vec{S}_2\right) e^{-r/\lambda}$, where $\lambda = \frac{\hbar c}{m_V}$ gives the range of the interaction and $\vec{S}_{1,2}$ denotes the spin of the interacting particles.

For the mass range of interest ($m_V$ between $10^{-4}$~eV and 10~eV, see sec.~\ref{sec:mediatormass}), the force operates over a range between $10$~nm to a few mm, with the shortest range of interactions being of most relevance for our strongly interacting scenario, whereas the full range is of interest for our weakly interacting scenario. Various probes of such spin-dependent forces between electrons have constrained the allowed values of $\alpha_e$ for different interaction ranges/mediator masses. A summary plot of the exclusion curves in the range of interest is shown in Fig.~\ref{fig:constraints}.

For masses $m_V \gtrsim 10$~eV, the strongest constraint is $\alpha_e \lesssim 6 \times 10^{-12}$ which is set by a precision measurement of the  hyperfine splitting interval of the positronium ground state~\cite{Karshenboim:2010cj}.
For the range $0.5 \lesssim m_V \lesssim 50$ eV, the most stringent constraint on $\alpha_e$ has been imposed by using Double Electron Electron Resonance (DEER) measurements of the coupling between two
electron spins located at two ends of a molecular ruler~\cite{Jiao:2019ikv}. The constraint set by these measurement is $\alpha_e \lesssim 4.9 \times 10^{-13}$.

\begin{figure}[htbp]
\begin{center}
\includegraphics[width=0.75\linewidth]{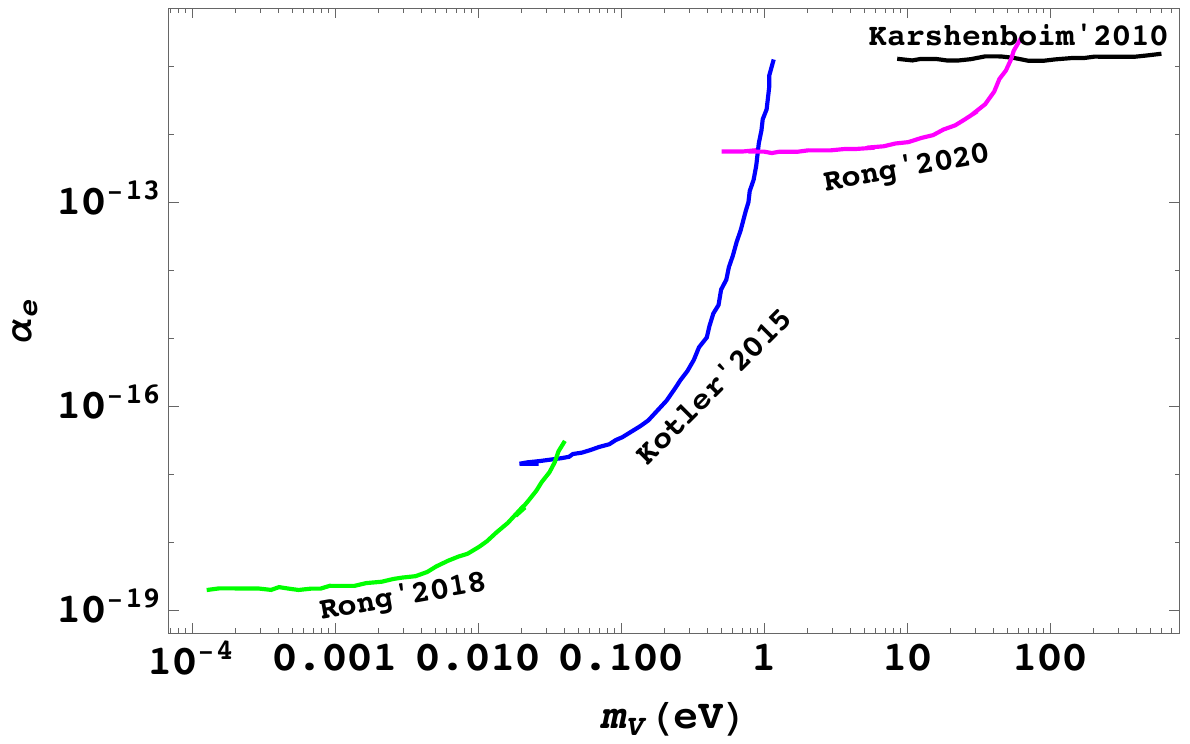}
\caption{Upper bounds from laboratory experiments on the interaction strength $\alpha_e$ of spin-dependent forces between electrons. The force is assumed to be mediated by an axial-vector boson with a mass $m_V$ in the range $10^{-4}$ eV - 100~eV. The black, magenta, blue and green lines represent bounds on $\alpha_e$  obtained in refs.~\cite{Karshenboim:2010cj,Jiao:2019ikv,Kotler:2015ura,Rong:2018yos}, respectively.}
\label{fig:constraints}
\end{center}
\end{figure}

 For  $0.001~{\rm eV} \le m_V \le $ 0.1 eV, the strongest constraints on exotic spin-dependent dipole-dipole interactions has been obtained by measuring the magnetic interaction between two trapped $^{88}{\rm Sr}^{+}$ ions~\cite{Kotler:2015ura}, and this sets a bound $\alpha_e \lesssim 1.2\times10^{-17}$. For a mass range $ 10^{-4} \lesssim m_V \lesssim 10^{-2}$~eV, the strongest constraint comes from single nitrogen valency centers in diamond which can be used as quantum sensors for detecting weak magnetic signals~\cite{Rong:2018yos}. The constraint in this range of mediator masses is $\alpha_e \lesssim  1.8\times10^{-19}$.

\item{\bf Collider searches:}
The Lagrangian of our effective theory in eq.~\ref{eq:lagrangian} must be UV completed at an effective scale $\Lambda \sim \frac{m_V}{\sqrt{\alpha_e}}/\left(\frac{3g^2}{16\pi^2}\right)$, where $g$ is the Standard Model (SM) $SU(2)_L$ gauge coupling~\cite{Preskill:1990fr}. In general if the axial-vector boson is coupled to a non-conserved SM fermion current, then in the simplest UV completions, additional fermions must be introduced below the cut-off scale to make the theory anomaly free. Integrating these fermions out of the effective theory generates anomalous interactions between the vector $V$ and the SM gauge bosons~\cite{Dror:2017ehi, Dror:2017nsg}. These anomalous interactions would lead to phenomenological signatures such as the exotic $Z$ boson decay $Z\rightarrow \gamma V$, which are enhanced by a factor of $(m_Z/m_V)^2$ due to the Goldstone equivalence relation, which dictates that the coupling of the $Z$ to $V$ is dominantly to the longitudinal mode of $V$.

The anomalous decay branching fraction then has the form,
\begin{equation}
\textrm{Br}(Z\rightarrow \gamma V) \simeq 10^{-7} \mathcal{A}^2 \left( \frac{\textrm{TeV}}{m_V/\sqrt{4\pi\alpha_e}}\right)^2,
\end{equation}
where $\mathcal{A}$ is in general an $\mathcal{O}(1)$ anomaly coefficient. If $V$ then decays invisibly or is long-lived and escapes the detector, then LEP searches for single photons at half the $Z$ energy limit this branching
ratio to be $\lesssim 10^{-6}$~\cite{Acciarri:1997im,Abdallah:2003np}. Applying this bound, we get the stringent limit,
\begin{equation}
\alpha_e \lesssim 10^{-24}\left(\frac{m_V}{1 ~\textrm{eV}}\right)^2.
\end{equation}
However, in deriving this limit we needed to assume that the cut-off $\Lambda$ (or the heavy fermion masses) are larger than the $Z$ mass, which is not true for the parameter space of interest. Thus, if additional fermions are introduced below the $Z$ mass scale, the most stringent constraints would likely arise from $Z$ decays to these exotic fermions, but the details of such a constraint would be model dependent.

While the constraint derived from anomalous $Z$ decays is strong and rules out the parameter space that we have focused on, it is also contingent on the UV completion. It is possible that more exotic UV completions of our model which violate the electroweak symmetry of the Standard Model may evade these constraints~\cite{Dror:2017ehi}.

\item{\bf{Constraints from stellar cooling}}:
The weakly interacting light mediator $V$ could be produced inside the hot and dense interior of stars and would consequently lead to anomalous cooling, which is strongly constrained. The constraints on axial-vector couplings can be inferred through constraints imposed on an equivalent axion ($a$) couplings, since the dominant production mode of the axial-vector is via the longitudinal mode which behaves like a pseudo-scalar axion~\cite{Dror:2017nsg}. The equivalence can be captured through a change of the effective interaction term in the Lagrangian,

\begin{equation}
{\cal L } \supset g_e {\bar e } \gamma^5 \gamma^\mu  e V_\mu \sim \frac{1}{2 f_a}\overline{e}\gamma^5 \gamma^\mu  e \, \partial_\mu a \rightarrow
 -i g_{a e} \overline{e} \gamma^5   e \,  a ,
\end{equation}
where the first relationship follows from the Goldstone equivalence principle and the identification of $g_e V_\mu\rightarrow  \frac{1 }{f_a} \partial_\mu a $,  where the equivalent axion-decay constant $f_a = \frac{m_V}{g_e}$. The second relation follows from an integration by parts and application of the equations of motion, followed by the identification $g_{a e} = \frac{m_e}{f_a}$, where $m_e$ is the mass of electron.

Limits from a combined analysis of the tip of the red-giant branch in the globular cluster M5, anomalous white dwarf cooling, and horizontal branch stars/red giants in globular clusters gives the most stringent constraint on $g_{a e} < 2.6 \times 10^{-13}$ at 95\% CL~\cite{Giannotti:2017hny,Irastorza:2018dyq}.We can convert this to a constraint on $\alpha_e$ as,
\begin{equation}
 \alpha_e \lesssim  10^{-38} \left(\frac{m_V}{1~\textrm{eV}}\right)^2.
 \end{equation}

Once again this is a strong bound which rules out our parameter space of interest, but the constraint is model dependent. For example, the mediator particles can remain trapped inside stars if they have strong self-interactions~\cite{Jain:2005nh,Masso:2005ym}, considerably weakening the constraints. Alternatively, with a chameleon-like mechanism, the mediator could acquire a heavy effective mass inside dense media which prevents it from being produced in the first place~\cite{DeRocco:2020xdt,Budnik:2020nwz,Bloch:2020uzh}.

\item {{\bf Constraints on extra radiation species:}}  The light axial-vector bosons can potentially contribute to the effective number of relativistic degrees of freedom in the early universe. The number of extra radiation species is usually parameterized in terms of extra neutrino species as $\Delta N_\textrm{eff}$. There exist strong constraints on $\Delta N_{\rm eff}\lesssim 0.1$ from both BBN~\cite{Fields:2019pfx} and CMB~\cite{Aghanim:2018eyx} data.

    These constraints can be evaded if a) the axial-vector $V$ is not in thermal equilibrium with the SM plasma and is colder than the neutrino temperature b) $V$ is short-lived and decays before BBN, and thus does not form a thermal bath. The first possibility is ruled out since, for our parameter space of interest, $V$ would be strongly kinetically coupled to the plasma. The second possibility could occur either through the loop-process $V\rightarrow \gamma \gamma \gamma$ or a neutrino decay process $V\rightarrow \nu \overline{\nu}$. The latter reaction might occur in generic gauge-invariant UV completions of our model.

    Another possibility is that the extra radiation species could potentially ameliorate the Hubble $H_0$ and $\sigma_8$ tensions if it is self-interacting~\cite{Kreisch:2019yzn}.

\item \textbf{Constraints from kinetic decoupling:}
    The pre-recombination physics determines whether the DM particle $\chi$ is kinetically coupled to the gas at recombination. A fraction of the DM greater than $f\sim 0.01$ which is tightly coupled to the plasma at recombination is ruled out by observations of the CMB~\cite{Dubovsky:2003yn,Boddy:2018kfv,Boddy:2018wzy}.
    Thus, the strong and intermediate coupling scenarios for $f=1,0.1$ are ruled out.
     However, for smaller fractions $f=0.01, 0.001$, in the strong coupling scenario, the additional tightly coupled DM is indistinguishable from a small additional baryon content in terms of its effect on the CMB.

\item \textbf{Constraints on self-interaction coupling of DM:}

The light axial-vector mediator can also mediate self-interaction between DM particles.  In our case, the self-scattering DM cross-section has the form (appendix.~\ref{sec:appendixd}),
\begin{equation}
\sigma(\chi \chi \rightarrow \chi \chi) = 24 \pi \frac{\alpha^2_\chi}{m^2_\chi v^4} {\rm log} \left(\frac{m^2_\chi v^2}{m^2_V}\right),
\end{equation}
with $v$ being the relative velocity for scattering. Observations of the Bullet-cluster
and other halo shape observations set strong bounds on DM self-interactions (see ref.~\cite{Tulin:2017ara} for a review),
\begin{equation}
\frac{\sigma}{m_\chi} \lesssim (1-10)~{\rm cm^2}/{\rm g}.
\end{equation}
For $v = 1000$ km/s in a typical galaxy cluster, the coupling is constrained to be,
\begin{equation}
 \alpha_{\chi} \lesssim 10^{-6} \times \left(\frac{m_\chi}{0.1~\textrm {GeV}}\right)^{3/2}.
\end{equation}
 However, this constraint would only apply if the self-interacting DM is an $\mathcal{O}(1)$ fraction of the whole DM of the universe.

 \item \textbf{Dark Matter relic abundance and freeze-out:} We have assumed that our DM is asymmetric, but if we had assumed it were produced symmetrically, we would have strong constraints on the coupling $\alpha_\chi$ from the condition that the annihilation process $\chi \chi \rightarrow V V$ does not deplete the relic abundance, i.e. the cross-section for this process should be smaller than the freeze-out cross-section needed to yield a relic density with a fraction $f$ of the present day DM density, $\sigma v \lesssim \frac{3 \times 10^{-26} }{f} {{\rm cm}^3 \, \rm s}^{-1}$. This would lead to a bound $\alpha_\chi \lesssim 10^{-5} \times \left(\frac{m_{\chi}}{0.1 \, \textrm{GeV}}\right) \times \frac{1}{\sqrt{f}}$.

\end{enumerate}

\subsection{Summary of constraints:}
To summarize, there are two main, robust constraints -- a) the constraint from kinetic decoupling and b) the constraint from laboratory experiments on light mediators. Constraints from colliders, stellar cooling, and $\Delta N_{\textrm{eff}}$ can be relaxed completely in extensions of our effective theory. The self-interaction constraint only applies for $f=1$. Also, the freeze-out constraint does not apply, since we assume that $\chi$ is produced asymmetrically.

The kinetic decoupling constraint rules out the benchmark parameter space for the strong and intermediate coupling scenarios for $f=\, 0.1, 1$. For $f=1$, in the weak coupling regime, $m_\chi\sim 10^{-2}$~GeV, and therefore the upper end of the allowed mediator mass is $m_V \sim 1$~eV (see eq.~\ref{eq:medmass}). For this mediator mass, laboratory constraints set a limit $\alpha_e \lesssim 10^{-13}$~(see fig.~\ref{fig:constraints}). Also, self-interacting DM constraints yield $\alpha_\chi \lesssim 10^{-7}$. Thus, the combination of these two constraints also rules out the benchmark parameter space of the weak coupling scenario for $f=1$. However, since the self-interaction bound does not apply for $f=0.1$, the weak coupling parameter space is viable for $f=0.1$ with moderate values of $\alpha_\chi \sim 0.1$.

For all scenarios with $f=0.01, 0.001$ and for a mediator mass near the upper end of our allowed range, with moderate values of $\alpha_\chi \sim 0.1$, we would find the entire region of couplings in our benchmark signal parameter space of fig.~\ref{fig:parameterspaceplot} to be consistent with the laboratory constraints on light mediators in fig.~\ref{fig:constraints}.

\section{Summary, conclusions, and future directions}
\label{sec:conclusions}
Standard cosmology predicts two relatively weak and distinct, band-limited absorption features in the global 21 cm signal with the first minima near 20~MHz and the second minima at higher frequencies between $50-110$~MHz due to collisional gas dynamics and Ly-$\alpha$ photons from the first stars, respectively. Excess gas cooling models invoked to explain the anomalous EDGES absorption signal also predict the same distinct band-limited absorption features, although these features are predicted to be deeper than those of the standard cosmology.

In the current work, we have explored an alternative prediction of the global 21 cm signal in a model where dark matter interacts with electrons through a light axial vector mediator. This interaction leads to two distinct cosmological effects, the first is a predicted coupling $x_D$ of the gas spin temperature to a new effective temperature scale $\Teff = \mu  \left (\frac{T_\chi}{m_\chi} + \frac{T_K}{m_H} \right)$, and the second is a coupling of the gas kinetic temperature to the DM temperature. Through an explicit Born level calculation of these interaction rates, we have found that the spin-flip rate is larger than the kinetic energy transfer rate, which leads to characteristic predictions of our model which distinguish it from the excess gas cooling models.

We have found, generically, that our model leads to predictions of a single, strong, broadband absorption feature which is unlike that of either the standard cosmology or excess gas cooling models. The signal is strong because of the low temperature scale $\Teff$ and it is broadband because of the dominance of the spin-flip coupling $x_D$ over the other couplings (collisional, CMB) of the spin temperature for much of the post-recombination cosmological history.

As a benchmark, we have focussed on regions of parameter space in our model which lead to an absorption signal with strength $\delta T_b \simeq - 500$~mK at $z=17$, consistent with the magnitude of the EDGES absorption signal at this redshift. However, this was only used as a benchmark to pin our absorption signal, and we have made no demands on the shape of the spectrum.

In different regions of our model parameter space, we have found, through numerical studies backed by analytic estimates, different predictions for the predicted global 21~cm signal, depending upon the cosmological relevance of the kinetic energy transfer rate. We classified our parameter space of interest into three scenarios of strong, intermediate and weak coupling.
While all scenarios predict a single, strong broadband absorption signal, they differ in the detailed predictions.

In the strong coupling scenario, the kinetic energy transfer rate is important over most of the cosmological history, from recombination to the cosmic dawn, and it ensures a tight coupling between the DM and the gas. The spin temperature is also tightly coupled to the adiabatically cooling DM-gas temperature during this same period. This scenario thus leads to a prediction of a strong absorption signal that begins at $5$~MHz and extends up to the epoch of cosmic dawn, where it is expected to rise due to a combination of $X$-ray heating and Ly-$\alpha$ coupling.

In the intermediate coupling scenario, the kinetic energy transfer rate starts off as cosmologically relevant near recombination, but becomes less relevant at lower redshifts. Once again this scenario leads to a strong broadband signal, but in this scenario, the absorption signal begins at lower frequencies than 5~MHz, with a minima at frequencies well before those that correspond to the epoch of cosmic dawn.

In the weak coupling scenario, the kinetic energy transfer rate is completely irrelevant over the cosmological history. In this scenario, the spin-temperature couples to the adiabatically cooling DM temperature and thus leads to very strong absorption signals at high redshifts ($z\sim1000$ or $\nu = 1.4$~MHz). As the coupling $x_D$ becomes weaker than the coupling to the CMB temperature, $\delta T_b$ rises, leading to a minimum in the absorption signal deep in the cosmic dark ages.

At high frequencies, in both the intermediate and weak coupling scenarios, depending upon the history of $X$-ray heating of the gas and Ly-$\alpha$ coupling induced by the first stars, $\delta T_b$ is expected to either rise rapidly, or a second absorption dip might possibly be seen before the rise.

Our calculation of the DM induced hyperfine transition rate showed that the relevant cross-section scales as $1/\Delta^2$, where $\Delta$ is the hyperfine splitting. This allows for a large transition rate even for relatively small couplings between the DM and electrons.
We explored several constraints on these couplings from terrestrial experiments and astrophysical and cosmological probes. While collider and stellar cooling constraints are strong and would naively rule out the regions of parameter space that we have explored, these constraints may possibly be evaded under extensions of our effective field theory. The more robust constraints demand that the DM particle responsible for spin-flip interactions is asymmetric, and makes up a fraction $f\lesssim0.1$ of the total DM relic density. We also found that only the weak coupling scenario is viable for $f=0.1$, based upon CMB constraints of kinetic coupling of the DM to the plasma at recombination. However, for smaller values of $f = 0.01, 0.001$, we found that the strong/weak/intermediate couplings are all viable.

Besides the global 21 cm absorption signal, there are several secondary signatures that can be tested in both cosmic and laboratory settings. We list these secondary tests below:

Astrophysical and cosmological probes:
\begin{itemize}
\item While we have not made detailed predictions of the stochastic 21~cm signal, it would be interesting to see the differences between the power spectrum of 21~cm fluctuations predicted in our model and standard cosmological models. The anisotropy signal could be tested by future experiments such as SKA or even more futuristic space based anisotropy measurements.

\item
We have seen in the weak and intermediate coupling scenarios that the gas spin temperature and kinetic temperature may have different evolutions. While the 21~cm absorption signal would probe the spin temperature, if we had an independent probe of the gas kinetic temperature we could measure the deviation between $T_s$ and $T_K$, this would be an additional test of the spin-flip mechanism that we have proposed. The challenge would be to find a probe that is independently sensitive to the gas kinetic temperature at a redshift before reionization. One example of such a probe could be a measurement of the pressure-smoothing scale~\cite{2015ApJ...812...30K} which is sensitive to the integrated thermal history of the intergalactic medium (IGM)~\cite{2017Sci...356..418R}. However, this particular probe is mostly sensitive to the low redshift thermal history of the IGM post-reionization.
\end{itemize}

Particle physics probes:
\begin{itemize}
\item Since the axial-vector $V$ couples to an anomalous current, generically UV extensions of our model that are anomaly free would lead to decay $Z\rightarrow \gamma V$. In order to evade the strong constraints on such anomalous $Z$ decays, one would need to build a UV model with broken SM gauge symmetries. Such models could lead to interesting testable predictions at collider experiments.
\item In the strong coupling regime, the mediator mass is tightly constrained to be around 1~eV in order to evade constraints from laboratory searches for spin-dependent interactions between electrons. However, increasing the sensitivity of these laboratory searches by an order of magnitude could rule out the strong/intermediate coupling scenarios proposed in our work, or potentially discover the mediator particle.
\end{itemize}

The strong broadband signal that we have proposed in this work has exciting implications for global 21~cm signal experiments. Searching for this signal would require experiments to change their search strategies in order to discover the signal as compared to the typical band-limited search strategies motivated by standard cosmology. Specific examples of such changes would be use of alternative templates to extract the cosmological signal from the foreground dominated map, and comparing the extracted signal across experiments probing different regions of the radio spectrum. We have also suggested several secondary tests that could validate the particle physics origin of such a cosmological signal. We leave a more detailed exploration of these tests to future work.

\section*{Acknowledgments}
We thank Girish Kulkarni and Shikhar Mittal for helpful comments on the draft. We also acknowledge useful discussions with Varun Bhalerao, Subhendra Mohanty, Surhud More, Nadav Joseph Outmezguine, Arun Thalapillil, and Himanshu Verma. MD would like to acknowledge support through Inspire Faculty Fellowship of the Department of Science and Technology (DST), Government of India under the Grant Agreement number: IFA18-PH215. VR is supported by a DST-SERB Early Career Research Award (ECR/2017/000040) and an IITB-IRCC seed grant. VR would like to express a special thanks to the GGI Institute for Theoretical Physics for its hospitality and support.

\appendix

\section{Rate for excitation of hydrogen from the singlet to triplet state via dark matter scattering}
\label{sec:appendix}
In this appendix we will work out the rate for excitation and de-excitation of neutral hydrogen from the ground state ($H_{0}$) to the excited state ($H_{1}$), i.e. for the process
\begin{align}
\chi+H_{0}\rightleftarrows \chi +H_{1}.
\end{align}
In appendix~\ref{sec:appendixa}, we will work out the amplitude for excitation and de-excitation including the details of the bound state wave function of the electron. In appendix~\ref{sec:appendixb}, we use these amplitudes to compute the reaction rates for the forward and backward spin-flip reactions $D_{01}$ and $D_{10}$. In appendix~\ref{sec:appendixc}, we use these amplitudes to compute the energy transfer rate or equivalently the temperature equilibration time scale between the dark matter and gas.
\subsection{Amplitude for excitation and de-excitation}
\label{sec:appendixa}
Let us first consider the excitation process with specific spin states,
\begin{align}
\chi_{s}+H_{0}\rightarrow \chi_{{s}^{\prime}} +H_{1, {s^{\prime}_H}},
\end{align}
where $s$ and $s^\prime$ denote the spin state of the dark matter particle $\chi$, and $s^\prime_H$ denotes the spin state of the final state triplet.
\subsubsection{Bound state wave functions}
We first express the bound state wave function of a hydrogen atom in either the singlet or triplet states moving with velocity $\vec{v}$ in terms of free proton and electron states as,
\begin{align}
\vert H_{0}(\vec{v})\rangle=\sqrt{\frac{2m_{H}}{2m_{e}2m_{p}}}\int\frac{d^{3}k}{(2\pi)^{3}}\tilde{\psi}_{1s}(\vec{k})\frac{1}{\sqrt{2}} [\vert p(-\vec{k}+m_{p}\vec{v},\downarrow)\rangle \otimes\vert e(\vec{k}+m_{e}\vec{v},\uparrow)\rangle  \nonumber \\
-\vert p(-\vec{k}+m_{p}\vec{v},\uparrow)\rangle \otimes \vert e(\vec{k}+m_{e}\vec{v},\downarrow)\rangle ],
\end{align}
for the singlet state, and
\begin{align}
\vert H_{1}(\vec{v})\rangle=\sqrt{\frac{2m_{H}}{2m_{e}2m_{p}}}\int\frac{d^{3}k ^{}}{(2\pi)^{3}}\tilde{\psi}_{1s}(\vec{k} ^{})\frac{1}{\sqrt{2}}[\vert p(-\vec{k}^{}+m_{p}\vec{v},\downarrow)\rangle\otimes \vert e(\vec{k} ^{}+m_{e}\vec{v},\uparrow)\rangle  \nonumber  \\
+\vert p(-\vec{k} ^{}+m_{p}\vec{v},\uparrow)\rangle\otimes \vert e(\vec{k} ^{}+m_{e}\vec{v},\downarrow)\rangle],
\end{align}
for the triplet state which has spin $s^\prime_H=0$ along a chosen spin quantization $z$-axis. Note that the choice of spin quantization axis here is arbitrary. One can similarly write down the $s^\prime_H=\pm1$ states for the triplet by suitable choice of the proton and electron spins ($\uparrow \uparrow$ for the $s^\prime_H=1$  state and $\downarrow \downarrow$ for the $s^\prime_H=-1$ state). In these expressions, $m_H$, $m_e$ and $m_p$ denote the masses of hydrogen, the electron, and proton, respectively. $\vec{k}$ denotes the relative internal momentum between the proton and electron. We have neglected the mass difference of the singlet and triplet states in the normalization of the wave-functions.
The Fourier transform of the hydrogen $1s$ state wave-function is given by,
\begin{equation}
    \widetilde{\psi}_{1s}(\Vec{k}) =\frac{ 8 \sqrt{\pi} }{a_{0}^{5/2}} \frac{1}{\left( k^{2}+ \frac{1}{a^2_{0}} \right)^{2}},
\end{equation}
where $a_0$ is the Bohr radius.
The free particle $p$ and $e$ states have the standard relativistic normalization. For example, for the proton state we have,
\begin{align}
\langle p(q_{1},s_{1})\vert p(q_{0},s_{0})\rangle=2m_{p}\delta_{s_{1}s_{0}}(2\pi)^{3}\delta^{3}(\vec{q}_{1}-\vec{q}_{0}),
\end{align}
where $\vec{q}$ is the momentum of the proton state and $s_0$ ($s_1$) denotes the $z$ component of spin of the initial (final) state.
\subsubsection{Amplitude for excitation}
We take the reaction,
\begin{align}
\chi_{s}+H_{0}\rightarrow \chi_{{s} ^{\prime}} +H_{1}.
\end{align}
and assign a velocity $\vec{v}_{0}$ and $\vec{v}_{1}$ to the singlet and triplet hydrogen atoms, respectively.

In order to evaluate the amplitude for this process, we first need to find the amplitude for the free particle process
\begin{align}
\chi_{s}+e^-_{s_e} \rightarrow \chi_{{s} ^{\prime}} +e^-_{s^{\prime}_e},
\end{align}
which occurs through the $t$-channel exchange of the light mediator, and then insert it between the bound state wave functions discussed previously. Here, $s_e$ and $s^{\prime}_e$ denote the spin state of the initial and final state electrons.
Using the Lagrangian in eq.~\ref{eq:lagrangian}, this amplitude can be evaluated at tree-level as
\begin{align}
\mathcal{M}= &\frac{g_{\chi}g_{e}}{(q^{2}-m_{V}^{2})}\bar{u}(p_{\chi ^{\prime}}, {s} ^{\prime})\gamma^{\mu}\gamma^{5}u(p_{\chi}, s)\bar{u}(p ^{\prime}_{e}, s^{\prime}_e)\gamma_{\mu}\gamma^{5}u(p_{e}, s_e)
\label{eq:amplitude}
\end{align}

We have assigned 4-momenta $p_{\chi}$ ($p^\prime_{\chi}$) and $p_e$ ($p^\prime_{e}$) to the dark matter and electron, respectively for the initial state (final state). We have also defined the momentum transfer $q$ as,
\begin{equation}
q^2 \equiv (p ^{\prime}_{\chi}-p_{\chi})^2= (p_{e}-p ^{\prime}_{e})^2. % \simeq -m_H^2 (\vec{v}_{\mathrm{in}}-\vec{v}_{\mathrm{out}})^2,
\end{equation}
In the non-relativistic limit we can show that the amplitude eq.~\ref{eq:amplitude} can be written as~\cite{DelNobile:2013sia},
\begin{align}
\mathcal{M} \simeq - 16 \frac{g_{\chi}g_{e}}{(q^{2}-m_{V}^{2})} m_\chi m_e \langle {s} ^{\prime}, s^{\prime}_e\vert \vec{S}_\chi \cdot \vec{S}_e\vert   {s} , s_e\rangle,
\end{align}
where $\vec{S} = \frac{1}{2}\vec{\sigma}$ denotes the spin projection operator and $\vec{\sigma}$ are Pauli matrices.
We can further write
\begin{align}
 \vec{S}_{\chi}.\vec{S}_{e}=\frac{1}{4}\left(2({\sigma^{+}_{\chi}\sigma^{-}_{e}}+\sigma^{-}_{\chi}\sigma^{+}_{e})+\sigma^z_{\chi}\sigma^z_{e} \right),
\end{align}
where $\sigma^z$ is the diagonal Pauli matrix, and we have used the spin raising and lowering operators written in terms of the Pauli matrices $\sigma^{\pm}=\frac{1}{2}(\sigma^{x}\pm i \sigma^{y})$.
Thus, the matrix element in the scattering amplitude can be evaluated as,
\begin{align}
\langle {s} ^{\prime}, s^{\prime}_e\vert \vec{S}_\chi \cdot \vec{S}_e\vert   {s} , s_e\rangle =
 \begin{blockarray}{*{4}{c} l}
    \begin{block}{*{4}{>{$\footnotesize}c<{$}} l}
     $\uparrow \uparrow$ & $\uparrow \downarrow$ & $\downarrow \uparrow$ & $\downarrow \downarrow$ & \\
    \end{block}
    \begin{block}{(*{4}{c})>{$\footnotesize}l<{$}}
 \bigstrut[tb] \, \,\,\frac{1}{4} & \, \, \, 0 & \, \, \, 0 & 0 \bigstrut[tb]& $\uparrow \uparrow$ \\
  0 & -\frac{1}{4} & \, \, \,\frac{1}{2} & 0 & $\uparrow \downarrow$  \\
  0 & \, \, \, \frac{1}{2} & -\frac{1}{4} & 0  & $\downarrow \uparrow$ \\
 \bigstrut[tb] \, \,\, \,0 & \, \, \, 0 & \, \, \, 0 & \frac{1}{4} \bigstrut[tb] & $\downarrow \downarrow$ \\
    \end{block}
  \end{blockarray},
\end{align}
where the column (row) labels refer to the spin of the outgoing (incoming) dark matter and electron respectively.
Using the $e \chi \rightarrow e \chi$ scattering amplitude, we can now evaluate the hydrogen excitation amplitude by using the bound state wave functions. In the non-relativistic limit we can identify the 4-vector momentum transfer as $q \simeq \left (0 , m_H (\vec{v}_0- \vec{v}_1) \right ) \equiv (0,\vec{q} \,) $.

We can then write the resulting amplitudes for all spin combinations in terms of the following master amplitude which we denote as $\mathcal{M}_0$,
\begin{equation}
\mathcal{M}_0 = 4\sqrt{2} g_\chi g_e \frac{m_H m_\chi}{q^2-m_V^2 } F(q^2)
\end{equation}
where $F(q^2)$ is the hydrogen atom form factor,
\begin{align}
F(q^2) &= \int\frac{d^{3}k}{(2\pi)^{3}}\tilde{\psi}_{1s}(\vec{k})\tilde{\psi}_{1s}^{\ast}(\vec{k}+\vec{q}), \\
           & = \left(\frac{1}{1+\frac{q^2 a_0^2}{4}}\right )^2,\\
           &\rightarrow 1 \textrm{\, for \,} q^2 \ll \frac{1}{a_0^2}.
\end{align}
The full amplitudes are non-zero only for the following spin combinations,
\begin{align}
\mathcal{M}(\chi (s=\uparrow)+H_{0}\rightarrow \chi(s^\prime=\downarrow)+H_{1}(s ^{\prime}_H=1)) & =    \mathcal{M}_0, \\
\mathcal{M}(\chi (s=\downarrow)+H_{0}\rightarrow \chi(s^\prime=\uparrow)+H_{1}(s ^{\prime}_H=-1)) &=  - \mathcal{M}_0, \\
\mathcal{M}(\chi (s=\uparrow)+H_{0}\rightarrow \chi(s^\prime=\uparrow)+H_{1}(s ^{\prime}_H=0)) &=  -\frac{1}{\sqrt{2}}  \mathcal{M}_0, \\
\mathcal{M}(\chi (s=\downarrow)+H_{0}\rightarrow \chi(s^\prime=\downarrow)+H_{1}(s ^{\prime}_H=0)) &=  \frac{1}{\sqrt{2}}  \mathcal{M}_0,
\end{align}
where $s ^{\prime}_H$ denotes the $z$ component of the spin of the final state triplet.
Squaring the amplitude and taking a spin-sum over final states and averaging over initial states we get,
\begin{align}
\frac{1}{(2 S_\chi+1)(2 S_0+1)} \sum\limits_{\{\textrm{spins} \}} \vert \mathcal{M} \vert^2 = \frac{3}{2} \vert \mathcal{M}_0 \vert^2,
\label{eq:ex_amp_squared}
\end{align}
where $S_\chi=1/2$, $S_0=0$ are the spins of $\chi$ and $H_0$ respectively.
\subsubsection{Amplitude for de-excitation}
For the de-excitation process, we can similarly show that the spin-summed and averaged amplitude is given by
\begin{align}
\frac{1}{(2 S_\chi+1)(2 S_1+1)} \sum\limits_{\{\textrm{spins} \}} \vert \mathcal{M} \vert^2 = \frac{1}{2} \vert \mathcal{M}_0 \vert^2,
\end{align}
where  $S_1=1$ is the spin of the triplet $H_1$.

\subsection{Excitation and de-excitation rates}
\label{sec:appendixb}
We will now use the amplitudes that we have computed to work out the excitation and de-excitation rates.
The rate for excitations $\chi+H_{0}\rightarrow \chi +H_{1}$ is given by,
\begin{align}
D_{01}  \equiv & \,   n_\chi \langle  \sigma_{01} v_{\textrm{rel}}\rangle  \nonumber \\
 = & \,   n_\chi  (2\pi)^6 \int \frac{d^3p_{i}}{(2\pi)^3 2 E_{i}} \frac{d^3p_{0}}{(2\pi)^3 2 E_{0}} \frac{d^3p_{f}}{(2\pi)^3 2 E_{f}}  \frac{d^3p_{1}}{(2\pi)^3 2 E_{1}} f(p_{i}) f(p_0)
 \nonumber \\
& \,  \times \frac{1}{(2 S_\chi + 1)(2S_0 + 1)} \sum\limits_{\{\textrm{spins} \}} \vert \mathcal{M} \vert^2 (2\pi)^4 \delta^{(4)}(p_i + p_0 - p_f - p_1),
\label{eq:excitationrate}
\end{align}
where we have taken $p_i$ and $p_0$ to be the momenta of the initial dark matter particle and singlet $H_0$, respectively and $p_f$ and $p_1$ to be the momenta of the final state dark matter particle and triplet $H_1$. The distribution functions for the initial particles are taken to be of the form $f(p_{i}) = \frac{1}{(2\pi m_\chi T_\chi)^{3/2}} e^{-\frac{p_i^2}{2 m_\chi T_\chi}}$ and $f(p_{0}) = \frac{1}{(2\pi m_H T_K)^{3/2}} e^{-\frac{p_0^2}{2 m_H T_K}}$, where $T_\chi$ is the temperature of the dark matter and $T_K$ the gas temperature.
We can now change integration variables for the initial state particles by defining the relative momenta $\vec{p}$ and a conjugate momentum variable $\vec{p}_{m}$,
\begin{align}
\vec{p} \equiv \mu \vec{v}_{\textrm{rel}}  = \mu \left( \frac{\vec{p}_0}{m_H} - \frac{\vec{p}_i}{m_\chi} \right ), \\
\vec{p}_{m} = M \left (\frac{\frac{\vec{p}_0}{T_K} + \frac{\vec{p}_i}{T_\chi}}{\frac{m_H}{T_K} + \frac{m_\chi}{T_\chi} }   \right ),
\end{align}
 where $\vec{v}_{\textrm{rel}}$ is the relative velocity between the initial state particles, $ \mu = m_H m_\chi /(m_H + m_\chi)$ is the reduced mass of dark matter and hydrogen and $M = m_H +m_\chi$ is their mass sum. Then we can replace,
\begin{align}
\int d^3 p_{i} d^3 p_{0}  f(p_{i}) f(p_0) \rightarrow  \int  d^3 p d^3 p_{m}  f(p) f({p}_{m})
\end{align}
where the effective distribution functions for $p$ and $p_m$ are given respectively by,
\begin{align}
f(p) =  \frac{1}{(2\pi \mu \Teff)^{3/2}} e^{-\frac{p^2}{2 \mu \Teff}}, \textrm{\, and} \\
f(p_m) =   \frac{1}{(2\pi M T_m)^{3/2}} e^{-\frac{p_{m}^2}{2 M T_m}},
\end{align}
 where we have defined $\Teff = \mu  \left (\frac{T_\chi}{m_\chi} + \frac{T_K}{m_H} \right)$ and $T_m = \frac{T_K T_\chi}{\Teff}$.
Making this substitution in eq.~\ref{eq:excitationrate} and using the fact that the excitation cross-section only depends on the relative momentum $p$ and not on $p_m$, we can trivially perform the integration over $p_m$ and write
\begin{align}
D_{01}  =  n_\chi \langle \sigma_{01} v_{\textrm{rel}} \rangle = n_\chi \int  d^3p  f(p) \sigma_{01} v_{\textrm{rel}}.
\label{eq:exrate}
\end{align}
We can evaluate the cross-section in the center-of-momentum frame (COM) and then identify $p$ as the magnitude of incoming momentum of either particle in this frame. In terms of the COM frame scattering angle $\theta$, one can evaluate the cross-section as,
\begin{align}
\sigma_{01} v_{\textrm{rel}} = \frac{1}{2 E_i}\frac{1}{2 E_0}\frac{1}{16 \pi} \int d(\cos\theta) \frac{2 p^\prime}{E_i + E_0} \frac{1}{2} \sum\limits_{\{\textrm{spins} \}} \vert \mathcal{M} \vert^2,
\end{align}
where the final state momentum in the COM frame is $p^\prime \simeq \sqrt{p^2 - p_{\textrm{th}}^2}$  and  $p_{\textrm{th}} = \sqrt{2 \Delta \mu}$ is the excitation threshold momentum, with $\Delta$ being the energy splitting between the singlet and triplet states.
The momentum transfer $q^2$ can be written as $q^2 \simeq -p^2 -  p^{\prime 2} + 2 p p^\prime \cos \theta$. Substituting our expression for the amplitude-squared worked out in the previous sub-section eq.~\ref{eq:ex_amp_squared}, we obtain
\begin{align}
\sigma_{01} v_{\textrm{rel}} \simeq \frac{3}{2 \pi} g_\chi^2 g_e^2\mu p^\prime  \int_{-1}^{1}  d(\cos\theta) \left (\frac{1}{q^2-m_V^2} \right)^2 F^2(q^2).
\end{align}
The scattering process is dominated by forward scattering which has a low momentum transfer and thus we can approximate $q^2 \ll 1/a_0^2$, and we can then take the hydrogen form factor $F(q^2) \simeq 1$. The angular integral then simplifies to,

\begin{align}
\mathcal{I} & = \int_{-1}^{1}  d(\cos\theta) \left (\frac{1}{-p^2 -  p^{\prime 2}  - m_V^2 + 2 p p^\prime \cos \theta } \right)^2 ,\\
 &= \frac{2}{( p_{\textrm{th}}^2 - m_V^2)^2 + 4 m_V^2 p^2} \, ,
 \label{eq:spinflipcrossintegrand}
\end{align}
where we have also substituted for $p^\prime$ in terms of $p$ and $p_{\textrm{th}}$ in the last line.
Note that the angular integration would be singular in the limit that $m_V \rightarrow 0$ and  $p_{\textrm{th}}\rightarrow 0$. This is due to the usual forward elastic-scattering singularity with a massless mediator. The divergence in the angular integration is cut-off by both the finite mediator mass, as well as by the threshold momentum of the inelastic reaction.

We choose to work in the limit $m_V \rightarrow 0$, where the divergence in the angular integral is dominantly cut-off by the threshold momentum for excitations rather than the mediator mass. To be precise, we need to work in the limit $m_V^2 \ll  p_{\textrm{th}}^4 /p_c^2$, where $p_c^2= 2\mu \Teff$ is the characteristic relative momentum in the thermal distribution\footnote{This condition is more stringent than the condition $m_V^2 \ll  p_{\textrm{th}}^2$ which allows us to neglect the mediator mass in the first term in brackets in the denominator of eq.~\ref{eq:spinflipcrossintegrand}, but not in the second.}. This condition can be expressed as,

\begin{align}
m_V \ll \sqrt{\frac{2\mu}{\Teff}} \Delta \simeq 2.3~\textrm{eV} \sqrt{\left(\frac{1000~\textrm{K}}{\Teff}\right)\left( \frac{\mu}{\textrm{0.1~GeV}}\right)}.
\label{eq:medmass_upper}
\end{align}
In this limit, we obtain the excitation cross-section as,
\begin{align}
\sigma_{01} v_{\textrm{rel}} \simeq \frac{3}{4\pi} \frac{g_\chi^2 g_e^2}{\Delta^2} \frac{p^\prime}{\mu}.
\end{align}
Upon performing thermal averaging, we find
\begin{align}
\langle \sigma_{01} v_{\textrm{rel}} \rangle  &\simeq \int  d^3p  f(p) \sigma_{01} v_{\textrm{rel}}, \\
 &= \frac{3}{4\pi} \frac{g_\chi^2 g_e^2}{\Delta^2} \sqrt{\frac{8 \Teff}{\pi \mu}} e^{-\frac{\Delta}{\Teff}}.
\end{align}
Here,  $\langle v_{\textrm{rel}} \rangle = \sqrt{\frac{8 \Teff}{\pi \mu}}$ is the thermal average relative velocity and the exponential suppression factor is due to thermal suppression of the excitation reaction.
Similarly, for the de-excitation cross-section in the limit of negligible mediator mass we obtain,
\begin{align}
\langle  \sigma_{10} v_{\textrm{rel}} \rangle  \simeq  \frac{1}{4\pi} \frac{g_\chi^2 g_e^2}{\Delta^2} \sqrt{\frac{8 \Teff}{\pi \mu}}.
\end{align}
Note the key differences of absence of a factor of 3 (due to fewer final states) and absence of an exponential suppression factor (due to lack of a threshold energy for de-excitation).
The excitation and de-excitation rates $D_{01}$ and $D_{10}$ are found by simply multiplying these thermal cross-sections with the number density of dark matter particles at the relevant red-shift. The ratio of rates is given by,
\begin{align}
\frac{D_{01}}{D_{10}} = 3 e^{-\frac{\Delta}{\Teff}}.
\end{align}

The characteristic low momentum DM spin-flip interaction cross-section is given by,
\begin{align}
\sigma_{01}  \simeq 3 \sigma_{10}  &\simeq  12\pi    \frac{\alpha_\chi \alpha_e}{\Delta^2} \\
             &\simeq  4.2\times 10^{-14}\left( \frac{\alpha_\chi}{10^{-2}} \right) \left(\frac{\alpha_e}{10^{-14}}\right) ~\textrm{cm}^2 \simeq 4.2 \times10^{10}\left( \frac{\alpha_\chi}{10^{-2}} \right) \left(\frac{\alpha_e}{10^{-14}}\right)~\textrm{barns},
\end{align}
where we have rewritten the coupling constants in terms of $\alpha_\chi = g_\chi^2/4\pi$ and $\alpha_e = g_e^2/4\pi$. The typical relative velocity between the DM and hydrogen is given by  $\langle v_{\textrm{rel}} \rangle =  \sqrt{8\Teff/\pi\mu } \simeq 1.4~\textrm{km/s}\sqrt{\left(\frac{\Teff}{10~\textrm{K}} \right) \left ( \frac{0.1 \textrm{~GeV}}{\mu}\right )}$.

These cross-sections are huge by particle physics standards, and this is all the more surprising given the benchmark couplings we have assume are so weak. The reason for the large cross-section is easy to understand however: \textbf{\textit{Since the mass splitting between the singlet and triplet states is tiny, the t-channel scattering of dark matter with these states is almost elastic and therefore has a nearly divergent scattering cross-section driven by the large probability for forward scattering. This divergence is cut-off by the tiny hyperfine mass-splitting between the singlet and triplet states and leads to a large cross-section for the spin-flip interaction.}}

The number density of dark matter at a red-shift $z$ can be written as\footnote{Here, for simplicity we are considering the homogenous dark matter distribution only. At low redshifts there will be a boost to the interaction rate (which can be parameterized as an effective scaling of the number density of dark matter) due to the formation of halos.}
\begin{align}
n_\chi  &=  \frac{f \Omega_{\textrm{DM}} \rho_c}{m_\chi} (1+z)^3, \\
             &\simeq 1.5 \times 10^{-3} \left( \frac{f}{0.1} \right) \left( \frac{0.1~\textrm{GeV}}{m_\chi} \right) \left( \frac{1+z}{1+10} \right)^3~\textrm{cm}^{-3}.
\end{align}
where $\rho_c = 3.63 \times 10^{-47}$~GeV$^4$ is the critical density and $\Omega_{\textrm{DM}}  = 0.26$ is the present day DM relic density fraction and we have allowed for the possibility that the species we are considering makes up only a fraction $f$ of the dark matter.
Thus, we can write the excitation and de-excitation rates as,
\begin{align}
\label{eq:excitationanddeexrate}
D_{10}  &\simeq \frac{1}{3}D_{01}, \nonumber \\
& = n_\chi 4 \pi \frac{\alpha_\chi \alpha_e}{\Delta^2} \sqrt{\frac{8 \Teff}{\pi \mu}} \nonumber \\
&= 3.01\times 10^{-12}\left( \frac{f}{0.1} \right)  \left( \frac{0.1~\textrm{GeV}}{m_\chi} \right)  \left( \frac{\alpha_\chi}{10^{-2}} \right) \left(\frac{\alpha_e}{10^{-14}}\right)    \left(\frac{0.1~\textrm{GeV}}{\mu} \right)^{\frac{1}{2}} \left(\frac{\Teff}{10~\textrm{K}} \right)^{\frac{1}{2}} \left( \frac{1+z}{1+10} \right)^3~\textrm{s}^{-1}.
\end{align}

\subsection{Kinetic energy transfer rate}
\label{sec:appendixc}
The same spin-flip interactions that we had considered $\chi_{s}+H_{0}\leftrightarrows \chi_{{s}^{\prime}} +H_{1, {s^{\prime}_H}}$, can lead to kinetic energy transfer from the DM to neutral hydrogen and vice-versa. It is also possible to have kinetic energy transfer via the elastic reaction $\chi_{s} + H_{1} \rightarrow \chi_{{s}^{\prime} }+ H_{1}$\footnote{This reaction proceeds only through the $s^{\prime}_H=\pm1$ states of $H_1$ without spin-flip. Both the other possible elastic scattering processes, involving the $s^{\prime}_H=0$ state of $H_1$ and the process $\chi + H_0 \rightarrow \chi + H_0$, have vanishing amplitudes.}, however we will argue that this contribution is sub-dominant.

In this sub-section we will compute the kinetic energy transfer rate and the effect on temperature evolution of the dark matter and neutral hydrogen species.

The temperature evolution of a species can be computed from the second moment of the Boltzmann equation assuming that scattering does not distort the distribution far away from thermal. For the DM species for example, we can write down the temperature evolution equation as,
\begin{equation}
\frac{d T_\chi}{dt} = - 2 H T_\chi + \frac{2}{3} \dot{Q}_\chi,
\end{equation}
where $\dot{Q}_\chi$ is the dark matter heating rate due to kinetic energy transfer from the hotter hydrogen gas.
We can write an expression for $\dot{Q}_\chi$  as,
\begin{align}
\dot{Q}_\chi  &\equiv \dot{Q}^{01}_\chi + \dot{Q}^{10}_\chi,\\
                         &\equiv    n_0 R_{01}  +   n_1 R_{10} ,
\label{eq:energytransferrate}
\end{align}
where the energy transfer weighted rate coefficient for excitation is $R_{01}$ and for de-excitations is $R_{10}$.
Symbolically, we may express the rate coefficient for excitations as,
\begin{align}
R_{01} \sim \langle E_T \overline{\sigma}_{01} v_{\textrm{rel}}\rangle,
\end{align}
where $E_T$ is the energy transfer and $\overline{\sigma}_{01}$ is the ``energy-transfer'' cross-section.
A more precise definition is as follows,
\begin{align}
R_{01}    =& \,  (2\pi)^6\int \frac{d^3p_{i}}{(2\pi)^3 2 E_{i}} \frac{d^3p_{0}}{(2\pi)^3 2 E_{0}} \frac{d^3p_{f}}{(2\pi)^3 2 E_{f}}  \frac{d^3p_{1}}{(2\pi)^3 2 E_{1}} f(p_{i}) f(p_0) \times (E_f-E_i) \nonumber \\
 &\times \left( \frac{1}{(2 S_\chi + 1)(2S_0 + 1)} \sum\limits_{\{\textrm{spins} \}} \vert \mathcal{M} \vert^2  (2\pi)^4 \delta^{(4)}(p_i + p_0 - p_f - p_1) \right ).
\end{align}

A similar expression holds for the de-excitation energy transfer rate coefficient.
It is instructive to explicitly show how to evaluate the spin-excitation energy-transfer rate, which we will do for excitations and we will just state the result for de-excitations. We begin by making the change of variables from $p_0$, $p_i$ $\rightarrow$ $p$, $p_m$ as before.
Then, taking the relative momentum to be along the $z$-axis without loss-of-generality, and defining $\theta$ as the scattering angle in the COM frame, the energy transfer can be written as,
\begin{equation}
 (E_f-E_i) \simeq   \left(\frac{T_K - T_\chi}{\Teff} \right) \left(\frac{p^2}{M}\right) \left(  1 - \frac{p^\prime}{p} \cos \theta    \right) + \textrm{linear corrections in $p_m$}.
\end{equation}
Once again $p$ is the relative momentum in the COM frame of the incoming particles and $p^{\prime 2} \simeq p^2 - 2\mu \Delta$ is the outgoing momentum in the COM frame. The linear correction in $p_m$ to the energy transfer is unimportant since it does not contribute once we integrate over all possible directions of $p_m$.
The integration over $p_m$ can once again be performed trivially and we are thus left with an integral over relative momentum $p$,
\begin{align}
R_{01}  =  \int d^3p f(p) E_T \overline{\sigma}_{01} v_{\textrm{rel}}
\end{align}
where $f(p)=   \frac{1}{(2\pi \mu \Teff)^{3/2}} e^{-\frac{p^2}{2 \mu \Teff}} $ and we have schematically written the form of the integrand.
The energy transfer weighted rate before thermal averaging is given by,
\begin{align}
E_T \overline{\sigma}_{01} v_{\textrm{rel}} & \simeq  \left(\frac{T_K - T_\chi}{\Teff} \right)  \left(\frac{p^2}{M}\right)3 \frac{g_\chi^2 g_e^2}{2 \pi} \mu p^\prime  \int_{-1}^{1} d\cos\theta \frac{\left( 1-\frac{p^\prime}{p} \cos \theta \right ) }{\left (-p^2 - p^{\prime 2} -m_V^2 + 2 p p^{\prime} \cos \theta \right )^2} \\
& \simeq  \left(\frac{T_K - T_\chi}{\Teff} \right)  \left(\frac{p^2}{M}\right)3 \frac{g_\chi^2 g_e^2}{2 \pi} \mu p^\prime \left\{ \frac{1}{2\mu\Delta p^2}  + \textrm{Log}\left( \frac{4 p^4}{\mu^2 \Delta^2} \right)\right \}  \\
 &\rightarrow   \left(\frac{T_K - T_\chi}{2} \right)  \frac{3}{4 \pi }  \frac{g_\chi^2 g_e^2}{ \Delta^2}\left(\frac{2\Delta \mu}{M \Teff}\right)  \frac{p^\prime}{\mu}. \label{eq:inelasticrate}
\end{align}
In the second line above, we see that the forward scattering divergence of the integral is cut-off by the inelastic hyperfine splitting parameter $\Delta$. The divergence has two parts, the leading divergence scales as $1/\Delta$ and a sub-leading piece which scales as $\textrm{Log} \, \Delta$. In the last line above we have dropped the sub-leading contribution.

Note that in the second line in the equation above we have neglected the mediator mass which could also cut-off the forward divergence. This can be justified if we assume the same upper bound on the mediator mass that we had assumed as in eq.~\ref{eq:medmass_upper}, when calculating the spin-flip rate. However, for the elastic scattering process $\chi + H_1 \rightarrow \chi + H_1$, the forward scattering divergence is cut-off only by the mediator mass. In that case, the corresponding integrand only has a logarithmic divergence, scaling as $\textrm{Log} \, m_V$, i.e. there is no $1/m_V$ type divergence (see a derivation in  Appendix~\ref{sec:appendixd}). Thus, the elastic scattering energy transfer cross-section is suppressed relative to the inelastic cross-section that we have considered here, and hence we will ignore this contribution to the energy transfer.

Now performing the thermal averaging of the rate in eq.~\ref{eq:inelasticrate} we get,
\begin{align}
R_{01}  & \simeq \left(\frac{T_K - T_\chi}{2} \right) \left(\frac{2\Delta \mu}{M \Teff}\right) \frac{3}{4 \pi }  \frac{g_\chi^2 g_e^2}{ \Delta^2} \sqrt{\frac{8\Teff}{\pi \mu}} e^{-\frac{\Delta}{\Teff}}.
\label{eq:excitationenergytransfer}
\end{align}
We can similarly evaluate the rate for de-excitation reactions as,
\begin{align}
R_{10}  & \simeq \left(\frac{T_K - T_\chi}{2} \right) \left(\frac{2\Delta \mu}{M \Teff}\right) \frac{1}{4 \pi }  \frac{g_\chi^2 g_e^2}{ \Delta^2} \sqrt{\frac{8\Teff}{\pi \mu}}.
\end{align}
Combining the energy transfer rates from both excitations and de-excitations we get,
\begin{equation}
\dot{Q}_\chi =  \Gamma_\chi  \left(T_K - T_\chi \right),
\end{equation}
where we have defined $\Gamma_\chi$ as the characteristic energy transfer rate. This would correspond to the inverse time-scale to transfer an $\mathcal{O}(1)$ fraction of the baryon kinetic energy to the dark matter.

If we further make the approximation that $n_1 \simeq 3 n_0$ (for spin temperature $T_s \gg \Delta$) and using $n_H = n_0 + n_1$, we can write the energy transfer rate $\Gamma_\chi$ as,
\begin{align}
\label{eq:gammchianalytic}
\Gamma_\chi \simeq n_H \left(\frac{\Delta \mu}{2 M \Teff}\right) 12 \pi   \frac{\alpha_\chi \alpha_e}{ \Delta^2} \sqrt{\frac{8\Teff}{\pi \mu}}.
\end{align}
Upon comparing this with the expression for the rate $D_{01}$ for excitations (eq.~\ref{eq:excitationanddeexrate}) we see that the energy transfer rate is suppressed by a factor of $S\equiv\left(\frac{\Delta \mu}{2M \Teff}\right)  = 3.42\times 10^{-4} \left(\frac{10 \, \textrm{K}}{\Teff} \right ) \left(\frac{\mu}{0.1\textrm{~GeV}}\right) \left( \frac{1\textrm{~GeV}}{M} \right )$. This is because the energy transfer per collision is not of $\mathcal{O}(T_K - T_\chi)$, but rather  since the scattering process is dominantly forwards, the energy transfer is of the order of the mass splitting between the singlet and triplet states, i.e. it is $\mathcal{O}(\Delta)$. Thus, the timescale to transfer an $\mathcal{O}(1)$ fraction of the kinetic energy of the gas to the DM is longer than the interaction time scale by a factor of $\mathcal{O}(\Teff/\Delta)$.

Now we can use,
\begin{align}
n_H  &=  \frac{ \Omega_b \rho_c}{m_H} (1+z)^3, \\
             & =  3.0\times 10^{-4}  \left( \frac{1+z}{1+10} \right)^3~\textrm{cm}^{-3}.
\end{align}
Here, we have used the present-day baryon density fraction $\Omega_b = 0.05$ and we have assumed for simplicity that all the baryons are in the form of neutral hydrogen at the relevant red-shifts.
Thus, we can write an expression for the rate $\Gamma_\chi$ as,
\begin{align}
\Gamma_\chi &=  6.02 \times 10^{-16} \left( \frac{\alpha_\chi}{10^{-2}} \right) \left(\frac{\alpha_e}{10^{-14}}\right) \left( \frac{1~\textrm{GeV}}{M} \right) \left(\frac{\mu}{0.1~\textrm{GeV}} \right)^{\frac{1}{2}}  \left(\frac{10~\textrm{K}}{\Teff} \right)^{\frac{1}{2}}   \left( \frac{1+z}{1+10} \right)^3~\textrm{s}^{-1}.
\end{align}

We can also similarly work out the temperature evolution of the hydrogen kinetic temperature,
\begin{equation}
\frac{d T_K}{dt} = - 2 H T_\chi +\Gamma_c (\Tcmb-T_K) + \frac{2}{3} \dot{Q}_H,
\end{equation}
where the second term is the heating due to CMB and $\Gamma_c$ is the compton rate, which depends on the free electron fraction.
In the last term $\dot{Q}_H$ is the energy transfer rate from the dark matter fluid to the gas, which can be related to the heating rate of the DM $\dot{Q}_\chi$ as,
\begin{align}
\dot{Q}_H =& -\frac{n_\chi}{n_H} \dot{Q}_\chi, \\
                    =& \, \Gamma_H (T_\chi-T_K),
\end{align}
where we have defined the rate constant $\Gamma_H$ as,
\begin{align}
\Gamma_H =&\,  n_\chi \left(\frac{\Delta \mu}{2M \Teff}\right) \frac{3}{4 \pi }  \frac{g_\chi^2 g_e^2}{ \Delta^2} \sqrt{\frac{8\Teff}{\pi \mu}},  \nonumber \\
=& \, 3.11\times 10^{-15} \left( \frac{f}{0.1} \right) \left( \frac{0.1~\textrm{GeV}}{m_\chi}\right ) \left( \frac{\alpha_\chi}{10^{-2}} \right) \left(\frac{\alpha_e}{10^{-14}}\right)  \left( \frac{1~\textrm{GeV}}{M} \right) \left(\frac{\mu}{0.1~\textrm{GeV}} \right)^{\frac{1}{2}}  \left(\frac{10~\textrm{K}}{\Teff} \right)^{\frac{1}{2}}   \left( \frac{1+z}{1+10} \right)^3~\textrm{s}^{-1}.   \nonumber \\
\end{align}

\section{Elastic scattering energy-transfer cross-section}
\label{sec:appendixd}
We discuss here the forward scattering divergence of elastic scattering for processes such as $\chi + H_1 \rightarrow \chi + H_1$, $\chi + e^- \rightarrow \chi + e^-$ and $\chi + \chi \rightarrow \chi+ \chi$. The energy transfer cross-section for each of these processes can be defined as follows,
\begin{equation}
\overline{\sigma} = \int d\Omega \frac{d\sigma}{d\Omega} (1-\cos \theta),
\end{equation}
the integral over the polar scattering angle takes the form,
\begin{align}
\mathcal{I} & =  \int_{-1}^{1} d\cos\theta \frac{\left( 1- \cos \theta \right ) }{\left (1 -\cos \theta + \frac{m_V^2}{2 p^2} \right )^2} \\
& =\textrm{Log} \left( 1 +\frac{4p^2}{m_V^2}\right) -\frac{1}{1 +\frac{m_V^2}{4p^2}} \\
&\underset{m_V \to 0}{\longrightarrow} \textrm{Log} \left(\frac{4p^2}{m_V^2}\right),
\end{align}
where $p = \mu v$ is the relative momentum of the incoming particles, $\mu$ being the reduced mass and $v$ the relative velocity of the incoming particles. In the last expression we have taken the limit of small mediator mass, $m_V \rightarrow 0$. For elastic scattering, the forward divergence in the energy transfer cross-section is cut-off by the mediator mass and this gives rise to a logarithmic divergence. Similar expressions can be found in the literature, for example see ref.~\cite{Tulin:2013teo} for DM elastic scattering through a light mediator.

For completeness, we present here the various energy transfer cross-sections for elastic scattering processes.

The process $\chi + e^- \rightarrow \chi + e^-$ is important for determining the pre-recombination initial conditions on the DM temperature. The energy transfer cross-section for this process is given by,
\begin{align}
\overline{\sigma}(\chi + e^- \rightarrow \chi + e^-) &= \frac{3}{8 \pi} \frac{g_\chi^2 g_e^2}{\mu^2_{\chi e}v^4} \textrm{Log} \left( \frac{4 \mu^2_{\chi e}v^2 }{m_V^2}\right),
\end{align}
where $\mu_{\chi e}$ is the reduced mass of the DM and electron.

DM self-scattering $\chi+ \chi \rightarrow \chi +\chi$ is important when considering astrophysical self-interaction constraints. For this process the energy transfer cross-section is given by,
% \blue{this formula only takes into account t-channel but not u-channel}
\begin{align}
\overline{\sigma}(\chi + \chi \rightarrow \chi +\chi) &\sim \frac{3}{2\pi} \frac{g_\chi^4}{m^2_{\chi}v^4} \textrm{Log} \left( \frac{m^2_{\chi}v^2 }{m_V^2}\right).
\end{align}

Finally, for elastic $\chi+ H_1 \rightarrow  \chi + H_1$ scattering, which only proceeds through the $s^{\prime}_H=\pm1$ states of $H_1$ without flipping the spin, we have the energy transfer cross-section of the form,
\begin{align}
\overline{\sigma}(\chi + H_1 \rightarrow \chi + H_1) &= \frac{1}{24 \pi} \frac{g_\chi^2 g_e^2}{\mu^2v^4} \textrm{Log} \left( \frac{4 \mu^2v^2 }{m_V^2}\right),
\end{align}
where $\mu$ is the reduced mass of the DM and hydrogen.

%%%%%%%%%%%%%%%%%%%%%%%%%%%%%
\bibliography{broadband}
\end{document}